\documentclass[sigconf]{acmart}

\AtBeginDocument{%
  }

\copyrightyear{2025}
\acmYear{2025}
\setcopyright{cc}
\setcctype{by-nc}
\acmConference[CHI '25]{CHI Conference on Human Factors in Computing
Systems}{April 26-May 1, 2025}{Yokohama, Japan}
\acmBooktitle{CHI Conference on Human Factors in Computing Systems (CHI
'25), April 26-May 1, 2025, Yokohama,
Japan}\acmDOI{10.1145/3706598.3714106}
\acmISBN{979-8-4007-1394-1/25/04}

\usepackage{soul}

\begin{document}

\title{Briteller: Shining a Light on AI Recommendation for Children}


\author{Xiaofei Zhou}
\authornote{Both authors contributed equally to this research.}
\email{xzhou50@ur.rochester.edu}
\orcid{0000-0002-0802-0722}
\affiliation{%
  \institution{University of Rochester}
  \city{Rochester}
  \state{New York}
  \country{USA}
}
\author{Yi Zhang}
\authornotemark[1]
\email{yiz70@uci.edu}
\orcid{}
\affiliation{%
  \institution{University of California, Irvine}
  \city{Irvine}
  \state{California}
  \country{USA}
}


\author{Yufei Jiang}
\email{yjiang68@u.rochester.edu}
\orcid{}
\affiliation{%
 \institution{University of Rochester}
  \city{Rochester}
  \state{New York}
 \country{USA}}



\author{Yunfan Gong}
\email{ygong18@u.rochester.edu}
\orcid{}
\affiliation{%
 \institution{University of Rochester}
  \city{Rochester}
  \state{New York}
 \country{USA}}

\author{Alissa N. Antle}
\orcid{0000-0001-9427-8772}
\email{aantle@sfu.ca}
\affiliation{%
  \institution{Simon Fraser University}
  \city{Vancouver}
  \country{Canada}}

\author{Zhen Bai}
\email{zhen.bai@rochester.edu}
\orcid{0000-0002-3258-0228}
\affiliation{%
  \institution{University of Rochester}
  \city{Rochester}
  \state{New York}
  \country{USA}
}

\renewcommand{\shortauthors}{Zhou et al.}



\begin{abstract}
Understanding how AI recommendations work can help the younger generation become more informed and critical consumers of the vast amount of information they encounter daily. However, young learners with limited math and computing knowledge often find AI concepts too abstract.
To address this, we developed Briteller, a light-based recommendation system that makes learning tangible. By exploring and manipulating light beams, Briteller enables children to understand an AI recommender system's core algorithmic building block, the dot product, through hands-on interactions. Initial evaluations with ten middle school students demonstrated the effectiveness of this approach, using embodied metaphors, such as  "merging light" to represent addition. To overcome the limitations of the physical optical setup, we further explored how AR could embody multiplication, expand data vectors with more attributes, and enhance contextual understanding.
Our findings provide valuable insights for designing embodied and tangible learning experiences that make AI concepts more accessible to young learners.
\end{abstract}


\begin{teaserfigure}
  \includegraphics[width=\textwidth]{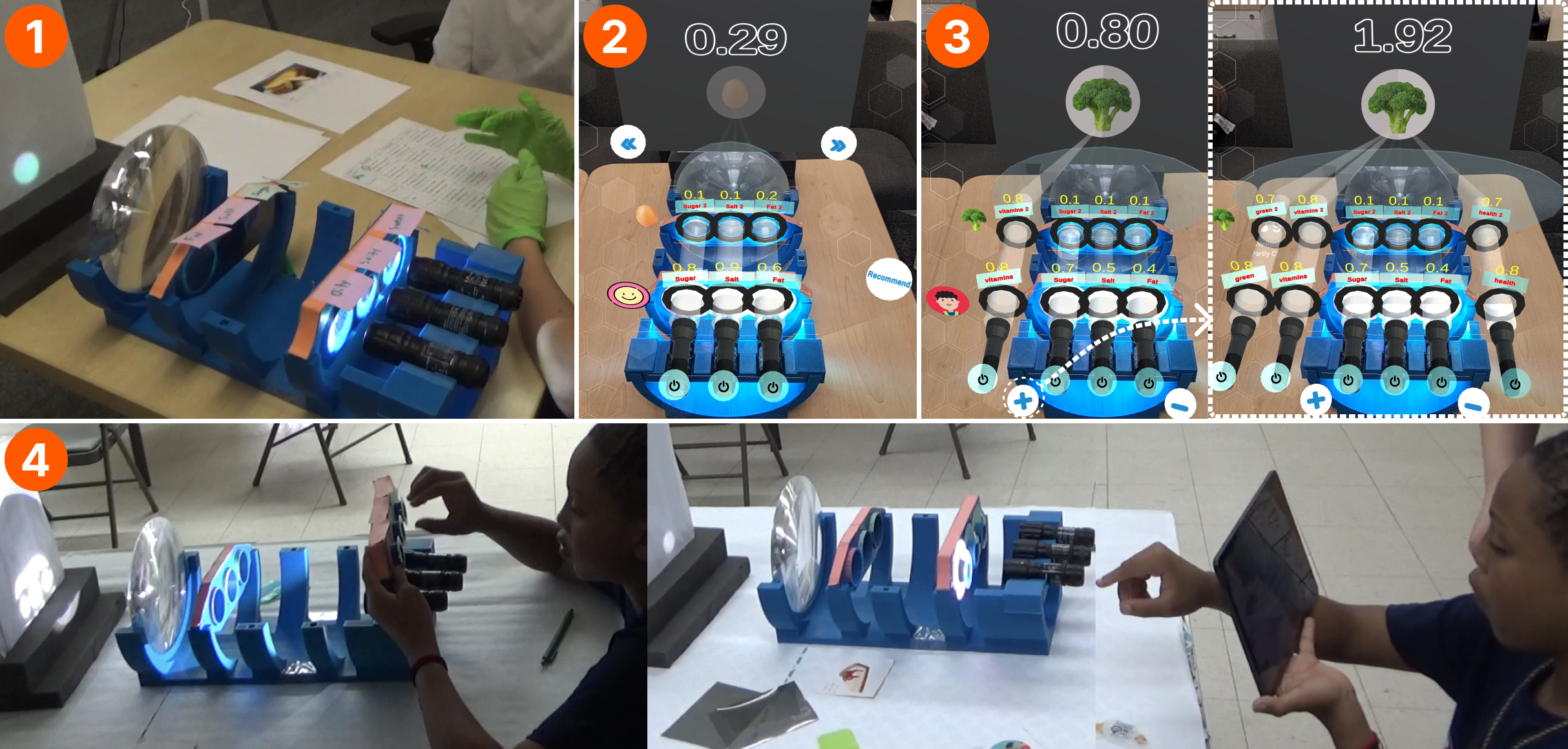}
  \caption{Comparison between Briteller and AR-enhanced Briteller: \textbf{(1)} a middle-school student interacted with Briteller; \textbf{(2)} item visibility that reflects the recommendation output, numeric value labels for individual data attributes, and visible light beams augmented in AR-enhanced Briteller; \textbf{(3)} the capability of editing and adding data attributes to user and item vectors in AR-enhanced Briteller; \textbf{(4)} a middle-school student interacted with AR-enhanced Briteller.}
  \Description{}
  \label{fig:teaser}
\end{teaserfigure}


\maketitle

\section{Introduction}
Artificial intelligence (AI) recommendation systems are ubiquitous in the digital world, delivering content such as videos on TikTok and YouTube or product recommendations on Amazon~\cite{lu2015recommender}.
Given children's underdeveloped critical thinking skills and impulsive behaviors, 
these AI systems significantly impact their opinions and behaviors~\cite{alshamrani2020detecting,lee2023fostering,wang2024chaitok}, while the inner workings generally remain mysterious. 
This raises concerns about children's data privacy, online well-being, bias, and narrowed perspectives~\cite{akgun2022artificial,lee2023fostering,reid2016children}.

However, increasing awareness of ethical issues in AI is insufficient. First, more technical details underlying AI algorithms could empower young users to take action in real-life problems about AI applications~\cite{zhou2024bee,gong2024approaching,long2020ai}, instead of feeling ``powerless''~\cite{schaper2023five}.
For instance, young learners still feel a lack of control over online datafication and privacy issues without sufficient algorithmic literacy~\cite{wang2022don,wang2023treat}.
Second, oversimplified mental models of AI, AI-related folk theories, and uninformed algorithm-driven decisions are common and hard to correct once formed~\cite{hitron2019can,lee2023fostering,alvarado2021exploring}. Therefore, young students need to develop more technical AI competencies.

Such technical AI literacy could be challenging for young learners, especially the underrepresented groups in STEM. 
Without sufficient math and computational knowledge, children tend to perceive concepts like data vectors and the mechanisms behind AI outputs as overly complex and intimidating~\cite{kahn2018ai}. 

To tackle the challenges above, Graspable AI~\cite{ghajargar2022graspable}, using physical forms or tangible interactions~\cite{ishii1997tangible,fitzmaurice1995bricks} as an explanation modality, could be a promising solution.
\citeauthor{alvarado2022towards} explored tangible movie recommender algorithms to enhance transparency and user control.
Studies on embodied cognition have shown that such physical experiences aid children's cognitive processes and information communication~\cite{lakoff2000mathematics,goldin2009gesture},
and may improve their understanding and memory retention~\cite{papert1991situating,black2012embodied}.
Tangible interfaces leverage the benefits of multiple human senses to bridge abstract AI concepts with concrete physical objects and movements~\cite{ghajargar2022graspable,ghajargar2021explainable}.
Initial research indicates that tangibility lifts hidden assumptions towards AI~\cite{murray2024metaphor}, helps adult users understand AI explanations, and encourages reflection on their trust in AI models~\cite{colley2023exploring,alvarado2022towards}. 
However, three research gaps remain. 
First, rarely are technical details revealed through tangible or graspable AI~\cite{murray2024metaphor}.
Second, more research efforts are needed to ground the design of tangible interfaces in the theory of embodied cognition and metaphors~\cite{li2022meta}.
Third, it is under-explored how tangible and graspable AI can support young users to grasp and control AI systems in an informed and effective manner~\cite{kim2024tangible,kaspersen2021machine}. 

We propose using light beams and other optical materials to make AI graspable. 
Through an iterative design process, we created Briteller, a novel content-based recommendation system.
Briteller utilizes image schemas---mental structures formed from recurring sensorimotor experiences~\cite{johnson2013body,antle2009body,hurtienne2011image}---related to light representation and manipulations. 
For instance, light intensity (DARK-BRIGHT) represents the value of a specific attribute of a data vector.
The transmission of light beams in Briteller (BLOCKAGE, MERGING) embodies the operations in the dot product.
The dot product is a core algebraic operation that takes two equal-length data vectors and returns a value representing the similarity between the data vectors.
It was chosen because it is a fundamental building block for various recommendation algorithms (e.g., content-based recommendation algorithm, collaborative filtering)~\cite{sun2022survey} and numerous AI applications (e.g., matrix factorization and neural networks)~\cite{rendle2020neural}.
In this design-based research, we investigated the research questions:
\begin{description}
 \item[RQ1] Can light serve as a new embodied learning medium for children to understand AI recommendation systems?
 \item [RQ2] What are the unique affordances and limitations of using light to support the understanding of AI recommendation mechanisms?
\end{description}

Ten middle school students interacted with Briteller, demonstrating the effectiveness of its tangible interface with six embodied metaphors in delivering key AI concepts. 
For example, eight students successfully controlled and debugged the recommendation system by manipulating data bars and light through contrastive learning. However, constraints are identified in (1) forming quantitative connections due to inaccuracies in perceiving light intensity, (2) transferring knowledge from light-based interactions to recommendation contexts~\cite{lindgren2022learning}, and (3) scaling to higher-dimensional feature spaces due to physical limitations.

To address the above constraints, we developed new features through tablet-based Augmented Reality (AR). Virtual elements augment the dot product operation, contextualize prediction output, and enable the scalability of data vectors, resulting in an AR-enhanced Briteller.
We invited ten students from underrepresented backgrounds to evaluate the AR-enhanced Briteller through hands-on interaction. 
Pre- and post-tests showed statistically significant learning gains across all target AI concepts. 
Findings indicate that students struggling with math may benefit from reasoning through tangible manipulation and embodied metaphors. 
New features effectively clarify how values change at different algorithmic steps, contextualize AI output, and deepen understanding of datafication by expanding to higher-dimensional feature spaces.
Future work can integrate the features using alternative methods beyond tablet-based AR.

In summary, our work presents a light-based tangible interface that reveals AI's inner workings to young learners with diverse backgrounds. With the two empirical studies and an iterative design process, our main contributions have three folds:
\begin{enumerate}
    \item a light-based recommendation system, Briteller, a new form of graspable AI for children to develop AI literacy;
    \item a list of benefits and limitations of light-based tangible interactions for delivering abstract AI and math concepts;
    \item \textcolor{black}{design implications to create graspable AI with embodied metaphors and tangible interactions for young users.}
\end{enumerate}

\section{Related Work}
\subsection{Revealing AI Recommendation Systems to Young Students}
Recent research highlights the importance of promoting AI literacy among the young generation, particularly those without a technical background~\cite{long2020ai,casal2023ai}. 
For K-12 students, AI literacy includes recognizing different stakeholders and end-users, understanding the broader societal implications of AI~\cite{ali2019constructionism,lee2022ai,touretzky2021artificial}, and enhancing their control over AI systems to address real-world challenges~\cite{schaper2023five}.

However, the opaqueness and complexity of AI systems stem from their reliance on abstract mathematical models, intricate algorithms, and vast amounts of data, which require a solid foundation in computer science and mathematics to comprehend~\cite{angelov2021explainable}. 
Many students, thereby are excluded from gaining access to powerful knowledge about AI~\cite{long2020ai}. 
Educators also face difficulties in designing learning experiences that make AI tangible and relatable for children, particularly because AI applications are both abstract and rapidly evolving.

Identifying applications relevant to young people's daily lives, such as recommendation systems used in e-shopping, e-learning, and social media, is one strategy to bridge this gap~\cite{lu2015recommender,wang2024chaitok}. 
Existing curricula and technologies primarily focus on introducing young students to ethical considerations related to AI recommendation systems. For example, \citeauthor{zhou2024bee} employed analogies and embodied metaphors to teach children about the loss of information diversity caused by similarity-based recommendations and how to diversify recommendations. Similarly, \citeauthor{lee2023fostering} fostered critical thinking by engaging youth in artistic, first-person experiences of AI's impacts. Other approaches include design activities where children redesign YouTube’s recommendation systems to be more ethical~\cite{dipaola2020decoding}, structured curricula and programs addressing AI ethics~\cite{garrett2020more,solyst2023investigating}, and co-design workshops encouraging young people to explore AI's effects on their lives while learning strategies for managing online datafication~\cite{wang2022don,wang2023treat}.

Our work aims to demystify the technical inner workings of AI recommendation systems for children without technical backgrounds by making foundational AI concepts accessible and tangible. By interacting with user vectors, item vectors, and dot product operations in a tangible content-based recommendation system~\cite{shu2018content}, students gain a clearer understanding of how these systems function.

\subsection{Embodied Learning for Computational Thinking and Math}
Piaget’s~\cite{piaget1976piaget} constructivism emphasizes the body as a major factor in learning. Building on this, Papert’s constructionist learning approach~\cite{papert1991situating} advocates for the use of tangible tools to foster knowledge acquisition through hands-on engagement and social interaction. Similarly, \citeauthor{lakoff2000mathematics} introduced the embodied mathematics theory, suggesting that our bodies, brains, and daily interactions with the world shape human concepts and reasoning, including mathematical concepts and mathematical reasoning.
Rooted in philosophy and cognitive science, embodied learning plays a significant role in education by allowing students to engage with learning material physically, thereby connecting abstract concepts to real-world experiences~\cite{papert1991situating, abrahamson2020future}.

Empirical evidence from numerous studies supports the positive impact of embodied learning on computational thinking and mathematics, with many comparative studies reporting greater learning gains in embodied learning conditions~\cite{georgiou2019embodied,de2024data,yannier2016adding}. In summary, research indicates that embodied interaction enhances learning in computational thinking and math by grounding abstract concepts through concrete representations~\cite{howison2011mathematical, chatain2022grasping, carbonneau2013meta}, reducing cognitive load~\cite{esteves2013physical,borner2015tangible,de2024data}, simplifying complex tasks~\cite{esteves2013physical,ma2016designing}, promoting creativity~\cite{im2021draw2code}, and encouraging self-exploration and self-regulated learning~\cite{howison2011mathematical,tancredi2022balance}. 

However, \citeauthor{li2022meta} noted that some studies overly focused on embodied interaction and ignored their metaphor meanings. Tangible interaction should integrate metaphors~\cite{markova2012tangible}, as these enhance learning by embedding intrinsic relationships into interactive processes~\cite{li2022meta}. \citeauthor{de2024data} found that interaction metaphors influence understanding, interpretation, memorability, and affective engagement with data. They both highlight the need for further research on designing and evaluating interaction metaphors for their cognitive and educational effects~\cite{de2024data,li2022meta}. 
Our research aims to extend and complement existing research by designing embodied metaphors using light. Towards that goal, we leverage image schemas~\cite{hurtienne2007image,hurtienne2015designing} as embodied metaphors to represent abstract knowledge and investigate how these metaphors support learners' reasoning and conceptualization in math and recommendation systems.

\subsection{Tangible Representation and Interaction of AI and AI Literacy}
Emerging research explores tangible interaction as a means to enhance understanding and interaction with AI systems. For example, \citeauthor{ghajargar2022graspable} introduced ``Graspable AI,'' a design space that seeks to make AI more understandable by making it tangible, drawing insights from Explainable AI (XAI), Data Physicalization, and Tangible User Interfaces (TUIs)~\cite{ghajargar2022graspable,ghajargar2021explainable,ghajargar2022making}. Similarly, \citeauthor{colley2022tangible} proposed the ``TangXAI'' framework, combining Hornecker and Buur’s tangible interaction framework~\cite{hornecker2006getting} with Belle and Papantonis’s four general approaches to explainability: simplified rule extraction, feature relevance, local explanations, and visual explanations~\cite{belle2021principles}. These frameworks suggest that the affordances of tangible forms can significantly enhance AI explainability~\cite{colley2022tangible,ghajargar2022graspable}. 

Although empirical studies and prototypes are limited, preliminary evidence shows that tangible interactions can improve user awareness and control over AI systems, enhance understanding of AI concepts, and increase motivation and engagement in learning AI~\cite{alvarado2022towards,colley2023exploring,kim2024tangible,kaspersen2021machine,long2023fostering}. For instance, \citeauthor{alvarado2022towards} introduced Recffy, a TUI designed to improve awareness and control in movie recommendation systems. Participants reported that physical interactions improved their sense of ownership and control. Other studies have explored TUIs in AI education for children, such as Any-Cubes for object recognition~\cite{scheidt2019any} and the ``Tangible AI'' curriculum using Microbit and Google Teachable Machine~\cite{kim2024tangible}, both of which significantly boosted student motivation, engagement, and comprehension of AI concepts.

However, a comprehensive understanding of how tangible interactions support AI learning remains lacking~\cite{ghajargar2022graspable, colley2022tangible}. Many tangible interfaces do not reveal the inner workings of AI systems, with some relying on metaphorical representations that provide only superficial understanding~\cite{alvarado2022towards, murray2024metaphor}. Additionally, most prototypes do not incorporate data physicalization, meaning data is not physically represented in the objects or their behaviors~\cite{bae2022making}. Moreover, while frameworks like Graspable AI and TangXAI have been proposed, there is limited empirical research on implementing these elements effectively. Only \citeauthor{colley2023exploring} has introduced tangible XAI interface concepts within these frameworks. Our design and evaluation of Briteller aim to fill these gaps by (1) enhancing the comprehensibility of AI recommendations, (2) creating datafied representations that offer learners greater control over data, and (3) contributing empirical evidence to the field of graspable AI.

\subsection{Light for Embodied Learning}
Light, as a dynamic, familiar, and easily accessible material, has been utilized in various embodied learning environments and data physicalization~\cite{li2020tangible}, supporting the learning of additive color theory and artistic creation~\cite{qin2024kaleidolight}, stage lighting~\cite{nicholson2021tangible}, systems thinking~\cite{kikin2006light}, energy harvesting~\cite{ryokai2014energybugs}, light behavior~\cite{price2009effect}, and serving as an interactive symbol for collaborative problem-solving in classrooms~\cite{alavi2012ambient}. Light is particularly effective for visualizing abstract concepts~\cite{ryokai2014energybugs, houben2016physikit}. \citeauthor{houben2016physikit} found that among their four designed cubes, PhysiLight was perceived as the most useful visualization due to its high output bandwidth. Moreover, light’s dynamic nature promotes changes in the environment and peer interactions, allowing students to explore concepts through hands-on activities and observations, thereby deepening their understanding~\cite{price2009effect}.

Most existing work uses light for visualization and aesthetics design~\cite{li2020tangible} without leveraging its inherent properties to represent values and computation, and there is limited exploration into how learners perceive these representations. Researchers have called for more empirical work on physical variables in data physicalization and tangible interface design, noting that the choice of material is often overlooked~\cite{li2022meta}, which can lead to ineffective designs~\cite{jansen2015opportunities, bae2022making}. For instance, studies show that using volume instead of surface area in data physicalization leads to larger errors in people’s accuracy when perceiving sphere sizes~\cite{jansen2015psychophysical}. Recent advancements in Optical Neural Networks (ONNs), which are physical implementations of artificial neural networks using optical components~\cite{sui2020review}, inspire us to explore light as a material for teaching AI literacy by physicalizing it. Our work aims to extend existing research by (1) expanding the use of light in embodied learning to AI literacy education and providing a prototype Briteller, and (2) conducting empirical observations to understand how students perceive information encoded in light and optical materials, thereby gaining more insights into the effectiveness of light in data physicalization and embodied learning.

\section{Methodology Overview}

\begin{table*}[]
    \centering
    \caption{The iterative design and evaluation procedure.}
    \label{tab:procedure-overview}
    \resizebox{0.84\textwidth}{!}{
    \begin{tabular}{l|l}
    \toprule
    \textbf{Study 1}: Briteller & \textbf{Study 2}: AR-Enhanced Briteller \\
    \textbf{Goal}: Conduct a preliminary evaluation of Briteller. & \textbf{Goal}: Evaluate \textcolor{black}{AR-enhanced} Briteller with diverse learners. \\
    \textbf{Prototype}: Tangible system & \textbf{Prototype}: Tangible system and AR enhancement \\
    \textbf{Participant}: 10 middle school students & \textbf{Participant}: 10 middle school students from the underrepresented groups \\
    \bottomrule
    \end{tabular}
    }
\end{table*}

\subsection{Design-Based Research}
Our study employed a design-based research (DBR) approach~\cite{anderson2012design}, in which theory and design evolved together through iterative cycles. This approach is not a specific methodology or assessment type but rather a framework for conducting empirical research. 
The DBR framework was chosen to develop and refine Briteller and its AR enhancement through cycles of design, implementation, and evaluation.
Such an iterative process generated practical design principles for embodied learning in AI education, while advancing theoretical understanding of how embodied metaphors can support children in comprehending complex AI concepts.

\subsection{Study Procedure Overview}
We employ image schemas~\cite{antle2009body,hurtienne2011image} as six core embodied metaphors for learning (Section~\ref{core-metaphor}), leveraging light as an embodied learning medium. These metaphors are intended to enable students to experience and internalize abstract concepts through physical engagement with Briteller (Section ~\ref{tangible-briteller}). In Study 1, ten middle school students engaged in Briteller activities (Section ~\ref{tangible-activity}). We analyzed their learning gains in target AI concepts and their tangible interactions (Section ~\ref{tangible-measure}). Findings confirmed the preliminary effectiveness of the light-based embodied metaphors in making AI recommendation systems more graspable for young learners. Based on students' remaining misconceptions and feedback, we identified constraints in the light-based interface (Section ~\ref{tangible-findings}).

Informed by Study 1, we proposed an extended list of embodied metaphors, leading to new features in an AR-enhanced Briteller (Section~\ref{ar-rationale}). In Study 2, we invited ten students from low socioeconomic and underrepresented backgrounds in STEM to evaluate the system's effectiveness with more diverse learners (Section~\ref{study2-rm}). The results indicate the effectiveness of AR-enhanced embodied metaphors in enhancing mathematical understanding of the recommendation algorithm, contextualizing prediction outputs, and expanding data vectors (Section ~\ref{ar-findings}). Future work includes integrating the validated new features into Briteller without reducing meaningful tangible manipulation (Section~\ref{future-work}).

\begin{figure*}
  \includegraphics[width=0.84\textwidth]{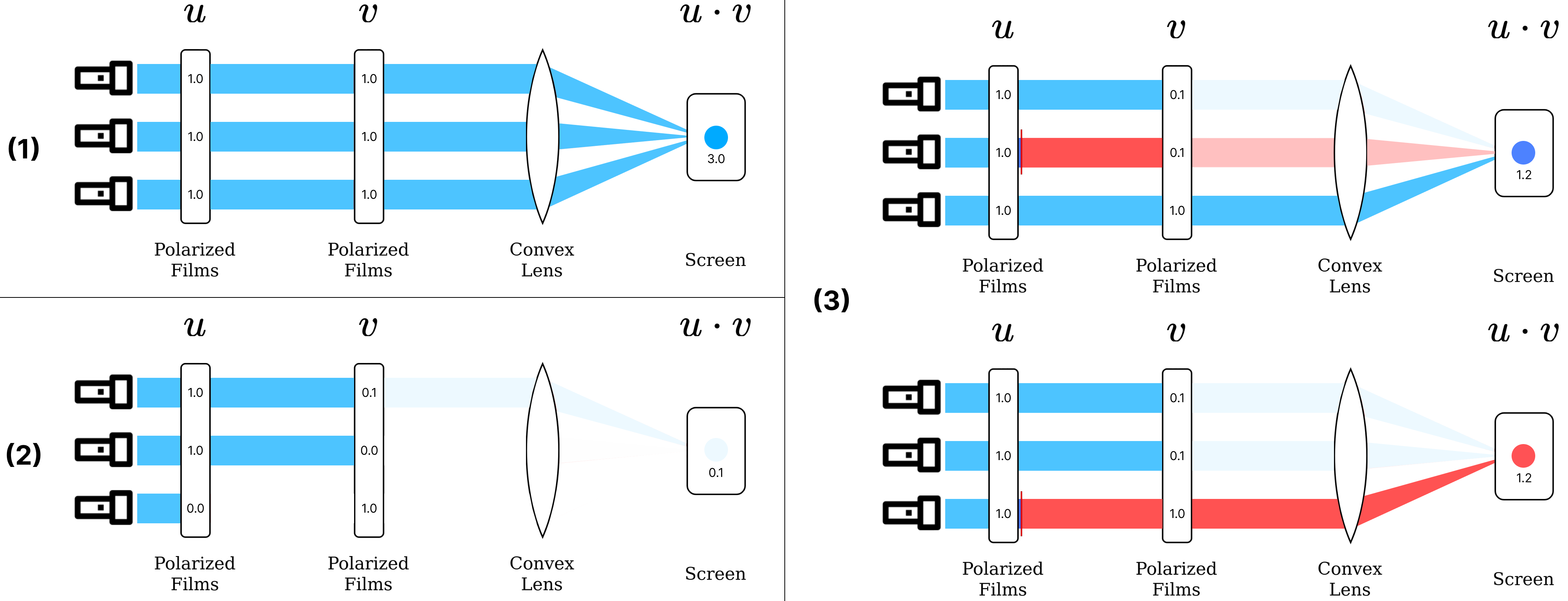}
  \caption{Briteller examples for demonstrating \textbf{(1)} the dot product between a user vector \emph{u} = (1, 1, 1) and an item vector \emph{v} = (1, 1, 1) for the content-based recommendation system; \textbf{(2)} the dot product between a user vector \emph{u} = (1, 1, 0) and an item vector \emph{v} = (0.1, 0, 1); \textbf{(3)} placing a red filter in front of the third attribute of the user vector, the final light dot turns red because the second attribute has a higher contribution to the final output; placing it to the second attribute, the color of the final light dot doesn't change much because the second attribute has a relatively lower impact. Please note that the blue color in the figure is for figure demonstration. We used white light in the study.}
  \label{fig:example}
  \Description{Three examples of different settings of the polarized films in Briteller. The first example demonstrates the dot product between the data vector (1, 1, 1) and the data vector (1, 1, 1). The second example demonstrates the dot product between the data vector (1, 1, 0) and the data vector (0.1, 0, 1). The third example demonstrates the dot product between the data vector (1, 1, 1) and the data vector (0.1, 0.1, 1). The light sources in the third example are blue. A red filter is placed in front of the second attribute of the first data vector on the upper side of the third example. The light dot on the last screen is purple, a mixture of red light and blue light. The red filter is in front of the third attribute of the first data vector on the lower side, which leads to a red light dot at the end.}
\end{figure*}

\section{Design 1: Briteller} 
\subsection{Learning Goals}
\label{learninggoals}
A content-based recommendation system is a suitable starting point for beginners to grasp due to its simplicity~\cite{lops2019trends} and it forms the foundation for more advanced techniques like collaborative filtering~\cite{sun2022survey,shu2018content}. Key concepts in content-based recommendation systems include user vectors, item vectors, prediction results, and similarity metrics. These systems typically have three main components: (1) a Content Analyzer that extracts item representations, (2) a Profile Learner that constructs user preference profiles, and (3) a Filtering Component that matches user profiles with item representations using similarity metrics~\cite{lops2011content}. Users and items are described by attributes and represented as vectors, with the recommendation processes matching user attributes to item attributes~\cite{pazzani2007content, lops2011content}. The dot product, an algebraic operation that can calculate a preference score by multiplying and then summing corresponding values of user and item vectors, is commonly used to measure user-item similarity in recommendation systems~\cite{rendle2020neural, koren2009matrix}. It is also foundational in many AI models~\cite{rendle2020neural} and simpler for children to understand, making it a more suitable choice than cosine similarity in alignment with XAI's model simplification approach~\cite{belle2021principles}. We identified a list of key AI concepts (Table~\ref{tab:metaphor-recommendation}(1)) as target learning content.

\subsection{A Taxonomy of Using Light-Based Embodied Metaphor for a Graspable Recommendation System}
\label{core-metaphor}
This section provides an overview of the core embodied metaphors for AI concepts in the design of a light-based recommendation system, with detailed definitions in Section~\ref{tangible-briteller}.

\subsubsection{Light-Based Embodied Metaphors}
For the target AI concepts, we selected a set of image schemas~\cite{hurtienne2011image,loffler2014population,hurtienne2009sad} as light-based embodied metaphors (Table~\ref{tab:metaphor-recommendation}(4)). For example, DARK-BRIGHT denotes light intensity, BLOCKAGE denotes blocking light to adjust its intensity, MERGING refers to combining beams via refraction or reflection, and HUE represents different colors of light.

We chose light-based representation and interaction (Table~\ref{tab:metaphor-recommendation}(2) \& (3)) for three reasons. First, young learners are familiar with optical phenomena, such as light intensity and colors, through their daily experience and the STEM curriculum~\cite{states13}. 
Second, light offers immediate visual feedback and supports hands-on manipulation, making it popular in STEM museums.
Third, recent advancements in optical computing (e.g., ONNs~\cite{goodman1978fully,hayasaki1992optical}) suggest light as a promising medium to demystify AI algorithms such as vector operations.

\subsubsection{Mapping between Target AI Concepts and Embodied Metaphors}
Embodied metaphors can be categorized into two types. 
Grounding metaphors convey basic, intuitive ideas by mapping everyday experiences to abstract concepts (e.g., grouping items like addition), while linking metaphors yield sophisticated ideas by blending different metaphors (e.g., numbers as points on a line)~\cite{lakoff2000mathematics}.
We developed a taxonomy of light-based embodied metaphors for a graspable recommendation system (Table~\ref{tab:metaphor-recommendation}). 
The basic concepts (e.g., data value, data vector, addition) are mapped to grounding metaphors.
The more complicated concepts (e.g., decimal multiplication and the dot product for AI recommendation) are mapped with linking metaphors.
In Section ~\ref{tangible-briteller}, we introduce Briteller, which leverages this taxonomy.

\subsection{The Design of Briteller}
\label{tangible-briteller}

\begin{table*}
\centering
\caption{A taxonomy of light-based embodied metaphors to map with core components in a graspable AI: (1) Recommendation system concepts, (2) tangible representation, (3) tangible interaction, (4) light-based embodied metaphors.}
\label{tab:metaphor-recommendation}
\resizebox{0.91\textwidth}{!}{
\begin{tabular}{p{0.1\linewidth}p{0.27\linewidth}p{0.26\linewidth}p{0.3\linewidth}p{0.2\linewidth}}
\toprule
\multicolumn{1}{l}{AI Concept} & \textbf{(1)} Recommendation Concept & \textbf{(2)} Tangible Representation & \textbf{(3)} Tangible Interaction & \textbf{(4)} Embodied Metaphor\\
\midrule
Data vector
 & \textbf{User vector}: a set of attributes representing the user's preference. \textbf{Item vector}: representing the characteristic of an item. & \textbf{Data bar}: an array of \textbf{light beams} & \textbf{Turn on/off} individual flashlights; \textbf{mount/unmount} data bar. & PART-WHOLE\\
  & & & \textbf{Rotate} knobs to set values as the percentage of light passing through. & BLOCKAGE \\
 \midrule
 Prediction 
 & \textbf{Prediction} of the likelihood that the user likes a given item & \textbf{Intensity} of the light dot represents the prediction result & & DARK-BRIGHT \\
  & \textbf{Differentiate which attribute} impacts the output & The \textbf{color shade} of the final light dot & \textbf{Place/remove} color filters (HUE) in front of a light beam; the intensity of the colored light (DARK-BRIGHT) indicates the contribution of the specific light source. & HUE, DARK-BRIGHT, PART-WHOLE \\ 
 \midrule
 Mechanisms
 & \textbf{Dot product} is a basic method to compute similarity of two vectors: (1) multiply two values (\textbf{decimal multiplication}) & \textbf{The intensity of light beams} after passing through two data bars & \textbf{Rotate} knobs to block a certain amount of light (BLOCKAGE) passing through (LINKAGE) to change the light intensity (DARK-BRIGHT). & BLOCKAGE, LINKAGE, DARK-BRIGHT \\
 & (2) add individual multiplication results (\textbf{addition}) & \textbf{A convex lens} converges an array of light beams. & \textbf{Mount/unmount} convex lens: add/remove the addition operation & MERGING \\
\bottomrule
\end{tabular}}
\end{table*}

\subsubsection{Embodied Metaphors for User and Item Vector}
Data representation in AI recommendation systems typically consist of multiple attributes describing the characteristics of the data point.
To make the concept of a data vector more accessible for children, we use an array of light beams as the embodied metaphor for an array of attributes. Individual light beams in the array represent individual data attributes in a vector (Fig. ~\ref{fig:vector_design}(1)) (PART-WHOLE). 

An inspector is created with color filters for learners to differentiate data attributes in a vector (Fig. ~\ref{fig:vector_design}(5)). Placing a color filter in front of a light beam changes the light to that color, helping learners see how each beam contributes differently to the final light dot (HUE and DARK-BRIGHT) (see Fig. ~\ref{fig:example}.3 for an example). 
For instance, if a learner places a red filter in front of the first polarized film in a set, they observe a more saturated red dot on the projection screen (Fig.~\ref{fig:vector_design}(5)), whereas placing it in front of the second film results in a less saturated red dot. This helps the learner understand that the first attribute in the data vector contributes more to the final output.

\subsubsection{Embodied Metaphors for the Dot Product}
There are two major operations in the dot product. First, multiply corresponding attribute values from the same position in both vectors. Second, sum the products from the first step. Polarized film is an optical material that allows only a certain amount of polarized light to pass through. We calibrated the knob scale so that no light passes through the polarized films at a knob value of zero, and all light passes through at a knob value of one.
To change the value of a specific data attribute, learners can rotate a knob attached to a polarized film overlaid with another polarized film, which blocks a certain amount of light (BLOCKAGE) of the specific light beam passing through (LINKAGE) and controls the light intensity (DARK-BRIGHT).
Briteller unveils the two-step dot product through BLOCKAGE, MERGING, and LINKAGE (Table~\ref{tab:metaphor-recommendation}):
\begin{description}
    \item[Multiplication] The amount of light changes after passing through two sets of polarized films, representing the multiplication of corresponding values in two vectors (Fig. ~\ref{fig:vector_design}(b)) (BLOCKAGE). For example, Fig. ~\ref{fig:example}.2 shows multiplications of three values pairs in vectors. Additionally, PART-WHOLE embodies decimal.
    \item[Addition] Light beams converge after passing through a convex lens, representing the sum of the products (Fig. ~\ref{fig:vector_design}(c)) (MERGING). Fig. ~\ref{fig:example}.1 illustrates the maximum light intensity from merging three light beams.
\end{description}

The light beams passing through the polarized films and convex lens embody the connection between attributes being multiplied and the resulting sums. 

\subsubsection{Embodied Metaphor for the AI Recommendation Output}
\begin{figure*}
\centering
\includegraphics[width=0.84\textwidth]{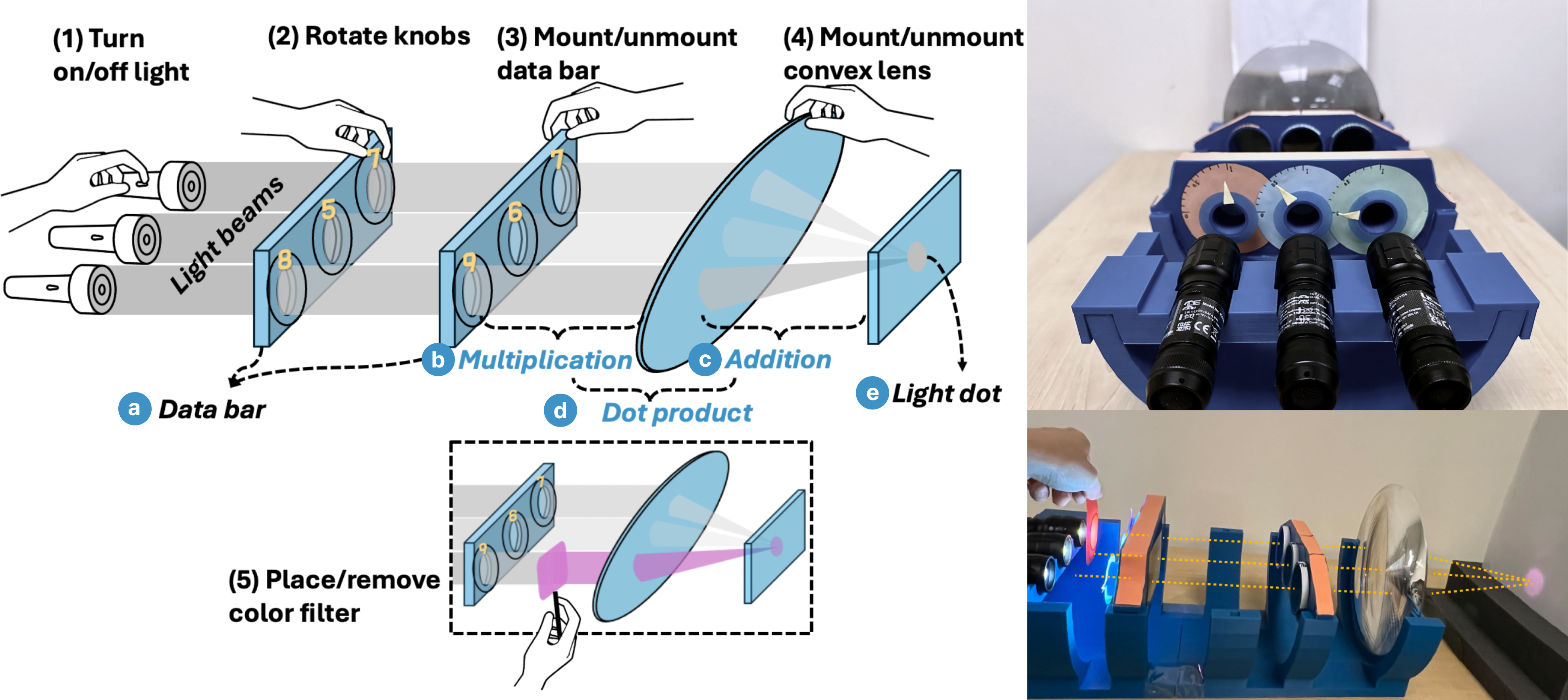}
\caption{Embodied representations and interactions in Briteller.}
\label{fig:vector_design}
\Description{The set-up of the Briteller system from different views. More details can be found in the figure caption.}
\end{figure*}

In the end, the projection of the focused beam represents the output of the dot product and the light intensity embodies the value (Fig. ~\ref{fig:vector_design}(e)) (DARK-BRIGHT).

\subsection{Briteller Learning Activity}
\label{tangible-activity}
We designed Briteller into a snack recommendation system as snacks provide a familiar and inclusive topic to engage children with varying levels of AI exposure and can also extend to nutrition science, an essential knowledge domain for children~\cite{riehmann2014visualizing,wimer2024beyond,cai2024see}. Based on standard nutrition labels~\cite{wimer2024beyond}, we identified three data attributes for snacks: sugar, saltiness, and fat. 
To evaluate the effectiveness of the embodied metaphors in Briteller, we create three learning activities, each lasting for about 10-20 minutes.

\subsubsection{Explore the Recommendation System by Predicting Your Snack Preference}
Learners rotate the first set of polarized films to create a user vector, reflecting their sweetness, salinity, and greasiness tastes on a scale from 0 (``strongly dislike'') to 1 (``strongly like''). Next, they set the item into the sugar, salt, and oil content of a chocolate bar. 
Learners read the light dot intensity on the final screen as Briteller's prediction, compare it with their actual snack preference, and discuss prediction accuracy and mechanism.

\subsubsection{Explore the Recommendation System by Predicting Different User Preferences for One Snack}
Learners rotate knobs for a given user named Sammy and a cake. Learners read Briteller's prediction that Sammy likes the cheesecake. Then they change the user vector to Jessie and compare the outputs.

\subsubsection{Manipulate One Value to Alternate the Prediction Result}
This activity supports understanding the dot product by engaging learners in an interactive exploration of Briteller. Learners adjust knobs representing potato chips, producing a bright final light dot. They then make one change in Briteller to shift the output from ``strongly like'' to ``strongly dislike''. 
Building on prior activities that involved reading and adjusting light, students now use color filters to differentiate the knobs' effects on the output. 
By tinkering with color filters, knobs, and a convex lens, they gain insights into the multiplication and addition operations underlying the dot product.

\section{Study 1: Evaluate and Iterate Briteller}
Study 1 is a proof-of-concept case study assessing Briteller's preliminary effectiveness and exploring students' interactions and feedback to inform future iterations. 
The findings led to AR-enhanced Briteller in Study 2.

\subsection{Participant}
\textcolor{black}{Using study flyers and snowball sampling, we recruited 10 middle school students (six males and four females) aged 11 to 12 (Mean = 11.5, SD = 0.53) for the evaluation.} The Institutional Research Subjects Review Board approved the study.

\subsection{Study Procedure and Data Collection}
\label{study1-procedure}
\begin{figure*}[h]
\centering
\includegraphics[width=0.84\textwidth]{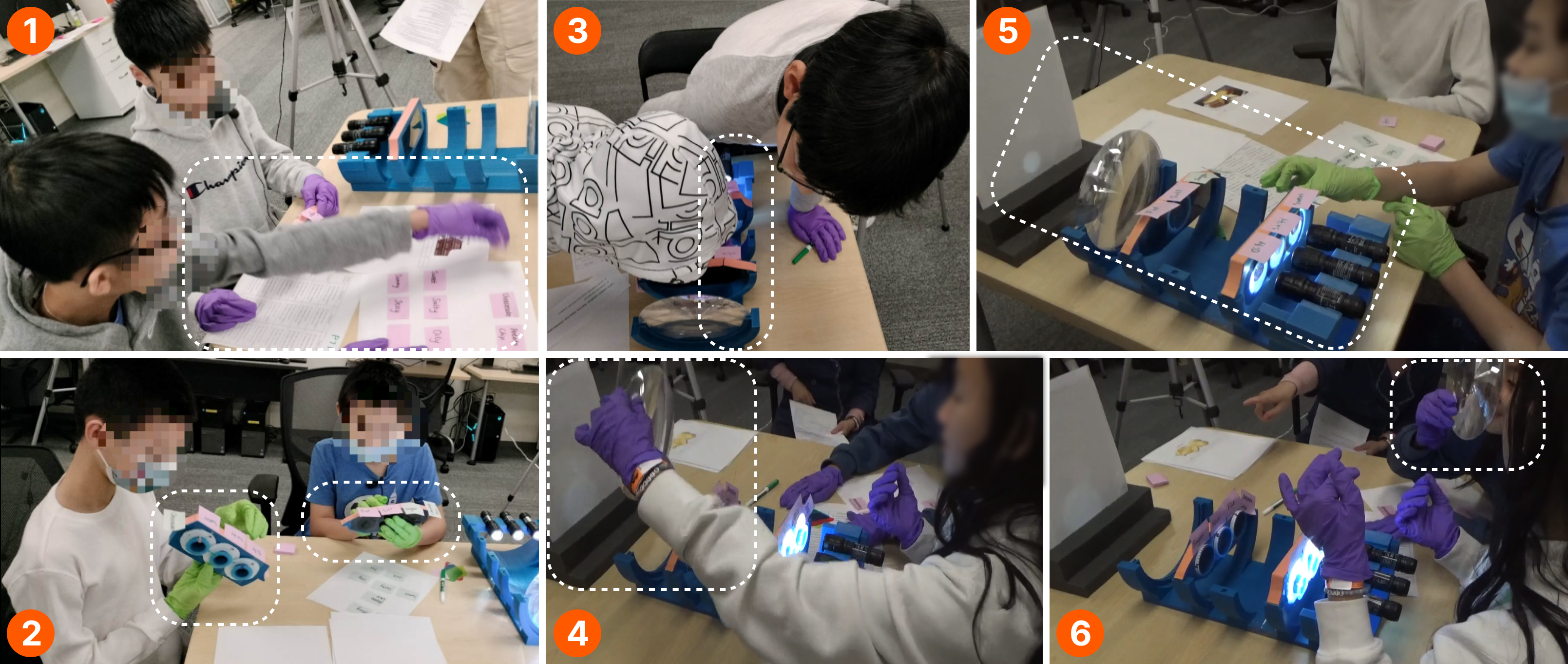}
\caption{Evaluation of Briteller with middle school students (N=10): \textbf{(1)} Students tagged data attribute labels to corresponding parts of the physical data bars (PART-WHOLE); \textbf{(2)} students rotated the knobs to block a specific amount of light for setting the value of a specific data attribute (BLOCKAGE); \textbf{(3)} students observed how individual light beams changed while passing through the light-based recommendation system (LINKAGE); \textbf{(4)} a student unmounted and mounted the convex lens to add three light beams together (MERGING); \textbf{(5)} students rotated the knobs to block a certain amount of light (BLOCKAGE) and compared the output of light intensity (DARK-BRIGHT) under different conditions; \textbf{(6)} a student got curious about the optical principles applied in Briteller.}
\label{fig:user}
\Description{The figure shows four pairs of children interacting with Briteller during Study 1.}
\end{figure*}

The study took place in an on-campus lab and lasted about one hour, facilitated by three researchers (Fig. ~\ref{fig:user}).
Students first completed a survey collecting basic demographic information and pre-testing of target AI concepts. Individual questions assessed different concepts involved in AI recommendation's inner workings, such as data vector and the dot product (Table~\ref{tab:lg}(1) \& (2)).
Participants first became more familiar with Briteller through free exploration and observation, then engaged in three scaffolded learning activities (Section~\ref{tangible-activity}), working collaboratively in pairs, fostering deeper learning and discussion~\cite{laal2012benefits}. Researchers explained each activity, provided printed instructions, encouraged participants to think aloud, observed the entire session, and offered facilitation when needed. While students worked collaboratively, learning gains and feedback were assessed on an individual basis. After interacting with Briteller, students completed a post-test with the same assessment questions. 
To have a more in-depth and qualitative understanding of students' embodied learning experience with Briteller, the researchers conducted post-study interviews by asking participants what they liked and disliked about Briteller.

We collected (1) video recordings from five cameras located at different places, (2) audio recordings of learners' discussions during the learning experience.
For the video recordings, five cameras covered four different angles and the top view of Briteller.

\subsection{Measure and Data Analysis}
\label{tangible-measure}
\subsubsection{RQ1: Can Light Serve as a New Embodied Learning Medium for Children to Understand AI Recommendation Systems?}
\paragraph{Learning Gains and Misconceptions}
We measured students' learning gains on the target AI concepts (Table~\ref{tab:lg}(1)) based on the learning goals (Section~\ref{learninggoals}).
Assessment questions in pre- and post-tests were designed based on the widely cited literature on content-based recommendation systems~\cite{pazzani2007content, lops2011content}. 
\textcolor{black}{A rubric was developed to capture students' understanding of key concepts including (1) attributes and values in data vectors, (2) the definition of user preference in AI recommendation systems, (3) the two mathematical operations involved in the dot product, and (4) the dot product calculation.
The rubric used a 0-3 scale to assess student understanding: 0 for no attempt or understanding, 1 for basic or partial understanding, 2 for solid understanding with some gaps, and 3 for full mastery with no significant errors.}
Two researchers assigned scores to the answers independently, reaching a near-perfect agreement (Cohen’s Kappas = 0.87). 
Then researchers discussed and resolved the disagreements in the coding results.

\paragraph{Learner Feedback}
For the post-study interviews, three researchers conducted a thematic analysis~\cite{braun2006using} of students' subjective learning experiences. 
The analysis focused on students' feedback on what they liked and disliked about Briteller, and what they found challenging and confusing.
Researchers discussed and confirmed the data-supporting themes by comparing each category and connecting categories derived during open coding~\cite{lazar2017research}.
Combining the analysis of students' learning performance and their feedback, we derived an initial understanding of Briteller’s effective design elements and needed design iterations.

\begin{table*}[]
    \centering
    \caption{Major embodied learning experiences that Briteller aims to support: \textbf{(A)} embodied interactions and visualizations and the co-occurred \textbf{(B)} verbal learning behaviors demonstrating students' understanding.}
    \label{tab:behaviors}
    \resizebox{0.91\textwidth}{!}{
    \begin{tabular}{p{0.05\textwidth}p{0.52\textwidth}p{0.39\textwidth}}
    \toprule
        & \textbf{(A)} Embodied interactions & \textbf{(B)} Verbal learning behaviors \\
        \midrule
        Low level & (1) Read the light intensity & (1) Describe the value as low, medium, or high \\
        & (2) Rotate knobs & (2) Describe values for a data attribute \\
        & (3) Mount or unmount the convex lens & (3) Describe the addition operation \\
        & (4) Interact with light beams by using hands to trace or block light paths & (4) Describe corresponding relationships between values \\
        High level & (5) Exploration with flashlights, knobs, the lens, and color filters to make observations and discoveries without a clear understanding of the inner workings in mind & (5) Describe observations while exploring \\
        & (6) Experimentation with flashlights, knobs, the lens, and color filters to test out the inner workings & (6) Describe the hypothesis, observe the outcomes, and compare with hypothesis \\
        & (7) Explanation with hands-on demonstration & (7) Clear communication of the AI and math concepts to others \\
        \bottomrule
    \end{tabular}
    }
\end{table*}

\subsubsection{RQ2: What Are the Unique Affordances and Limitations of Using Light to Support the Understanding of AI Recommendation Mechanisms?}
\label{study1rq2measure}
With the video recordings of the study session, we analyzed the occurrence of low- and high-level embodied interactions that helped students grasp AI concepts during the learning activities.
Low-level embodied interactions are defined by grounding metaphors (e.g., BRIGHT is high user preference; addition is MERGING light).
We annotated their co-occurrences with verbal learning behaviors (Table~\ref{tab:behaviors}(1)-(4)).
High-level embodied learning behaviors---exploration, experimentation, explanation (Table~\ref{tab:behaviors}(5)-(7))---are grounded in the existing framework of how inquiry-based learning occurs~\cite {pedaste2015phases}.
Exploration is when students freely manipulate Briteller to make observations and discoveries without a clear hypothesis. Experimentation is using flashlights, knobs, the convex lens, and color filters to collect evidence concerning how they think Briteller works. Explanation is when students communicate evidence-based reasoning with light and optical materials, which can occur along with exploration and experimentation.

\begin{table*}
\centering
\small
\caption{Paired t-test results for pre- and post-tests, with the scores ranging 0-3 (N=10).}
\label{tab:lg}
\resizebox{0.84\textwidth}{!}{
\begin{tabular}{p{0.2\textwidth}p{0.29\textwidth}*{6}{l}}
\toprule
\textbf{(1)} Knowledge Component & \textbf{(2)} Assessment Question & \multicolumn{2}{l}{Pre-Test} & \multicolumn{2}{l}{Post-Test} &
{\it t-test}&{{\it p}} \\ 
\cline{3-6}
 & & M & SD & M & SD & &\\
 \midrule
 The user representation in a recommendation system & How to describe a user for a recommendation system? & 0.1	& 0.3 & 2.3	& 0.8	& -7.89	& \textless 0.001	\\ 
\midrule
 The item representation in a recommendation system & How to describe an item for a recommendation system? & 0.1	& 0.3 & 2.1	& 0.8	& -6.8	& \textless 0.001\\ 
\midrule
 The output of a recommendation system & What does a recommendation system predict?	& 0.4	& 0.5		& 2.8	& 0.4 & -11.38	& \textless 0.001	\\ 
\midrule
 The dot product in recommendation systems & How does a recommendation system calculate the prediction result with a user vector and an item vector? & 0.4	& 0.9		& 2.2	& 1.3	& -3.48	& 0.0026 \\ 
\midrule
 Multiplication operation in the dot product: Multiply corresponding values from user \& item vectors & To calculate a user's preference (user vector = (1, 9, 2)) for a book (book vector = (7, 0, 1)), which of the multiplication below will happen? & 0.3	& 0.9		& 1.8	& 1.6	& -2.61	& 0.0177 \\ 
\midrule
 The dot product calculation for a recommendation system & What's the numeric prediction result of a user's preference (user vector = (1, 9, 2)) for a book (book vector = (7, 0, 1))? & 0.6	& 1.265		& 1.8	& 1.6	& -1.897	& 0.739 \\ 
\bottomrule
\end{tabular}
}
\end{table*}

\section{Study 1: Results}
\label{tangible-findings}
\subsection{RQ1: Can Light Serve as a New Embodied Learning Medium for Children to Understand AI Recommendation Systems?}
\subsubsection{Learning Gains with Light-Based Embodied Metaphors}
The data were normally distributed based on the Shapiro-Wilk test. 
A paired-sample t-test comparing pre- and post-test results (Table~\ref{tab:lg}) showed statistically significant increases in students' conceptual understanding of (1) user representation, (2) item representation, (3) recommendation system output, and (4) the dot product in recommendation systems, with p-values $<$ 0.005. Additionally, understanding of multiplication in the dot product increased significantly (p $<$ 0.05). 
Furthermore, the mean pre-test results showed that the participants had little prior knowledge of the AI concepts around recommendation systems (Table~\ref{tab:lg}).

However, the results indicated no significant learning gain in understanding the numeric output of the dot product (p = 0.739). 
In Briteller, this output is represented by the intensity of the final light dot, which may not clearly connect to the numerical values in the calculation. We discuss how to better support and contextualize this embodied interaction within the numeric context of recommendations in Section~\ref{discussion1}.

\subsubsection{Remaining Misconceptions}
\paragraph{Missing Quantitative Connections between Data Vectors}
Three out of ten students did not specify the user vector as a set of preference values corresponding to item features. They gave vague concepts instead, such as ``their likes and dislikes''.

\paragraph{Limited Context Transfer}
Three students still used terms related to light instead of AI in the post-test.
For example, for Q3 about recommendation output,
one student wrote: ``When you make the knobs higher, the circle gets brighter.''

\paragraph{Limited Scalability in Data Representation}
In two students' post-test answers, we identified a lack of generalizability in data representation. 
When being asked about general AI recommendation systems, they only answered the questions in the context of snack recommendations.
In the post-interview, 
students noted that Briteller was limited to representing snacks with only three attributes.
A10: ``There could be exceptions - if you really dislike sweet things, a lot of the time there is an exception. It doesn't tell everything about what you like. For example, I like sweet stuff but I don't like chocolate'';
A5: ``If you have something that has much more attributes, then you can't handle it''.


\subsubsection{Learner Feedback}
Students found the tangible recommendation system easy to understand. Seven out of ten students found predicting with light novel and interesting: ``I never thought of it'' (A10); ``It's a smart idea to use light to make predictions'' (A5, A6); ``It's very visual and fun to figure out the math behind it'' (A2). A3 and A4 noted the predictions were accurate and the dot product operation was transparent: ``The prediction is pretty accurate. It's nice to see the inner workings.'' They appreciated how color filters highlighted the attributes with the highest contribution.

Overall, students considered the light-based system a creative and engaging way to learn AI and math. A1 commented, ``It's very creative... It's just math, but visual.'' A2 added, ``It directly shows you why you like something or not.'' Some students were motivated to create their own puzzles with Briteller, like giving an output number and item vector to deduce user preferences (A1, A2). Others expressed curiosity about the optical principles involved, asking questions like ``How does the polarized film control opacity?'' (A1, A2) and ``Why do the circles get bigger when we remove the convex lens?'' (A9, A10).

\subsection{RQ2: What Are the Unique Affordances and Limitations of Using Light to Support the Understanding of AI Recommendation Mechanisms?}

\paragraph{Understanding the Multiplication Operation through Experimentation}
\label{multiplication}
We observed all groups rotating knobs and observing light changes to understand the multiplication of corresponding values through experimenting with contrastive cases (Table~\ref{tab:embodied-data2-1}). 
For instance, A5 and A6 experimented with PART-WHOLE, LINKAGE, and DARK-BRIGHT to figure out the multiplication and correct their own misunderstanding. First, they applied a similar strategy as A3 and A4: ``If you turn this one (the first set of knobs) to zero, then this one (the second set) doesn't get any light at all.'' Then A5 generated a hypothesis of the dot product: ``The first set of knobs has more power than the second set because the light hits the first set first.'' A5 turned two flashlights off, observed how the only flashlight left contributed to the output, and iterated the previous hypothesis: ``If the second one is zero, the first is not zero, still there will not make any light.
Now I have changed my hypothesis. I think both values need to be larger. Both are the highest value and will be the most preferable.
I guess they work the same. They work together.'' 
With the support of PART-WHOLE and DARK-BRIGHT, A5 understood the direction of the multiplication is decreasing: ``The maximum value is one, so multiplying a value less than one leads to decreasing.''
A7 and A8 confirmed their hypothesis by ruling out the possibility of addition: ``[Setting one knob as one and the corresponding knob as zero] the result is zero. But one plus zero is one. So it's not an addition. It's multiplication.'' 
Such a reasoning process is supported by LINKAGE, PART-WHOLE, and DARK-BRIGHT.

Two out of the five groups also utilized color filters (HUE) during their experimentation. For instance, A9 and A10 tested medium values instead of just extremes like one and zero to understand the multiplication: ``[Placing color filters in front of a pair of knobs with low values], the color doesn't work. The small numbers, after you multiply them together, the less vibrant. The knobs ruled it out.''

\paragraph{Visual Feedback in Supporting Efficient Debugging}
 In four out of five groups, students demonstrated efficient debugging behaviors when faced with unexpected outcomes during experimentation (Table~\ref{tab:embodied-data2-2}). For instance, when the focused light dot did not match their expectations after rotating the knobs, students quickly identified which knob required correction without relying on extensive trial and error. This efficiency was facilitated by the immediate visual feedback provided by light-based embodied metaphors.
At the very beginning, A1 set the user vector to represent his preferences for ``Sugar'', ``Salt'', and ``Fat'' and expected a high prediction value for the chocolate bar; but the focused light dot turned out to be very dim; without hesitation, A1 quickly identified that he set one preference incorrectly: ``instead of a teacher telling you, I can figure it out myself.'' A4 set the knob for the “Fat” preference in the user vector and placed the yellow filter in front of it; A4 thought the focused light dot would turn into yellow, but it did not; then A3 quickly rotated the ``Fat'' attribute in the item vector to a higher value which resulted in the desired outcome. 

\paragraph{Explore, Experiment, and Explain Recommendation Output with Color Filters} 
Except for one group (A1 and A2) directly confirming and explaining their hypothesis using color filters without free exploration, 
four out of the five groups followed the paths of first exploration and then experimentation.
For instance, A6 started with free exploration, stacking all color filters together in front of a light source and observing that the light dot became darker; then A6 switched to using a single color filter and explored the effect of removing any flashlight. Then A6 adjusted knobs one by one to dim or brighten the light.
Through systematic experiments, A6 discovered that setting any attribute in one data bar or its corresponding attribute in the other to zero would result in a zero output for that pair, ultimately eliminating the light.Meanwhile, A5 used the color filters to validate A6’s explanation.

All groups of students used color filters (HUE) along with a convex lens (MERGING) to explain the impact of attributes (Table~\ref{tab:embodied-data2-3}).
A9 and A10 set the left and the right knobs to zero: ``Now the color can be only added to the middle one''.
A7 and A8 explained: ``The darker shades mean zero, so the color won't make any changes to the end''.
A3 and A4 used one color filter to identify the attribute with the highest impact on the final result, then rotated the knob behind the filter to demonstrate its effect on the light dot with brighter or dimmer light: ``If you turn it (the knob) higher, it (the color) will work better.''
\textcolor{black}{To enhance their explanation, A3 and A4 further used two hues: placing a red filter in front of a high-impact attribute and a blue filter in front of a low-impact attribute. They observed that the final light dot appeared predominantly red, reflecting the greater influence of the high-impact attribute. 
Finally, they placed one filter in front of each flashlight and demonstrated that all the light paths combined into a single output.}

\paragraph{Limitations of Light-Based Tangible Interface}
From students' misconceptions, feedback, and interaction behaviors, we identified four key limitations of the tangible Briteller: (1) low visibility of light beams within the tangible system, (2) difficulty in accurately measuring light intensity, (3) limited contextual transfer between light phenomena and AI recommendations, and (4) restricted scalability of data vectors for more generalized representations.

\section{Design 2: AR-Enhanced Briteller}
\label{ar-rationale}

Following an iterative design-based research approach, 
we explored how to represent abstract AI concepts while addressing the key constraints identified in Study 1.
We augmented Briteller with tablet-based AR, resulting in AR-enhanced Briteller. It combines the complementary strengths of tangible and AR interactions 
while engaging students with both.
The reasons are threefold:
\begin{enumerate}
    \item AR provides a highly visual and interactive experience by overlaying virtual 3D objects onto the real world~\cite{yuen2011augmented, lindgren2013emboldened}. This approach enhances reality rather than replacing it, offering access to otherwise unobservable phenomena~\cite{lindgren2013emboldened}. Previous research shows that a mixed-reality environment with physical objects results in greater learning~\cite{yannier2016adding}. Specifically, combining physical and virtual manipulatives enhances conceptual understanding of light more effectively than either method alone~\cite{olympiou2012blending}.
    \item While designing a fully tangible system to overcome all constraints identified in Study 1 is possible, it would require costly optical components, sensors, and other electronics, increasing manufacturing complexity and reducing scalability. Specifically, we considered utilizing the Tyndall effect to visualize light paths and introducing light sensors to detect the final light dot intensity, but these enhancements involve complex manufacturing (e.g., adding dense liquid or smoke) and face physical constraints, such as sensitivity to varying light conditions. The current design only uses affordable components like flashlights and convex lenses. Tablet-based AR is portable, cost-effective, and accessible through mobile devices~\cite{yuen2011augmented}. These materials make it easier to scale and reach more students.
    \item Using both the purely tangible Briteller and the tablet-based AR interface retains the benefits of embodied interaction observed in Study 1, rather than fully replacing it.
\end{enumerate}

\subsection{New Features}
\label{ar-briteller}
\subsubsection{Augmented Visibility and Quantifiable Intensity for Light}
In Briteller, light beams are nearly invisible, and it is difficult to perceive light intensity accurately. In AR, the light beams are visualized with 3D models (Fig.~\ref{fig:enhanced}). These beams link the two multipliers in decimal multiplication, with their intensity changing dynamically based on the values. This enhances the embodied metaphors LINKAGE, PART-WHOLE, and DARK-BRIGHT. Attribute values are labeled above the knobs for easier observation, and the dot product results are displayed on the final screen, helping students link light intensity to values. These designs aim to support a more quantitative understanding of the dot product.

\subsubsection{Contextualized Output for Better Transferability}
In AR, the recommended item is overlaid with the light dot on the final screen. After inputting the user vector, learners can click the ``Recommend’’ button to view recommendations. The visibility and order of items on the screen are determined by prediction scores (Fig.~\ref{fig:enhanced}(1)). 
This directly links the visibility of items with the visibility of items in an AI recommendation system.
Learners navigate through items using ``Next'' and ``Prev''. User and item images are labeled next to the data bars. Attributes are marked above the knobs. These designs consistently remind students of the problem context.

\subsubsection{Scalable Data Vector}
\begin{figure*}[h]
\centering
\includegraphics[width=0.7\textwidth]{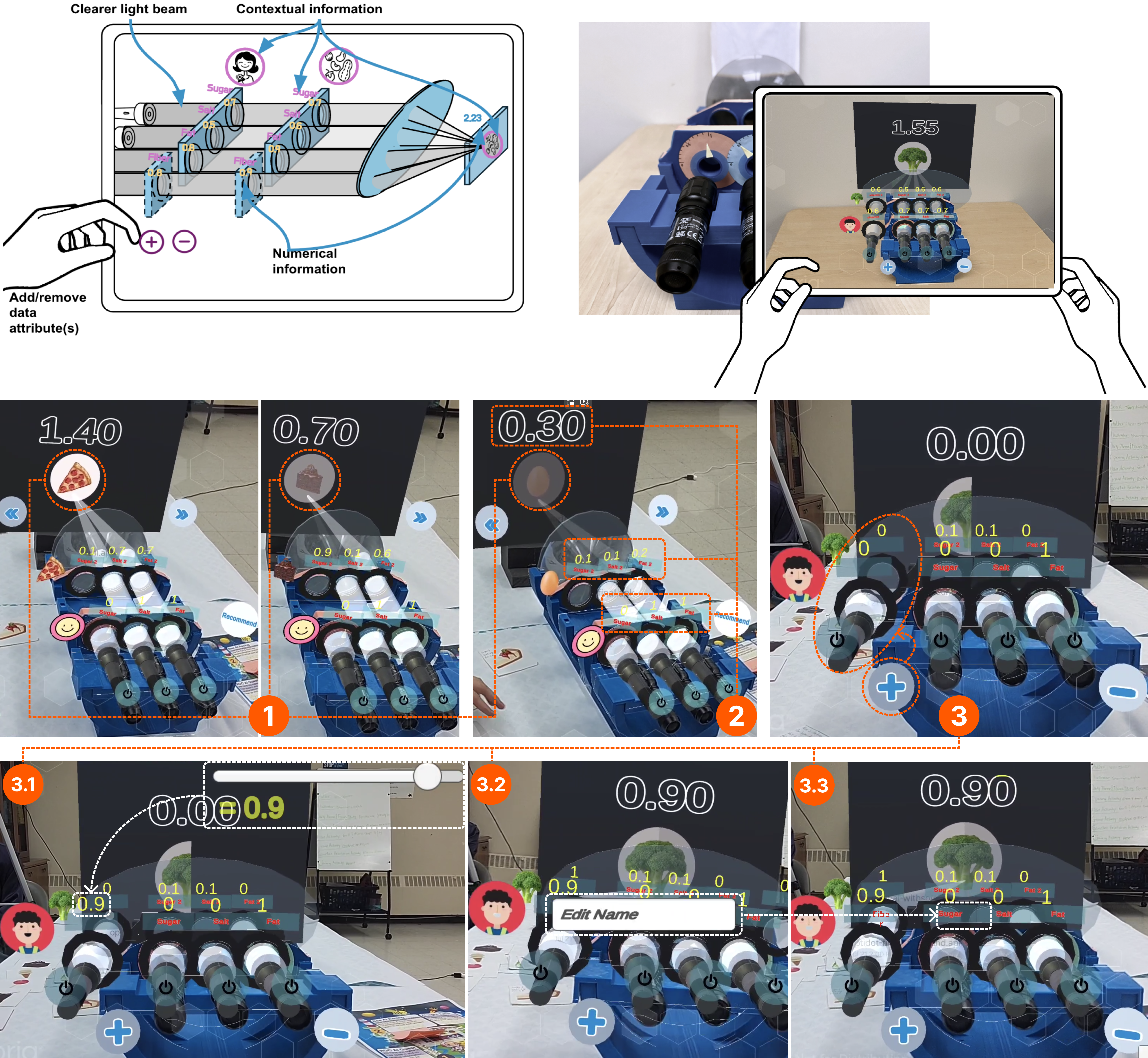}
\caption{In Study 2, the tangible interface of Briteller is augmented by virtual objects in AR: \textbf{(1)} the item visibility (DARK-BRIGHT, SUPERIMPOSITION) represents how likely the item should be shown/recommended to the target user; the user sees the item more clearly when the recommendation system predicts a higher user preference; \textbf{(2)} the data vectors and dot product in this light-based recommendation system is augmented by numeric values of individual data attributes; \textbf{(3)} a learner added a new flashlight to use describe users and items with a new pair of data attributes; the learner edited the attribute's name ``fiber'' and set the value for it.
}
\label{fig:enhanced}
\Description{.}
\end{figure*}

Only three attributes are available in Briteller. With AR, students can easily add or remove data attributes in the vectors and customize attribute names. The recommendation results are immediately visualized on the final screen (Fig. ~\ref{fig:enhanced}(3)), showing how changes in attributes affect the prediction.

\subsection{AR-Enhanced Briteller Learning Activity}
\label{ar-activity}
\textcolor{black}{In the improved learning experience, students begin with exploring Briteller to familiarize themselves with the optical setup, and then engage in purely tangible activities (adapted from Section~\ref{tangible-activity}) for about 20 minutes, followed by AR activities for another 20 minutes.}
\textcolor{black}{
Based on the learning gains and misconceptions from Study 1, we designed three AR activities to evaluate the new features detailed in Section~\ref{ar-briteller} for students to
(1) enhance their numeric understanding of the light-based dot product (Section ~\ref{explore-dotproduct}), (2) explore a more contextualized food recommender as informed AI users (Section ~\ref{user}), and (3) design and expand attributes in data vectors as an empowered AI designer (Section ~\ref{designer}).}

\subsubsection{Explore AI Recommendation in AR}
\label{explore-dotproduct}
Learners first freely slide value sliders to adjust values, observing how numerical results, light beams, and the intensity of the light dot change in real time.

\subsubsection{Read and Control Food Recommendation in AR}
\label{user}
Learners input the user vector based on their preferences, click the ``Recommend'' buttons for food recommendations, and navigate through different foods. Then learners recommend food for a user who prefers food with high fat and sugar (Fig.~\ref{fig:enhanced}(1)) and make changes to recommend less sweet food.

\subsubsection{Customize and Expand Attributes for Better Recommendation in AR}
\label{designer}
Learners read a given user David's description: he likes sports and cares about fitness. Learners design and set David’s user vector. Learners observe a mismatch between the intensity of the two semi-circles in the final light dot with the food image overlaid on the final screen (Fig. ~\ref{fig:enhanced}(3)). This represents an inaccurate prediction result. To increase the accuracy, learners come up with more user and item attributes for data vectors, add more flashlights, or change existing attributes that make sense to match the intensity in the two halves of the final light dot (Fig.~\ref{fig:enhanced}(3.1)-(3.3)).

\section{Study 2: Research Method}
\label{study2-rm}
\textcolor{black}{Following an iterative DBR process, Study 2 evaluates the effectiveness of AR-enhanced Briteller with two primary goals: (1) to assess whether the new features (Section~\ref{ar-briteller}), designed as add-ons to the tangible interface, address the limitations identified in Study 1, and (2) evaluate it with more diverse learners, including underrepresented groups missing in Study 1, as} increasing the persistence of minority students in STEM is crucial~\cite{estrada2016improving}, and AI literacy is particularly important for these groups who may be disproportionately impacted by algorithmic harms~\cite{solyst2023would}.

\subsection{Participants}
The participants consisted of ten students from two summer camps in an ethnically and economically diverse urban school district in Northern New York (Table~\ref{participant}). 
Most participants identified as black or African American, and some reported being biracial. 

\begin{table}[h!]
\centering
\caption{Basic demographic information of students who participated in the evaluation study.}
\label{participant}
\resizebox{0.49\textwidth}{!}{
\begin{tabular}{llllp{0.2\textwidth}}
\toprule
PID & Gender & Age & Grade & Race \\
\midrule
B1 & Female & 16 & 10th & Asian/ Pacific Islander \\
B2 & Female & 16 & 10th & Black or African American \\
B3 & Male & 15 & 9th & Black or African American \\
B4 & Male & 15 & 9th & Black or African American \\
B5 & Female & 16 & 10th & Black or African American; Hispanic \\
B6 & Prefer not to say & 16 & 11th & Asian/ Pacific Islander \\
B7 & Male & 13 & 8th & Black or African American \\
B8 & Male & 12 & 7th & Black or African American \\
B9 & Male & 11 & 7th & Amerian Indian or Alaskan Native; Black or African American; Hispanic \\
B10 & Male & 13 & 8th & Black or African American \\

\bottomrule
\end{tabular}
}
\end{table}

\subsection{Study Procedure}
The study took place on-site during two summer camps. The entire study procedure lasted 1 to 1.5 hours (Table~\ref{tab:procedure}). \textcolor{black}{Participants interacted with AR-enhanced Briteller (Section~\ref{ar-activity}) for about 40 minutes. Two researchers resided in one room to assist students, as in Study 1.} For each activity, two students collaborated, or one student completed it independently if no other students were available for pairing.

\begin{table}[h!]
\centering
\caption{The overview of the study procedure.}
\scalebox{0.98}{
\begin{tabular}{p{0.21\linewidth}p{0.56\linewidth}}
\toprule
Session & Activities \\
\midrule
Pre-survey & (1) demographic information; (2) pre-test on the target concepts about AI recommendation systems (i.e., data representation, the dot product). \\
Warm-up & Introduction to AI and its everyday application. \\
AR-enhanced Briteller & (1) tangible activities; (2) AR activities. \\
Post-survey & Post-test on the target concepts about AI recommendation systems. \\
\bottomrule
\end{tabular}
}
\label{tab:procedure}
\end{table}

\subsection{Data Collection}

\emph{Pre- and post-test.}
Before and after interacting with AR-enhanced Briteller, participants answered the same set of assessment questions from Study 1 (Table~\ref{tab:lg}(2)), plus an additional question about expanding data vectors: As an AI engineer, how can you make the recommendations more accurate?


\emph{Video \& screen recording.}
We used two cameras to record participants' verbal and nonverbal behaviors during activities, including their interactions with the AR-enhanced Briteller, explanations of their reasoning, and responses to tasks from the facilitator. We also recorded the screen while participants were using AR. This recording provided more detailed information about students' observations and interactions related to the virtual light paths, value sliders, reading the item visibility, and adding data attributes.

\subsection{Measure and Data Analysis}
\label{ar-measure}
\emph{For RQ1:}
We measured students' learning gains on target AI concepts using pre- and post-tests (Table~\ref{tab:test2}(1)).

\emph{For RQ2:}
We applied the same data analysis method described in Section \ref{study1rq2measure} to the tangible activities session. For the AR activities session, we defined major embodied interactions supported by the system (Table~\ref{tab:ar-behaviors}), which can help students learn the recommendation system and math. We analyzed the occurrence of embodied interactions with video recordings. We also annotated their co-occurrences with verbal learning behaviors (Table~\ref{tab:ar-behaviors}).

\begin{table*}[h!]
    \centering
    \caption{Major embodied learning experiences that AR enhancement aims to support: \textbf{(A)} embodied interactions and visualizations for learning and \textbf{(B)} verbal learning behaviors demonstrating students' understanding.}
    \label{tab:ar-behaviors}
    \begin{tabular}{p{0.56\textwidth}p{0.4\textwidth}}
    \toprule
        \textbf{(A)} Embodied interaction & \textbf{(B)} Verbal learning behaviors \\
        \midrule
        1. Slide value slider to adjust values & Explain the dot product process using numbers in AR\\
        2. Observe the intensity of the light dot, light beams, or the numerical result & Describe observations and explain the changes\\
        3. Slide value slider to adjust attribute's values & Explain the values for the data attributes \\
        4. Press flashlights to turn on and off light  & Describe how this impact the output\\
        5. Click to generate recommendations for different items & Describe observations, explain the output \\
        6. Click to navigate between different items & Describe items, explain the visibility of items \\
        7. Add more flashlights to expand data bars & Explain the new attributes, explain the understanding of data representation \\
        8. Press data bar labels to change the name of data attributes & Explain the data attributes \\
        \bottomrule
    \end{tabular}
\end{table*}

\section{Study 2: Results}
\label{ar-findings}

In this section, we present our findings organized around two key themes: (1) embodied learning behaviors and students’ understanding of the system, and (2) the role of AR in bridging the learning gap.
Overall, our design effectively supported students in learning AI concepts including data vectors, prediction, and the dot product.
\subsection{RQ1: Can Light Serve as a New Embodied Learning Medium for Children to Understand AI Recommendation Systems?}

\begin{table*}[h!]
\centering
\caption{Paired Wilcoxon results for pre-and post-tests (N=10).}
\label{tab:test2}
\resizebox{0.98\textwidth}{!}{
\begin{tabular}{p{0.63\textwidth}p{0.01\textwidth}*{6}{l}}
\toprule
\textbf{(1)} Knowledge Component &  & \multicolumn{2}{l}{Pre-Test} & \multicolumn{2}{l}{Post-Test} 
&{{\it p}} \\ 
\cline{3-6}
 & & M & SD & M & SD & &\\
 \midrule
 The user representation in a recommendation system & & 0.70	& 1.19		& 2.20	& 1.17	& 0.024	\\ 
\midrule
 The item representation in a recommendation system & & 0.10	& 0.30		& 2.10	& 1.37	& 0.011 \\ 
\midrule
 The output of a recommendation system & & 1.00	& 1.34		& 2.70	& 0.90	& 0.020	\\ 
\midrule
 The dot product in recommendation systems & & 0.40	& 0.92		& 3.00	& 0	& 0.004 \\ 
\midrule
Multiplication in the dot product: Multiply corresponding values from user \& item vectors &  & 1.50	& 1.50	& 3.00	& 0.00	& 0.025 \\
\midrule
 The dot product calculation for a recommendation system & & 0.40 & 0.92 & 2.70	& 0.90 & 0.007 \\
 \midrule
Feature selection for AI recommendation systems &  & 1.20	& 1.47 & 3.00 & 0.00 & 0.014 \\ 
\bottomrule
\end{tabular}
}
\end{table*}

The average differences of all questions between pre- and post-tests were not normally distributed. A paired-sample Wilcoxon signed-rank test, a non-parametric statistical test, was conducted (Table~\ref{tab:test2}). The results showed significant increases for all seven assessment questions, with p-values $<$ 0.05. 
Furthermore, the mean pre-test results showed that participants in \textcolor{black}{Study 2} had little prior knowledge of concepts related to AI recommendation systems (Table~\ref{tab:test2}).

\subsection{RQ2: What Are the Unique Affordances and Limitations of Using Light to Support the Understanding of AI Recommendation Mechanisms?}
\subsubsection{Effectiveness and Limitations of Embodied Metaphors in Briteller} 
Study 2 revalidated embodied metaphors in Briteller as effective in supporting students' exploration, experimentation, and explanation, consistent with Study 1. The evidence is presented in Table (\ref{tab:embodied2-data1}-\ref{tab:embodied2-data3}). The section focuses on the new findings of Study 2.

\paragraph{Challenges in Understanding Decimal Multiplication with BLOCKAGE} 
Students faced greater difficulties in understanding multiplication compared to addition. Seven students conceptualized MERGING as addition, using phrases like ``Add together'' (B1, B4), ``Add them all'' (B7), and ``Addition'' (B6, B9) to describe it. However, only one student (B6) recognized the multiplication from the beginning, indicating that the BLOCKAGE might be less intuitive for representing multiplication. Four students (B3, B7, B9, B10) initially thought it represented subtraction, and one student (B8) thought it was an addition. Four students (B1, B2, B4, B5) did not offer their understanding initially. The primary misunderstanding was subtraction, which is reasonable given that all students understood each knob could block part of the light, leading them to think the knobs were subtracting light, rather than realizing the operation involved two knobs. For example, B9 believed it was subtraction, rotating one knob, ``When I do this, I'm taking away from my light.'' B10 insisted it was subtraction: ``It's still subtraction, it's taking it away.'' Only B8 initially thought it was an addition, believing: ``These two had added on each and made it brighter. Added on track.'' Notably, four students (B3, B7, B8, B9) arrived at the correct answer through experimentation, particularly by using extreme values such as 0. The detailed evidence can be found in Table (\ref{tab:embodied2-data1}-\ref{tab:embodied2-data3}). It indicated the interplays between embodied metaphors still can support students’ understanding through experimentation.

\paragraph{With Briteller, Students Who Demonstrated Struggles with Calculating Math Verbally Can Reason with Light} 
We found that three students (B8, B9, B10) who made errors in simple oral arithmetic showed extensive exploration and experimentation using light. They used many gestures to explain their understanding; for example, all three used gestures to illustrate the light's journey. B8 appeared confused when performing oral calculations (e.g., ``One plus one equals 2... equals 5, wait, 6.'' ``It’s one plus 0.5, 1.5. That’s 0.51.''). However, he quickly finished the activities and did a lot of exploration with light. He understood decimal multiplication using a color filter. B8 rotated one knob to 1 and placed the color filter, and the final light dot appeared red. He then changed the value to 0 and placed the filter again, noticing the light disappeared: ``So 0 times 1 equals 0, right? So put it as 0, 0 going to reflect it on the board.'' B9, who made a simple error in oral calculation by saying, ``One times one equals 2,'' also displayed extensive exploration and understood multiplication by experimenting with light. B9 was the only student who discovered how Briteller used polarized films to control the light, and he also combined color filters to create new colors of light. B10, who did not grasp multiplication initially and said, ``0 and 1 will get 0 by subtraction,'' mentioned that he enjoyed experimenting with flashlights at home, which helped him quickly understand how the system worked.

\subsubsection{Affordances and Limitations of AR Enhancement}
\paragraph{Visible Light Beams and Quantifiable Light Intensity Enhance Students' Understanding of the Dot Product} 
The clearer light beams in AR aided observation, with four students (B7, B8, B9, B10) opting to view it from a side perspective, which is invisible in the tangible setting. Three students (B4, B5, B10) explicitly commented on the light beams in AR. B10 remarked, ``It’s much better.'' The numerical information helped scaffold students' transition from qualitative descriptions of the light—such as ``darker'', ``brighter'', ``higher'', or ``lower'' —to a more mathematical understanding. Nine students used AR to experiment with different values, observe the number changes, and explain the dot product process with the help of numerical information. Four students (B1, B2, B4, B5), who did not express their understanding of multiplication in the tangible setting, were able to explain the dot product with the assistance of the numbers in AR. For instance, B1 and B2 set the first data bar to (1, 1, 1) and the second to (0, 1, 1), with B1 explaining the final numerical result: ``0, 1, and 0, added together equals 1.'' B8 and B9, who understood the dot product in the tangible setting but made errors during oral calculations, were able to use the AR numerical information to do calculations, indicating that the AR numbers may have reduced their cognitive load. For example, B8 was initially unsure what result he would get if he rotated all the knobs to 1. When using AR with the numerical information, he quickly and correctly explained, ``To get 3, I just want to put all of them to 1.'' Similarly, B9 used AR as a calculator to help him. Evidence is detailed in Table~\ref{tab:ar-data1-1}.

\paragraph{Overlaying Item Visibility and Light Dot Contextualizes the Light Output}
AR enhancement provided an interactive experience with enhanced visual information, helping students better contextualize the recommendation system. Nine students articulated how the system worked, referencing the user and item vectors. During the AR activities, students input their preferences and generated personalized recommendations, which led to more engaged responses compared to merely observing light brightness. Students commented on their preferences and the quality of the recommendations. For example, B1 and B2 laughed upon seeing a lollipop recommended, while B4 adjusted his user vector after the first recommendation, saying, ``I want to switch a little bit''. B6 observed, ``Oh, so it's just based on my preference''.

In the tangible activities, students primarily commented on light brightness as an indicator of preference. Using AR, the visual contextual information helped them directly link the item visibility with the recommendation levels in the AI system. For example, B1 and B2 found that cake was the first and brightest item, with B1 saying, ``Because it's the most recommended.'' B3 noticed that the light was darker for certain items, and explained, ``It has zero fat and low sugar and salt. I didn't put those in, so it knows I don’t like it.'' B4 similarly noted, ``Keep decreasing. These are the ones I don't like.'' He pointed out that the broccoli’s low visibility was due to his preference settings. B5 immediately understood that dimmer items were not recommended, saying, ``Those were the ones that aren't recommended.'' B6 navigated between items and observed the brightness changes, stating, ``Some are dimmer, some are brighter.'' He also noticed the numerical values, interpreting them as percentages: ``Less recommended, higher recommended—it's a higher percentage based on the values I’ve inputted.'' B8 realized broccoli was invisible, explaining, ``Probably it’s not recommended. This is like the last thing recommended.'' The detailed evidence can be found in Table (\ref{tab:ar-data2-1}-\ref{tab:ar-data2-2}).

\paragraph{Expanding Attributes in AR Support Generalizable Understanding of Data Representation} 
All students were able to use structured data vectors to represent unstructured information and completed the learning activities. For example, students inferred that a person who loves sports would ``like fruit'', ``want to stay healthy'', and ``prefer less sugar'' and set the user’s sugar vector low. They also designed item vectors thoughtfully. When designing new attributes, students utilized their creativity. B6 listed attributes like calories, ingredients, and vitamins. B8 added several new attributes, such as ``Broccoli has a lot of vitamins, is healthy, fresh, and low cost,'' and even asked if more flashlights could be added to represent additional features. B9 changed the attributes to ``caffeine, sugar, water'' because he wanted the system to recommend coffee. However, three students (B1, B2, B5) found it somewhat confusing, and three others (B1, B2, B3) only came up with one new attribute. The detailed evidence can be found in Table (\ref{tab:ar-data3-1}-\ref{tab:ar-data3-2}).

Four students (B2, B3, B6, B9) extended their understanding of data representation beyond food recommendations and connected it to everyday experiences. B2 remarked that the system worked by ``taking your interests and recommending more stuff'' and that ``adding more interests'' could improve its accuracy. B3 observed that adding attributes helped the system ``know more'', relating it to how personal preferences inform machine recommendations in daily life: ``The information they wanted to know was about what I eat or do. The preference would be sent to the machine so it knows what I like and recommends it to me.'' B6 drew a parallel with TikTok’s data filtering, explaining that TikTok asks for a user’s age to filter content: ``It filters out if users are underage or show similar content by clicking `like'.'' B9 concluded, ``If we add more attributes, the light expands, and the recommendation system adds more things, so it can know more about what they like.''

\paragraph{Limitations of Tablet-Based AR} We observed that using tablet-based AR reduced students use of meaningful gestures, which often convey reasoning. With Briteller, students engaged in more extensive hands-on experimentation, but with AR, they were limited to clicking the tablet screen. Some students also struggled with tablet use—B10 had trouble sliding the slider and experienced unresponsive clicks. B8 asked for assistance with tapping new attributes. These issues may be more challenging for younger children.

\section{Discussion}

\subsection{Light's Benefits for Diverse Learners}
We propose that light-based tangible manipulation can enhance learning abstract AI concepts, offering benefits for diverse learners, especially those from historically underrepresented groups or those disengaged by traditional learning environments. This approach makes abstract AI concepts more familiar, interdisciplinary, and tangible for learners.

First, light fosters curiosity and engagement by linking familiar phenomena to abstract concepts. In our study, students quickly grasped visual metaphors like DARK-BRIGHT, as these concepts, along with common optical materials like convex lenses and flashlights, resonated with their daily experiences and existing knowledge. This familiarity reduced the intimidation often associated with advanced computing topics. Feedback from students highlighted that they found this approach creative and engaging for learning AI and math.

\textcolor{black}{Second, light provides interdisciplinary learning opportunities, blending AI, mathematics, physics, and color theory. The benefits of interdisciplinary approaches in computing education are well-documented, particularly in their appeal to underrepresented groups~\cite{carter2014interdisciplinary, kulkarni2018promoting}. Prior research on integrating textile fabrication, mathematics, and engineering through loom kits demonstrates how interdisciplinary methods broaden participation~\cite{speer2023speerloom}. Similarly, some students in our study expressed curiosity about the optical principles underlying Briteller. 
Notably, a student from the underrepresented group demonstrated a strong interest in optical phenomena and was the only participant to discover and explain how polarized films control light. 
Optical neural network research shows that extensive matrix operations
can be modeled through light propagation~\cite{fu2024optical}, highlighting the potential of light-based interaction for more advanced AI education.}

\textcolor{black}{Third, our findings indicate that tangible manipulation with light benefits diverse students by scaffolding reasoning.
In Study 2, students from underrepresented backgrounds engaged in self-directed exploration and active manipulation of optical materials.}
Students who struggled with math calculations appeared to benefit from tangible manipulation.
For example, B10 struggled to verbalize his understanding. Despite making errors in simple oral arithmetic, he frequently used gestures to convey his comprehension of Briteller and preferred experimenting with light. His embodied interactions during the learning activities were accurate and aligned with the correct outcomes. B8 and B9, who made errors in simple oral arithmetic also demonstrated active exploration and experimentation.
Such observation aligns with previous studies suggesting that tangible tools can reduce cognitive load by enabling students to utilize external resources~\cite{li2022meta, esteves2013physical} and gestures often reflect students’ thought processes~\cite{hostetter2008visible}. 
\textcolor{black}{Together, these findings underscore the potential of light-based tangible interactions to make abstract AI concepts more accessible and engaging for diverse learners.}
 
\subsection{Reflecting on Light-Based Tangible Interfaces in Computing Education}
\label{discussion1}
Although light-based tangible interfaces benefit learners, we also acknowledge their limitations. 
First, using light as a manipulative to represent AI concepts may hinder learning transfer (i.e., applying knowledge to new situations without using instructional manipulatives~\cite{nathan2012rethinking}). In our study, some learners focused on the light phenomena without generalizing their understanding to other recommendation systems. 
Second, purely tangible light-based systems lack numeric precision. Previous research also highlights a key criticism of manipulatives, noting that children cannot simply induce numerical meaning from physical representations~\cite{manches2012tangibles}. Students in our study interpreted light intensity as ``low'', ``medium'', or ``high'' but found it difficult to discern exact values. This lack of precision made it harder for them to grasp the quantitative relationships between user preferences and item attributes. 

Third, light-based tangible systems face challenges in supporting complex AI functions and may lead to higher manufacturing complexity and costs. For example, the number of attributes in a data vector is limited. It is also challenging to represent negative values, handle unstructured data, or simulate advanced AI functions such as learning from interactions, automatic profile updates, and fine-tuning.
\textcolor{black}{To address these physical constraints, we incorporated tablet-based AR, which mitigated some limitations but reduced opportunities for meaningful gestures. One potential solution to balance gesture use and digital information is the use of TUIs, which integrate the digital and physical worlds and allow digital information to be directly manipulated through physical interactions~\cite{ishii2008tangible}. Wearable devices, such as smart glasses with vision-based hand pose tracking techniques are another promising alternative~\cite{liang2015ar}.}

\textcolor{black}{Overall, in educational settings, alternative tools should be critically evaluated for their trade-offs between accessibility, practicality, and learning impact to ensure ecological validity. While our study highlights the benefits of light-based tangible interaction, designing physical systems to map AI concepts requires significantly more effort and resources compared to traditional methods. The HCI community has noted sustainability challenges associated with tangible artifacts, including the high costs of creation, modification, maintenance, and distribution~\cite{morais2024exploration, holmquist2023bits}. A simpler approach may achieve similar outcomes with less effort. Future research should assess whether the advantages of light-based physicalization justify these challenges and explore designs that balance learning impact with practicality.}

\subsection{Design Metaphors for AI and Math}
Metaphors are essential to mathematical thinking, allowing learners to reason about abstract domains using the inferential structure of concrete ones~\cite{lakoff2000mathematics, PRESMEG199825}. Our embodied metaphors effectively supported learners in understanding content-based recommendation systems. For instance, BRIGHT-DARK and HUE enabled students to intuitively link light intensity with prediction. Our findings also align with embodied mathematical theory, which distinguishes between grounding and linking metaphors~\cite{lakoff2000mathematics}. Grounding metaphors, like MERGING for addition, were intuitive, while linking metaphors which involve the blending of different metaphors, such as BLOCKAGE and LINKAGE for multiplication, required more explicit instruction ~\cite{lakoff2000mathematics}. 
A notable finding from Study 2 was students' difficulty with decimal multiplication, with nearly half initially interpreting it as subtraction. This may arise from two factors: (1) difficulty in extracting precise values from light brightness and (2) misconceptions regarding the effects of multiplying numbers between 0 and 1. Prior research highlights that fraction and decimal arithmetic are challenging for learners, partly due to the misconception that multiplication always results in a larger number, a misunderstanding largely attributed to the repeated-addition interpretation commonly taught in U.S. classrooms~\cite{lortie2015learning}.

\textcolor{black}{\citeauthor{abrahamson2012try} propose that metaphors act as discursive tools, enabling students and educators to create or refine shared meanings in learning environments. Our light-based metaphors can be internalized by students or referenced by educators in future teaching. For instance, our design provides a novel way to explain decimal arithmetic using light, offering a valuable complement to traditional approaches. }
Given the complexity of AI systems, which involve advanced algorithms and mathematics, revealing their inner workings to students without overwhelming them is a challenge. Future work should refine and expand metaphors to support both intuitive understanding and deeper engagement in learning math and AI, drawing from cognitive science insights.

\subsection{Connect Tangibility and Explainable AI Through Data Physicalization}
Briteller connects light-based tangible interaction with a selection of building blocks of content-based recommendation systems, including (1) data representation of users and items (user and item vectors), (2) the dot product operation between these vectors, and (3) the recommendation output. 
This fills in the gap of creating a datafied representation of tangible recommendation systems~\cite {alvarado2022towards}.
In this section, we discuss how such connections support students in grasping AI literacy by aligning with tangible XAI approaches~\cite{colley2022tangible,belle2021principles}, including feature relevance, local explanations, and simplified rule extraction.

First, the approach of \emph{feature relevance} suggests explaining AI output through visualizing the influence of each input parameter on the model's output~\cite{lundberg2017unified}. In Briteller, learners place a color filter at a specific part of the light. The intensity of the color in the final light dot demonstrates the influence of that specific data attribute on the output. 
Second, the approach of \emph{local explanations} closely connects individual input and output modes~\cite{ishii1997tangible,jorda2007reactable}. In Briteller, learners rotate a knob with a polarized film to block a specific amount of light passing through it and view the changes in the light output.
Third, the selection of content-based recommendation's building blocks (i.e., user vector, item vector, the dot product, recommendation output) provides a simpler transparent proxy model of the AI black box, following the approach of \emph{simplified rule extraction}~\cite{ishii1997tangible}. Such taxonomy of light-based tangible interfaces for embodying abstract AI concepts can be extended to create tangible XAI for other algorithms.

\textcolor{black}{\citeauthor{alvarado2022towards} note that while participants valued granular control in TUI, many still perceived their prototype as merely a remote control, without adding transparency. They suggest exploring datafied representation to improve algorithmic transparency. Our findings support this by demonstrating how light can be used as a material for data physicalization, offering dynamic visual feedback that enhances transparency. By manipulating data attributes, students gained insights into content-based recommendation systems. Tangible manipulation also holds the potential for handling more complex data processing tasks, as demonstrated by Physically Dynamic Bar Charts~\cite{taher2015exploring}. Given the numerous features and attributes of AI systems, future work should explore designing interfaces that integrate XAI principles, enabling learners to physically manipulate these features and foster a deeper understanding of AI algorithms.}

\section{Limitations and Future Work}
\subsection{Limitations}
\textcolor{black}{his study has several key limitations. 
\emph{First}, 
the AR-enhanced Briteller was developed based on insights from existing literature and practical requirements, serving as a research instrument to investigate our core ideas. 
This lacked the co-design process and limited the exploration of alternatives, such as tangible computing tools. Future studies should directly compare these approaches in terms of learning outcomes and user experience.
\emph{Second}, following an iterative design approach, Study 2 focused on evaluating the new AR-enhanced Briteller rather than comparing tangible and AR systems. Because students used both versions, the lack of a formal comparative study makes it difficult to isolate their unique contributions. 
\emph{Third}, while this study measured individual learning outcomes, most participants worked collaboratively. We observed that most pairs engaged in meaningful discussions and worked well together, which likely contributed positively to their learning gains. We also noticed group dynamics varied, as some students were more active than others. Researchers facilitated activities to ensure more balanced participation. However, the impact of collaboration on learning outcomes was not systematically analyzed, and future research should examine the impact of collaborative versus individual learning setups on outcomes and engagement.
\emph{Fourth}, the study adopts an exploratory design aligned with educational research in dynamic contexts~\cite{barab2016design}. However, the small sample size and flexible setting limit statistical analysis.
\emph{Lastly}, while conducting Study 2 in summer camps enhanced ecological validity, this study was conducted with researcher facilitation, which may not fully reflect how the tools would function in real classrooms. We did not examine how teachers might integrate Briteller into their teaching or its long-term impact. Future research should assess its effectiveness in real-world learning environments through longitudinal studies.}

\subsection{Future Directions} 
\label{future-work}
To address the aforementioned limitations, future work will focus on (1) exploring alternative methods for integrating validated features into the tangible system to maintain meaningful interactions that support learning, (2) conducting comparative studies, and (3) enhancing ecological validity through diverse educational settings, including real-world classroom applications. Additionally, we propose the following future directions. Deploying a light-based recommendation system in science museums is promising~\cite{amato2023recommender,trichopoulos2023large}. Immersive exhibits and workshops could use light-based tangible interfaces to visualize algorithms, making AI concepts engaging and accessible to the public. Lastly, Briteller can be iterated for greater scalability and versatility. The system's base could be made expandable for portability, and its components modularized to create a more powerful and adaptable toolbox for children. 

\section{Conclusion}
We developed Briteller, a light-based tangible interface, to help children grasp AI recommendation mechanisms. Our initial study revealed key affordances and limitations, which we addressed by integrating AR enhancements. We then evaluated the AR-enhanced Briteller with students from underrepresented backgrounds.
Findings show that light-based physicalization demystifies AI by linking metaphors with abstract concepts, fosters exploration and experimentation, and scaffolds reasoning through tangible interaction, benefiting diverse learners. 
Challenges remain in learning transfer, physical constraints, and manufacturing complexity. 
While tablet-based AR addressed some limitations, it limited meaningful gestures.
This work highlights light as a promising medium for graspable AI and offers insights for integrating embodied metaphors, tangible interaction, and visual augmentation to enhance AI literacy.

\begin{acks}
We appreciate the student participants. We also acknowledge the support of the National Science Foundation (NSF) RETTL program (Award No. 2238675) and the NSF NRT project (Award No. 1922591).
\end{acks}

\bibliographystyle{ACM-Reference-Format}
\bibliography{optidot}


\begin{thebibliography}{120}


\ifx \showCODEN    \undefined \def \showCODEN     #1{\unskip}     \fi
\ifx \showDOI      \undefined \def \showDOI       #1{#1}\fi
\ifx \showISBNx    \undefined \def \showISBNx     #1{\unskip}     \fi
\ifx \showISBNxiii \undefined \def \showISBNxiii  #1{\unskip}     \fi
\ifx \showISSN     \undefined \def \showISSN      #1{\unskip}     \fi
\ifx \showLCCN     \undefined \def \showLCCN      #1{\unskip}     \fi
\ifx \shownote     \undefined \def \shownote      #1{#1}          \fi
\ifx \showarticletitle \undefined \def \showarticletitle #1{#1}   \fi
\ifx \showURL      \undefined \def \showURL       {\relax}        \fi
\providecommand\bibfield[2]{#2}
\providecommand\bibinfo[2]{#2}
\providecommand\natexlab[1]{#1}
\providecommand\showeprint[2][]{arXiv:#2}

\bibitem[Abrahamson et~al\mbox{.}(2012)]%
        {abrahamson2012try}
\bibfield{author}{\bibinfo{person}{Dor Abrahamson}, \bibinfo{person}{Jos{\'e}~F Guti{\'e}rrez}, {and} \bibinfo{person}{Anna~K Baddorf}.} \bibinfo{year}{2012}\natexlab{}.
\newblock \showarticletitle{Try to see it my way: The discursive function of idiosyncratic mathematical metaphor}.
\newblock \bibinfo{journal}{\emph{Mathematical Thinking and Learning}} \bibinfo{volume}{14}, \bibinfo{number}{1} (\bibinfo{year}{2012}), \bibinfo{pages}{55--80}.
\newblock


\bibitem[Abrahamson et~al\mbox{.}(2020)]%
        {abrahamson2020future}
\bibfield{author}{\bibinfo{person}{Dor Abrahamson}, \bibinfo{person}{Mitchell~J Nathan}, \bibinfo{person}{Caro Williams-Pierce}, \bibinfo{person}{Candace Walkington}, \bibinfo{person}{Erin~R Ottmar}, \bibinfo{person}{Hortensia Soto}, {and} \bibinfo{person}{Martha~W Alibali}.} \bibinfo{year}{2020}\natexlab{}.
\newblock \showarticletitle{The future of embodied design for mathematics teaching and learning}. In \bibinfo{booktitle}{\emph{Frontiers in Education}}, Vol.~\bibinfo{volume}{5}. Frontiers Media SA, \bibinfo{pages}{147}.
\newblock


\bibitem[Akgun and Greenhow(2022)]%
        {akgun2022artificial}
\bibfield{author}{\bibinfo{person}{Selin Akgun} {and} \bibinfo{person}{Christine Greenhow}.} \bibinfo{year}{2022}\natexlab{}.
\newblock \showarticletitle{Artificial intelligence in education: Addressing ethical challenges in K-12 settings}.
\newblock \bibinfo{journal}{\emph{AI and Ethics}} \bibinfo{volume}{2}, \bibinfo{number}{3} (\bibinfo{year}{2022}), \bibinfo{pages}{431--440}.
\newblock


\bibitem[Alavi and Dillenbourg(2012)]%
        {alavi2012ambient}
\bibfield{author}{\bibinfo{person}{Hamed~S Alavi} {and} \bibinfo{person}{Pierre Dillenbourg}.} \bibinfo{year}{2012}\natexlab{}.
\newblock \showarticletitle{An ambient awareness tool for supporting supervised collaborative problem solving}.
\newblock \bibinfo{journal}{\emph{IEEE Transactions on Learning Technologies}} \bibinfo{volume}{5}, \bibinfo{number}{3} (\bibinfo{year}{2012}), \bibinfo{pages}{264--274}.
\newblock


\bibitem[Ali et~al\mbox{.}(2019)]%
        {ali2019constructionism}
\bibfield{author}{\bibinfo{person}{Safinah Ali}, \bibinfo{person}{Blakeley~H Payne}, \bibinfo{person}{Randi Williams}, \bibinfo{person}{Hae~Won Park}, {and} \bibinfo{person}{Cynthia Breazeal}.} \bibinfo{year}{2019}\natexlab{}.
\newblock \showarticletitle{Constructionism, ethics, and creativity: Developing primary and middle school artificial intelligence education}. In \bibinfo{booktitle}{\emph{International workshop on education in artificial intelligence k-12 (eduai’19)}}. \bibinfo{pages}{1--4}.
\newblock


\bibitem[Alshamrani(2020)]%
        {alshamrani2020detecting}
\bibfield{author}{\bibinfo{person}{Sultan Alshamrani}.} \bibinfo{year}{2020}\natexlab{}.
\newblock \showarticletitle{Detecting and measuring the exposure of children and adolescents to inappropriate comments in YouTube}. In \bibinfo{booktitle}{\emph{Proceedings of the 29th ACM International Conference on Information \& Knowledge Management}}. \bibinfo{pages}{3213--3216}.
\newblock


\bibitem[Alvarado et~al\mbox{.}(2021)]%
        {alvarado2021exploring}
\bibfield{author}{\bibinfo{person}{Oscar Alvarado}, \bibinfo{person}{Vero Vanden~Abeele}, \bibinfo{person}{David Geerts}, \bibinfo{person}{Francisco Guti{\'e}rrez}, {and} \bibinfo{person}{Katrien Verbert}.} \bibinfo{year}{2021}\natexlab{}.
\newblock \showarticletitle{Exploring tangible algorithmic imaginaries in movie recommendations}. In \bibinfo{booktitle}{\emph{Proceedings of the Fifteenth International Conference on Tangible, Embedded, and Embodied Interaction}}. \bibinfo{pages}{1--12}.
\newblock


\bibitem[Alvarado et~al\mbox{.}(2022)]%
        {alvarado2022towards}
\bibfield{author}{\bibinfo{person}{Oscar Alvarado}, \bibinfo{person}{Vero Vanden~Abeele}, \bibinfo{person}{David Geerts}, {and} \bibinfo{person}{Katrien Verbert}.} \bibinfo{year}{2022}\natexlab{}.
\newblock \showarticletitle{Towards tangible algorithms: Exploring the experiences of tangible interactions with movie recommender algorithms}.
\newblock \bibinfo{journal}{\emph{Proceedings of the ACM on Human-Computer Interaction}} \bibinfo{volume}{6}, \bibinfo{number}{CSCW2} (\bibinfo{year}{2022}), \bibinfo{pages}{1--30}.
\newblock


\bibitem[Amato(2023)]%
        {amato2023recommender}
\bibfield{author}{\bibinfo{person}{Alba Amato}.} \bibinfo{year}{2023}\natexlab{}.
\newblock \showarticletitle{Recommender Systems in the Museum Sector: An Overview}. In \bibinfo{booktitle}{\emph{International Conference on Advanced Information Networking and Applications}}. Springer, \bibinfo{pages}{251--260}.
\newblock


\bibitem[Anderson and Shattuck(2012)]%
        {anderson2012design}
\bibfield{author}{\bibinfo{person}{Terry Anderson} {and} \bibinfo{person}{Julie Shattuck}.} \bibinfo{year}{2012}\natexlab{}.
\newblock \showarticletitle{Design-based research: A decade of progress in education research?}
\newblock \bibinfo{journal}{\emph{Educational researcher}} \bibinfo{volume}{41}, \bibinfo{number}{1} (\bibinfo{year}{2012}), \bibinfo{pages}{16--25}.
\newblock


\bibitem[Angelov et~al\mbox{.}(2021)]%
        {angelov2021explainable}
\bibfield{author}{\bibinfo{person}{Plamen~P Angelov}, \bibinfo{person}{Eduardo~A Soares}, \bibinfo{person}{Richard Jiang}, \bibinfo{person}{Nicholas~I Arnold}, {and} \bibinfo{person}{Peter~M Atkinson}.} \bibinfo{year}{2021}\natexlab{}.
\newblock \showarticletitle{Explainable artificial intelligence: an analytical review}.
\newblock \bibinfo{journal}{\emph{Wiley Interdisciplinary Reviews: Data Mining and Knowledge Discovery}} \bibinfo{volume}{11}, \bibinfo{number}{5} (\bibinfo{year}{2021}), \bibinfo{pages}{e1424}.
\newblock


\bibitem[Antle et~al\mbox{.}(2009)]%
        {antle2009body}
\bibfield{author}{\bibinfo{person}{Alissa~N Antle}, \bibinfo{person}{Greg Corness}, {and} \bibinfo{person}{Milena Droumeva}.} \bibinfo{year}{2009}\natexlab{}.
\newblock \showarticletitle{What the body knows: Exploring the benefits of embodied metaphors in hybrid physical digital environments}.
\newblock \bibinfo{journal}{\emph{Interacting with Computers}} \bibinfo{volume}{21}, \bibinfo{number}{1-2} (\bibinfo{year}{2009}), \bibinfo{pages}{66--75}.
\newblock


\bibitem[Bae et~al\mbox{.}(2022)]%
        {bae2022making}
\bibfield{author}{\bibinfo{person}{S~Sandra Bae}, \bibinfo{person}{Clement Zheng}, \bibinfo{person}{Mary~Etta West}, \bibinfo{person}{Ellen Yi-Luen Do}, \bibinfo{person}{Samuel Huron}, {and} \bibinfo{person}{Danielle~Albers Szafir}.} \bibinfo{year}{2022}\natexlab{}.
\newblock \showarticletitle{Making data tangible: A cross-disciplinary design space for data physicalization}. In \bibinfo{booktitle}{\emph{Proceedings of the 2022 CHI Conference on Human Factors in Computing Systems}}. \bibinfo{pages}{1--18}.
\newblock


\bibitem[Barab and Squire(2016)]%
        {barab2016design}
\bibfield{author}{\bibinfo{person}{Sasha Barab} {and} \bibinfo{person}{Kurt Squire}.} \bibinfo{year}{2016}\natexlab{}.
\newblock \showarticletitle{Design-based research: Putting a stake in the ground}.
\newblock In \bibinfo{booktitle}{\emph{Design-based Research}}. \bibinfo{publisher}{Psychology Press}, \bibinfo{pages}{1--14}.
\newblock


\bibitem[Belle and Papantonis(2021)]%
        {belle2021principles}
\bibfield{author}{\bibinfo{person}{Vaishak Belle} {and} \bibinfo{person}{Ioannis Papantonis}.} \bibinfo{year}{2021}\natexlab{}.
\newblock \showarticletitle{Principles and practice of explainable machine learning}.
\newblock \bibinfo{journal}{\emph{Frontiers in big Data}}  \bibinfo{volume}{4} (\bibinfo{year}{2021}), \bibinfo{pages}{688969}.
\newblock


\bibitem[Black et~al\mbox{.}(2012)]%
        {black2012embodied}
\bibfield{author}{\bibinfo{person}{John~B Black}, \bibinfo{person}{Ayelet Segal}, \bibinfo{person}{Jonathan Vitale}, {and} \bibinfo{person}{Cameron~L Fadjo}.} \bibinfo{year}{2012}\natexlab{}.
\newblock \showarticletitle{Embodied cognition and learning environment design}.
\newblock In \bibinfo{booktitle}{\emph{Theoretical foundations of learning environments}}. \bibinfo{publisher}{Routledge}, \bibinfo{pages}{198--223}.
\newblock


\bibitem[B{\"o}rner et~al\mbox{.}(2015)]%
        {borner2015tangible}
\bibfield{author}{\bibinfo{person}{Dirk B{\"o}rner}, \bibinfo{person}{Bernardo Tabuenca}, \bibinfo{person}{Jeroen Storm}, \bibinfo{person}{Sven Happe}, {and} \bibinfo{person}{Marcus Specht}.} \bibinfo{year}{2015}\natexlab{}.
\newblock \showarticletitle{Tangible interactive ambient display prototypes to support learning scenarios}. In \bibinfo{booktitle}{\emph{Proceedings of the Ninth International Conference on Tangible, Embedded, and Embodied Interaction}}. \bibinfo{pages}{721--726}.
\newblock


\bibitem[Braun and Clarke(2006)]%
        {braun2006using}
\bibfield{author}{\bibinfo{person}{Virginia Braun} {and} \bibinfo{person}{Victoria Clarke}.} \bibinfo{year}{2006}\natexlab{}.
\newblock \showarticletitle{Using thematic analysis in psychology}.
\newblock \bibinfo{journal}{\emph{Qualitative research in psychology}} \bibinfo{volume}{3}, \bibinfo{number}{2} (\bibinfo{year}{2006}), \bibinfo{pages}{77--101}.
\newblock


\bibitem[Cai et~al\mbox{.}(2024)]%
        {cai2024see}
\bibfield{author}{\bibinfo{person}{Xueyan Cai}, \bibinfo{person}{Kecheng Jin}, \bibinfo{person}{Shang Shi}, \bibinfo{person}{Shichao Huang}, \bibinfo{person}{Ouying Huang}, \bibinfo{person}{Xiaodong Wang}, \bibinfo{person}{Jiahao Cheng}, \bibinfo{person}{Weijia Lin}, \bibinfo{person}{Jiayu Yao}, \bibinfo{person}{Yuqi Hu}, {et~al\mbox{.}}} \bibinfo{year}{2024}\natexlab{}.
\newblock \showarticletitle{" See, Hear, Touch, Smell, and,... Eat!": Helping Children Self-Improve Their Food Literacy and Eating Behavior through a Tangible Multi-Sensory Puzzle Game}. In \bibinfo{booktitle}{\emph{Proceedings of the 23rd Annual ACM Interaction Design and Children Conference}}. \bibinfo{pages}{270--281}.
\newblock


\bibitem[Carbonneau et~al\mbox{.}(2013)]%
        {carbonneau2013meta}
\bibfield{author}{\bibinfo{person}{Kira~J Carbonneau}, \bibinfo{person}{Scott~C Marley}, {and} \bibinfo{person}{James~P Selig}.} \bibinfo{year}{2013}\natexlab{}.
\newblock \showarticletitle{A meta-analysis of the efficacy of teaching mathematics with concrete manipulatives.}
\newblock \bibinfo{journal}{\emph{Journal of educational psychology}} \bibinfo{volume}{105}, \bibinfo{number}{2} (\bibinfo{year}{2013}), \bibinfo{pages}{380}.
\newblock


\bibitem[Carter(2014)]%
        {carter2014interdisciplinary}
\bibfield{author}{\bibinfo{person}{Lori Carter}.} \bibinfo{year}{2014}\natexlab{}.
\newblock \showarticletitle{Interdisciplinary computing classes: worth the effort}. In \bibinfo{booktitle}{\emph{Proceedings of the 45th ACM technical symposium on Computer science education}}. \bibinfo{pages}{445--450}.
\newblock


\bibitem[Casal-Otero et~al\mbox{.}(2023)]%
        {casal2023ai}
\bibfield{author}{\bibinfo{person}{Lorena Casal-Otero}, \bibinfo{person}{Alejandro Catala}, \bibinfo{person}{Carmen Fern{\'a}ndez-Morante}, \bibinfo{person}{Maria Taboada}, \bibinfo{person}{Beatriz Cebreiro}, {and} \bibinfo{person}{Sen{\'e}n Barro}.} \bibinfo{year}{2023}\natexlab{}.
\newblock \showarticletitle{AI literacy in K-12: a systematic literature review}.
\newblock \bibinfo{journal}{\emph{International Journal of STEM Education}} \bibinfo{volume}{10}, \bibinfo{number}{1} (\bibinfo{year}{2023}), \bibinfo{pages}{29}.
\newblock


\bibitem[Chatain et~al\mbox{.}(2022)]%
        {chatain2022grasping}
\bibfield{author}{\bibinfo{person}{Julia Chatain}, \bibinfo{person}{Virginia Ramp}, \bibinfo{person}{Venera Gashaj}, \bibinfo{person}{Violaine Fayolle}, \bibinfo{person}{Manu Kapur}, \bibinfo{person}{Robert~W Sumner}, {and} \bibinfo{person}{St{\'e}phane Magnenat}.} \bibinfo{year}{2022}\natexlab{}.
\newblock \showarticletitle{Grasping derivatives: Teaching mathematics through embodied interactions using tablets and virtual reality}. In \bibinfo{booktitle}{\emph{Proceedings of the 21st Annual ACM Interaction Design and Children Conference}}. \bibinfo{pages}{98--108}.
\newblock


\bibitem[Colley et~al\mbox{.}(2023)]%
        {colley2023exploring}
\bibfield{author}{\bibinfo{person}{Ashley Colley}, \bibinfo{person}{Matilda Kalving}, \bibinfo{person}{Jonna H{\"a}kkil{\"a}}, {and} \bibinfo{person}{Kaisa V{\"a}{\"a}n{\"a}nen}.} \bibinfo{year}{2023}\natexlab{}.
\newblock \showarticletitle{Exploring Tangible Explainable AI (TangXAI): A User Study of Two XAI Approaches}. In \bibinfo{booktitle}{\emph{Proceedings of the 35th Australian Computer-Human Interaction Conference}}. \bibinfo{pages}{679--683}.
\newblock


\bibitem[Colley et~al\mbox{.}(2022)]%
        {colley2022tangible}
\bibfield{author}{\bibinfo{person}{Ashley Colley}, \bibinfo{person}{Kaisa V{\"a}{\"a}n{\"a}nen}, {and} \bibinfo{person}{Jonna H{\"a}kkil{\"a}}.} \bibinfo{year}{2022}\natexlab{}.
\newblock \showarticletitle{Tangible explainable ai-an initial conceptual framework}. In \bibinfo{booktitle}{\emph{Proceedings of the 21st International Conference on Mobile and Ubiquitous Multimedia}}. \bibinfo{pages}{22--27}.
\newblock


\bibitem[Council(2013)]%
        {states13}
\bibfield{author}{\bibinfo{person}{National~Research Council}.} \bibinfo{year}{2013}\natexlab{}.
\newblock \showarticletitle{APPENDIX F: Science and Engineering Practices in the Next Generation Science Standards}.
\newblock \bibinfo{journal}{\emph{Next Generation Science Standards: For States, By States}} (\bibinfo{year}{2013}).
\newblock


\bibitem[de~Kreij et~al\mbox{.}(2024)]%
        {de2024data}
\bibfield{author}{\bibinfo{person}{Sander de Kreij}, \bibinfo{person}{Champika Ranasinghe}, {and} \bibinfo{person}{Auriol Degbelo}.} \bibinfo{year}{2024}\natexlab{}.
\newblock \showarticletitle{Data Physicalization and Tangible Manipulation for Engaging Children with Data: An Example with Air Quality Data}. In \bibinfo{booktitle}{\emph{Proceedings of the 23rd Annual ACM Interaction Design and Children Conference}}. \bibinfo{pages}{507--516}.
\newblock


\bibitem[DiPaola et~al\mbox{.}(2020)]%
        {dipaola2020decoding}
\bibfield{author}{\bibinfo{person}{Daniella DiPaola}, \bibinfo{person}{Blakeley~H Payne}, {and} \bibinfo{person}{Cynthia Breazeal}.} \bibinfo{year}{2020}\natexlab{}.
\newblock \showarticletitle{Decoding design agendas: an ethical design activity for middle school students}. In \bibinfo{booktitle}{\emph{Proceedings of the interaction design and children conference}}. \bibinfo{pages}{1--10}.
\newblock


\bibitem[Esteves et~al\mbox{.}(2013)]%
        {esteves2013physical}
\bibfield{author}{\bibinfo{person}{Augusto Esteves}, \bibinfo{person}{Elise Van Den~Hoven}, {and} \bibinfo{person}{Ian Oakley}.} \bibinfo{year}{2013}\natexlab{}.
\newblock \showarticletitle{Physical games or digital games? Comparing support for mental projection in tangible and virtual representations of a problem-solving task}. In \bibinfo{booktitle}{\emph{Proceedings of the 7th international conference on tangible, embedded and embodied interaction}}. \bibinfo{pages}{167--174}.
\newblock


\bibitem[Estrada et~al\mbox{.}(2016)]%
        {estrada2016improving}
\bibfield{author}{\bibinfo{person}{Mica Estrada}, \bibinfo{person}{Myra Burnett}, \bibinfo{person}{Andrew~G Campbell}, \bibinfo{person}{Patricia~B Campbell}, \bibinfo{person}{Wilfred~F Denetclaw}, \bibinfo{person}{Carlos~G Guti{\'e}rrez}, \bibinfo{person}{Sylvia Hurtado}, \bibinfo{person}{Gilbert~H John}, \bibinfo{person}{John Matsui}, \bibinfo{person}{Richard McGee}, {et~al\mbox{.}}} \bibinfo{year}{2016}\natexlab{}.
\newblock \showarticletitle{Improving underrepresented minority student persistence in STEM}.
\newblock \bibinfo{journal}{\emph{CBE—Life Sciences Education}} \bibinfo{volume}{15}, \bibinfo{number}{3} (\bibinfo{year}{2016}), \bibinfo{pages}{es5}.
\newblock


\bibitem[Fitzmaurice et~al\mbox{.}(1995)]%
        {fitzmaurice1995bricks}
\bibfield{author}{\bibinfo{person}{George~W Fitzmaurice}, \bibinfo{person}{Hiroshi Ishii}, {and} \bibinfo{person}{William~AS Buxton}.} \bibinfo{year}{1995}\natexlab{}.
\newblock \showarticletitle{Bricks: laying the foundations for graspable user interfaces}. In \bibinfo{booktitle}{\emph{Proceedings of the SIGCHI conference on Human factors in computing systems}}. \bibinfo{pages}{442--449}.
\newblock


\bibitem[Fu et~al\mbox{.}(2024)]%
        {fu2024optical}
\bibfield{author}{\bibinfo{person}{Tingzhao Fu}, \bibinfo{person}{Jianfa Zhang}, \bibinfo{person}{Run Sun}, \bibinfo{person}{Yuyao Huang}, \bibinfo{person}{Wei Xu}, \bibinfo{person}{Sigang Yang}, \bibinfo{person}{Zhihong Zhu}, {and} \bibinfo{person}{Hongwei Chen}.} \bibinfo{year}{2024}\natexlab{}.
\newblock \showarticletitle{Optical neural networks: progress and challenges}.
\newblock \bibinfo{journal}{\emph{Light: Science \& Applications}} \bibinfo{volume}{13}, \bibinfo{number}{1} (\bibinfo{year}{2024}), \bibinfo{pages}{263}.
\newblock


\bibitem[Garrett et~al\mbox{.}(2020)]%
        {garrett2020more}
\bibfield{author}{\bibinfo{person}{Natalie Garrett}, \bibinfo{person}{Nathan Beard}, {and} \bibinfo{person}{Casey Fiesler}.} \bibinfo{year}{2020}\natexlab{}.
\newblock \showarticletitle{More than" If Time Allows" the role of ethics in AI education}. In \bibinfo{booktitle}{\emph{Proceedings of the AAAI/ACM Conference on AI, Ethics, and Society}}. \bibinfo{pages}{272--278}.
\newblock


\bibitem[Georgiou and Ioannou(2019)]%
        {georgiou2019embodied}
\bibfield{author}{\bibinfo{person}{Yiannis Georgiou} {and} \bibinfo{person}{Andri Ioannou}.} \bibinfo{year}{2019}\natexlab{}.
\newblock \showarticletitle{Embodied learning in a digital world: A systematic review of empirical research in K-12 education}.
\newblock \bibinfo{journal}{\emph{Learning in a digital world: Perspective on interactive technologies for formal and informal education}} (\bibinfo{year}{2019}), \bibinfo{pages}{155--177}.
\newblock


\bibitem[Ghajargar and Bardzell(2022)]%
        {ghajargar2022making}
\bibfield{author}{\bibinfo{person}{Maliheh Ghajargar} {and} \bibinfo{person}{Jeffrey Bardzell}.} \bibinfo{year}{2022}\natexlab{}.
\newblock \showarticletitle{Making AI understandable by making it tangible: exploring the design space with ten concept cards}. In \bibinfo{booktitle}{\emph{Proceedings of the 34th Australian Conference on Human-Computer Interaction}}. \bibinfo{pages}{74--80}.
\newblock


\bibitem[Ghajargar et~al\mbox{.}(2021)]%
        {ghajargar2021explainable}
\bibfield{author}{\bibinfo{person}{Maliheh Ghajargar}, \bibinfo{person}{Jeffrey Bardzell}, \bibinfo{person}{Alison~Smith Renner}, \bibinfo{person}{Peter~Gall Krogh}, \bibinfo{person}{Kristina H{\"o}{\"o}k}, \bibinfo{person}{David Cuartielles}, \bibinfo{person}{Laurens Boer}, {and} \bibinfo{person}{Mikael Wiberg}.} \bibinfo{year}{2021}\natexlab{}.
\newblock \showarticletitle{From” explainable ai” to” graspable ai”}. In \bibinfo{booktitle}{\emph{Proceedings of the Fifteenth International Conference on Tangible, Embedded, and Embodied Interaction}}. \bibinfo{pages}{1--4}.
\newblock


\bibitem[Ghajargar et~al\mbox{.}(2022)]%
        {ghajargar2022graspable}
\bibfield{author}{\bibinfo{person}{Maliheh Ghajargar}, \bibinfo{person}{Jeffrey Bardzell}, \bibinfo{person}{Alison~Marie Smith-Renner}, \bibinfo{person}{Kristina H{\"o}{\"o}k}, {and} \bibinfo{person}{Peter~Gall Krogh}.} \bibinfo{year}{2022}\natexlab{}.
\newblock \showarticletitle{Graspable AI: Physical forms as explanation modality for explainable AI}. In \bibinfo{booktitle}{\emph{Proceedings of the Sixteenth International Conference on Tangible, Embedded, and Embodied Interaction}}. \bibinfo{pages}{1--4}.
\newblock


\bibitem[Goldin-Meadow(2009)]%
        {goldin2009gesture}
\bibfield{author}{\bibinfo{person}{Susan Goldin-Meadow}.} \bibinfo{year}{2009}\natexlab{}.
\newblock \showarticletitle{How gesture promotes learning throughout childhood}.
\newblock \bibinfo{journal}{\emph{Child development perspectives}} \bibinfo{volume}{3}, \bibinfo{number}{2} (\bibinfo{year}{2009}), \bibinfo{pages}{106--111}.
\newblock


\bibitem[Gong et~al\mbox{.}(2024)]%
        {gong2024approaching}
\bibfield{author}{\bibinfo{person}{Yunfan Gong}, \bibinfo{person}{Xiaofei Zhou}, \bibinfo{person}{Yushan Zhou}, \bibinfo{person}{April Luehmann}, \bibinfo{person}{Yujung Han}, {and} \bibinfo{person}{Zhen Bai}.} \bibinfo{year}{2024}\natexlab{}.
\newblock \showarticletitle{Approaching “Filter Bubble” in Recommendation Systems: A Transformative AI Literacy Learning Experience}. In \bibinfo{booktitle}{\emph{Proceedings of the 18th International Conference of the Learning Sciences-ICLS 2024, pp. 490-497}}. International Society of the Learning Sciences.
\newblock


\bibitem[Goodman et~al\mbox{.}(1978)]%
        {goodman1978fully}
\bibfield{author}{\bibinfo{person}{Joseph~W Goodman}, \bibinfo{person}{AR Dias}, {and} \bibinfo{person}{LM Woody}.} \bibinfo{year}{1978}\natexlab{}.
\newblock \showarticletitle{Fully parallel, high-speed incoherent optical method for performing discrete Fourier transforms}.
\newblock \bibinfo{journal}{\emph{Optics Letters}} \bibinfo{volume}{2}, \bibinfo{number}{1} (\bibinfo{year}{1978}), \bibinfo{pages}{1--3}.
\newblock


\bibitem[Hayasaki et~al\mbox{.}(1992)]%
        {hayasaki1992optical}
\bibfield{author}{\bibinfo{person}{Yoshio Hayasaki}, \bibinfo{person}{Ichiro Tohyama}, \bibinfo{person}{Toyohiko Yatagai}, \bibinfo{person}{Masahiko Mori}, {and} \bibinfo{person}{Satoshi Ishihara}.} \bibinfo{year}{1992}\natexlab{}.
\newblock \showarticletitle{Optical learning neural network using Selfoc microlens array}.
\newblock \bibinfo{journal}{\emph{Japanese journal of applied physics}} \bibinfo{volume}{31}, \bibinfo{number}{5S} (\bibinfo{year}{1992}), \bibinfo{pages}{1689}.
\newblock


\bibitem[Hitron et~al\mbox{.}(2019)]%
        {hitron2019can}
\bibfield{author}{\bibinfo{person}{Tom Hitron}, \bibinfo{person}{Yoav Orlev}, \bibinfo{person}{Iddo Wald}, \bibinfo{person}{Ariel Shamir}, \bibinfo{person}{Hadas Erel}, {and} \bibinfo{person}{Oren Zuckerman}.} \bibinfo{year}{2019}\natexlab{}.
\newblock \showarticletitle{Can children understand machine learning concepts? The effect of uncovering black boxes}. In \bibinfo{booktitle}{\emph{Proceedings of the 2019 CHI conference on human factors in computing systems}}. \bibinfo{pages}{1--11}.
\newblock


\bibitem[Holmquist(2023)]%
        {holmquist2023bits}
\bibfield{author}{\bibinfo{person}{Lars~Erik Holmquist}.} \bibinfo{year}{2023}\natexlab{}.
\newblock \showarticletitle{Bits Are Cheap, Atoms Are Expensive: Critiquing the Turn Towards Tangibility in HCI}. In \bibinfo{booktitle}{\emph{Extended Abstracts of the 2023 CHI Conference on Human Factors in Computing Systems}}. \bibinfo{pages}{1--8}.
\newblock


\bibitem[Hornecker and Buur(2006)]%
        {hornecker2006getting}
\bibfield{author}{\bibinfo{person}{Eva Hornecker} {and} \bibinfo{person}{Jacob Buur}.} \bibinfo{year}{2006}\natexlab{}.
\newblock \showarticletitle{Getting a grip on tangible interaction: a framework on physical space and social interaction}. In \bibinfo{booktitle}{\emph{Proceedings of the SIGCHI conference on Human Factors in computing systems}}. \bibinfo{pages}{437--446}.
\newblock


\bibitem[Hostetter and Alibali(2008)]%
        {hostetter2008visible}
\bibfield{author}{\bibinfo{person}{Autumn~B Hostetter} {and} \bibinfo{person}{Martha~W Alibali}.} \bibinfo{year}{2008}\natexlab{}.
\newblock \showarticletitle{Visible embodiment: Gestures as simulated action}.
\newblock \bibinfo{journal}{\emph{Psychonomic bulletin \& review}}  \bibinfo{volume}{15} (\bibinfo{year}{2008}), \bibinfo{pages}{495--514}.
\newblock


\bibitem[Houben et~al\mbox{.}(2016)]%
        {houben2016physikit}
\bibfield{author}{\bibinfo{person}{Steven Houben}, \bibinfo{person}{Connie Golsteijn}, \bibinfo{person}{Sarah Gallacher}, \bibinfo{person}{Rose Johnson}, \bibinfo{person}{Saskia Bakker}, \bibinfo{person}{Nicolai Marquardt}, \bibinfo{person}{Licia Capra}, {and} \bibinfo{person}{Yvonne Rogers}.} \bibinfo{year}{2016}\natexlab{}.
\newblock \showarticletitle{Physikit: Data engagement through physical ambient visualizations in the home}. In \bibinfo{booktitle}{\emph{Proceedings of the 2016 CHI conference on human factors in computing systems}}. \bibinfo{pages}{1608--1619}.
\newblock


\bibitem[Howison et~al\mbox{.}(2011)]%
        {howison2011mathematical}
\bibfield{author}{\bibinfo{person}{Mark Howison}, \bibinfo{person}{Dragan Trninic}, \bibinfo{person}{Daniel Reinholz}, {and} \bibinfo{person}{Dor Abrahamson}.} \bibinfo{year}{2011}\natexlab{}.
\newblock \showarticletitle{The Mathematical Imagery Trainer: from embodied interaction to conceptual learning}. In \bibinfo{booktitle}{\emph{Proceedings of the SIGCHI conference on human factors in computing systems}}. \bibinfo{pages}{1989--1998}.
\newblock


\bibitem[Hurtienne(2011)]%
        {hurtienne2011image}
\bibfield{author}{\bibinfo{person}{J{\"o}rn Hurtienne}.} \bibinfo{year}{2011}\natexlab{}.
\newblock \showarticletitle{Image schemas and design for intuitive use}.
\newblock  (\bibinfo{year}{2011}).
\newblock


\bibitem[Hurtienne and Israel(2007)]%
        {hurtienne2007image}
\bibfield{author}{\bibinfo{person}{J{\"o}rn Hurtienne} {and} \bibinfo{person}{Johann~Habakuk Israel}.} \bibinfo{year}{2007}\natexlab{}.
\newblock \showarticletitle{Image schemas and their metaphorical extensions: intuitive patterns for tangible interaction}. In \bibinfo{booktitle}{\emph{Proceedings of the 1st international conference on Tangible and embedded interaction}}. \bibinfo{pages}{127--134}.
\newblock


\bibitem[Hurtienne et~al\mbox{.}(2015)]%
        {hurtienne2015designing}
\bibfield{author}{\bibinfo{person}{J{\"o}rn Hurtienne}, \bibinfo{person}{Kerstin Kl{\"o}ckner}, \bibinfo{person}{Sarah Diefenbach}, \bibinfo{person}{Claudia Nass}, {and} \bibinfo{person}{Andreas Maier}.} \bibinfo{year}{2015}\natexlab{}.
\newblock \showarticletitle{Designing with image schemas: resolving the tension between innovation, inclusion and intuitive use}.
\newblock \bibinfo{journal}{\emph{Interacting with Computers}} \bibinfo{volume}{27}, \bibinfo{number}{3} (\bibinfo{year}{2015}), \bibinfo{pages}{235--255}.
\newblock


\bibitem[Hurtienne et~al\mbox{.}(2009)]%
        {hurtienne2009sad}
\bibfield{author}{\bibinfo{person}{J{\"o}rn Hurtienne}, \bibinfo{person}{Christian St{\"o}{\ss}el}, {and} \bibinfo{person}{Katharina Weber}.} \bibinfo{year}{2009}\natexlab{}.
\newblock \showarticletitle{Sad is heavy and happy is light: Population stereotypes of tangible object attributes}. In \bibinfo{booktitle}{\emph{Proceedings of the 3rd International Conference on Tangible and Embedded Interaction}}. \bibinfo{pages}{61--68}.
\newblock


\bibitem[Im and Rogers(2021)]%
        {im2021draw2code}
\bibfield{author}{\bibinfo{person}{Hyejin Im} {and} \bibinfo{person}{Chris Rogers}.} \bibinfo{year}{2021}\natexlab{}.
\newblock \showarticletitle{Draw2code: Low-cost tangible programming for creating ar animations}. In \bibinfo{booktitle}{\emph{Proceedings of the 20th Annual ACM Interaction Design and Children Conference}}. \bibinfo{pages}{427--432}.
\newblock


\bibitem[Ishii(2008)]%
        {ishii2008tangible}
\bibfield{author}{\bibinfo{person}{Hiroshi Ishii}.} \bibinfo{year}{2008}\natexlab{}.
\newblock \showarticletitle{The tangible user interface and its evolution}.
\newblock \bibinfo{journal}{\emph{Commun. ACM}} \bibinfo{volume}{51}, \bibinfo{number}{6} (\bibinfo{year}{2008}), \bibinfo{pages}{32--36}.
\newblock


\bibitem[Ishii and Ullmer(1997)]%
        {ishii1997tangible}
\bibfield{author}{\bibinfo{person}{Hiroshi Ishii} {and} \bibinfo{person}{Brygg Ullmer}.} \bibinfo{year}{1997}\natexlab{}.
\newblock \showarticletitle{Tangible bits: towards seamless interfaces between people, bits and atoms}. In \bibinfo{booktitle}{\emph{Proceedings of the ACM SIGCHI Conference on Human factors in computing systems}}. \bibinfo{pages}{234--241}.
\newblock


\bibitem[Jansen et~al\mbox{.}(2015)]%
        {jansen2015opportunities}
\bibfield{author}{\bibinfo{person}{Yvonne Jansen}, \bibinfo{person}{Pierre Dragicevic}, \bibinfo{person}{Petra Isenberg}, \bibinfo{person}{Jason Alexander}, \bibinfo{person}{Abhijit Karnik}, \bibinfo{person}{Johan Kildal}, \bibinfo{person}{Sriram Subramanian}, {and} \bibinfo{person}{Kasper Hornb{\ae}k}.} \bibinfo{year}{2015}\natexlab{}.
\newblock \showarticletitle{Opportunities and challenges for data physicalization}. In \bibinfo{booktitle}{\emph{proceedings of the 33rd annual acm conference on human factors in computing systems}}. \bibinfo{pages}{3227--3236}.
\newblock


\bibitem[Jansen and Hornb{\ae}k(2015)]%
        {jansen2015psychophysical}
\bibfield{author}{\bibinfo{person}{Yvonne Jansen} {and} \bibinfo{person}{Kasper Hornb{\ae}k}.} \bibinfo{year}{2015}\natexlab{}.
\newblock \showarticletitle{A psychophysical investigation of size as a physical variable}.
\newblock \bibinfo{journal}{\emph{IEEE transactions on visualization and computer graphics}} \bibinfo{volume}{22}, \bibinfo{number}{1} (\bibinfo{year}{2015}), \bibinfo{pages}{479--488}.
\newblock


\bibitem[Johnson(2013)]%
        {johnson2013body}
\bibfield{author}{\bibinfo{person}{Mark Johnson}.} \bibinfo{year}{2013}\natexlab{}.
\newblock \bibinfo{booktitle}{\emph{The body in the mind: The bodily basis of meaning, imagination, and reason}}.
\newblock \bibinfo{publisher}{University of Chicago press}.
\newblock


\bibitem[Jord{\`a} et~al\mbox{.}(2007)]%
        {jorda2007reactable}
\bibfield{author}{\bibinfo{person}{Sergi Jord{\`a}}, \bibinfo{person}{G{\"u}nter Geiger}, \bibinfo{person}{Marcos Alonso}, {and} \bibinfo{person}{Martin Kaltenbrunner}.} \bibinfo{year}{2007}\natexlab{}.
\newblock \showarticletitle{The reacTable: exploring the synergy between live music performance and tabletop tangible interfaces}. In \bibinfo{booktitle}{\emph{Proceedings of the 1st international conference on Tangible and embedded interaction}}. \bibinfo{pages}{139--146}.
\newblock


\bibitem[Kahn et~al\mbox{.}(2018)]%
        {kahn2018ai}
\bibfield{author}{\bibinfo{person}{Ken Kahn}, \bibinfo{person}{Rani Megasari}, \bibinfo{person}{Erna Piantari}, {and} \bibinfo{person}{Enjun Junaeti}.} \bibinfo{year}{2018}\natexlab{}.
\newblock \showarticletitle{AI programming by children using snap! Block programming in a developing country}. In \bibinfo{booktitle}{\emph{Thirteenth European Conference on Technology Enhanced Learning}}, Vol.~\bibinfo{volume}{11082}. Springer.
\newblock


\bibitem[Kaspersen et~al\mbox{.}(2021)]%
        {kaspersen2021machine}
\bibfield{author}{\bibinfo{person}{Magnus~H{\o}holt Kaspersen}, \bibinfo{person}{Karl-Emil~Kj{\ae}r Bilstrup}, {and} \bibinfo{person}{Marianne~Graves Petersen}.} \bibinfo{year}{2021}\natexlab{}.
\newblock \showarticletitle{The machine learning machine: A tangible user interface for teaching machine learning}. In \bibinfo{booktitle}{\emph{Proceedings of the fifteenth international conference on tangible, embedded, and embodied interaction}}. \bibinfo{pages}{1--12}.
\newblock


\bibitem[Kikin-Gil(2006)]%
        {kikin2006light}
\bibfield{author}{\bibinfo{person}{Erez Kikin-Gil}.} \bibinfo{year}{2006}\natexlab{}.
\newblock \showarticletitle{The Light-Wall: tangible user interfaces for learning systems thinking}.
\newblock \bibinfo{journal}{\emph{Personal and Ubiquitous Computing}} \bibinfo{volume}{10}, \bibinfo{number}{2} (\bibinfo{year}{2006}), \bibinfo{pages}{181--182}.
\newblock


\bibitem[Kim and Kwon(2024)]%
        {kim2024tangible}
\bibfield{author}{\bibinfo{person}{Keunjae Kim} {and} \bibinfo{person}{Kyungbin Kwon}.} \bibinfo{year}{2024}\natexlab{}.
\newblock \showarticletitle{Tangible computing tools in AI education: Approach to improve elementary students' knowledge, perception, and behavioral intention towards AI}.
\newblock \bibinfo{journal}{\emph{Education and Information Technologies}} (\bibinfo{year}{2024}), \bibinfo{pages}{1--32}.
\newblock


\bibitem[Koren et~al\mbox{.}(2009)]%
        {koren2009matrix}
\bibfield{author}{\bibinfo{person}{Yehuda Koren}, \bibinfo{person}{Robert Bell}, {and} \bibinfo{person}{Chris Volinsky}.} \bibinfo{year}{2009}\natexlab{}.
\newblock \showarticletitle{Matrix factorization techniques for recommender systems}.
\newblock \bibinfo{journal}{\emph{Computer}} \bibinfo{volume}{42}, \bibinfo{number}{8} (\bibinfo{year}{2009}), \bibinfo{pages}{30--37}.
\newblock


\bibitem[Kulkarni et~al\mbox{.}(2018)]%
        {kulkarni2018promoting}
\bibfield{author}{\bibinfo{person}{Anagha Kulkarni}, \bibinfo{person}{Ilmi Yoon}, \bibinfo{person}{Pleuni~S Pennings}, \bibinfo{person}{Kazunori Okada}, {and} \bibinfo{person}{Carmen Domingo}.} \bibinfo{year}{2018}\natexlab{}.
\newblock \showarticletitle{Promoting diversity in computing}. In \bibinfo{booktitle}{\emph{Proceedings of the 23rd Annual ACM Conference on Innovation and Technology in Computer Science Education}}. \bibinfo{pages}{236--241}.
\newblock


\bibitem[Laal and Ghodsi(2012)]%
        {laal2012benefits}
\bibfield{author}{\bibinfo{person}{Marjan Laal} {and} \bibinfo{person}{Seyed~Mohammad Ghodsi}.} \bibinfo{year}{2012}\natexlab{}.
\newblock \showarticletitle{Benefits of collaborative learning}.
\newblock \bibinfo{journal}{\emph{Procedia-social and behavioral sciences}}  \bibinfo{volume}{31} (\bibinfo{year}{2012}), \bibinfo{pages}{486--490}.
\newblock


\bibitem[Lakoff and N{\'u}{\~n}ez(2000)]%
        {lakoff2000mathematics}
\bibfield{author}{\bibinfo{person}{George Lakoff} {and} \bibinfo{person}{Rafael N{\'u}{\~n}ez}.} \bibinfo{year}{2000}\natexlab{}.
\newblock \bibinfo{booktitle}{\emph{Where mathematics comes from}}. Vol.~\bibinfo{volume}{6}.
\newblock \bibinfo{publisher}{New York: Basic Books}.
\newblock


\bibitem[Lazar et~al\mbox{.}(2017)]%
        {lazar2017research}
\bibfield{author}{\bibinfo{person}{Jonathan Lazar}, \bibinfo{person}{Jinjuan~Heidi Feng}, {and} \bibinfo{person}{Harry Hochheiser}.} \bibinfo{year}{2017}\natexlab{}.
\newblock \bibinfo{booktitle}{\emph{Research methods in human-computer interaction}}.
\newblock \bibinfo{publisher}{Morgan Kaufmann}.
\newblock


\bibitem[Lee et~al\mbox{.}(2022)]%
        {lee2022ai}
\bibfield{author}{\bibinfo{person}{Irene Lee}, \bibinfo{person}{Helen Zhang}, \bibinfo{person}{Kate Moore}, \bibinfo{person}{Xiaofei Zhou}, \bibinfo{person}{Beatriz Perret}, \bibinfo{person}{Yihong Cheng}, \bibinfo{person}{Ruiying Zheng}, {and} \bibinfo{person}{Grace Pu}.} \bibinfo{year}{2022}\natexlab{}.
\newblock \showarticletitle{AI Book Club: An Innovative Professional Development Model for AI Education}. In \bibinfo{booktitle}{\emph{Proceedings of the 53rd ACM Technical Symposium on Computer Science Education V. 1}}. \bibinfo{pages}{202--208}.
\newblock


\bibitem[Lee et~al\mbox{.}(2023)]%
        {lee2023fostering}
\bibfield{author}{\bibinfo{person}{Sunok Lee}, \bibinfo{person}{Dasom Choi}, \bibinfo{person}{Minha Lee}, \bibinfo{person}{Jonghak Choi}, {and} \bibinfo{person}{Sangsu Lee}.} \bibinfo{year}{2023}\natexlab{}.
\newblock \showarticletitle{Fostering Youth’s Critical Thinking Competency About AI through Exhibition}. In \bibinfo{booktitle}{\emph{Proceedings of the 2023 CHI Conference on Human Factors in Computing Systems}}. \bibinfo{pages}{1--22}.
\newblock


\bibitem[Li et~al\mbox{.}(2022)]%
        {li2022meta}
\bibfield{author}{\bibinfo{person}{Yanhong Li}, \bibinfo{person}{Meng Liang}, \bibinfo{person}{Julian Preissing}, \bibinfo{person}{Nadine Bachl}, \bibinfo{person}{Michelle~Melina Dutoit}, \bibinfo{person}{Thomas Weber}, \bibinfo{person}{Sven Mayer}, {and} \bibinfo{person}{Heinrich Hussmann}.} \bibinfo{year}{2022}\natexlab{}.
\newblock \showarticletitle{A meta-analysis of tangible learning studies from the tei conference}. In \bibinfo{booktitle}{\emph{Proceedings of the sixteenth international conference on tangible, embedded, and embodied interaction}}. \bibinfo{pages}{1--17}.
\newblock


\bibitem[Li et~al\mbox{.}(2020)]%
        {li2020tangible}
\bibfield{author}{\bibinfo{person}{Yanhong Li}, \bibinfo{person}{Beat Rossmy}, {and} \bibinfo{person}{Heinrich Hu{\ss}mann}.} \bibinfo{year}{2020}\natexlab{}.
\newblock \showarticletitle{Tangible interaction with light: A review}.
\newblock \bibinfo{journal}{\emph{Multimodal Technologies and Interaction}} \bibinfo{volume}{4}, \bibinfo{number}{4} (\bibinfo{year}{2020}), \bibinfo{pages}{72}.
\newblock


\bibitem[Liang et~al\mbox{.}(2015)]%
        {liang2015ar}
\bibfield{author}{\bibinfo{person}{Hui Liang}, \bibinfo{person}{Junsong Yuan}, \bibinfo{person}{Daniel Thalmann}, {and} \bibinfo{person}{Nadia~Magnenat Thalmann}.} \bibinfo{year}{2015}\natexlab{}.
\newblock \showarticletitle{Ar in hand: Egocentric palm pose tracking and gesture recognition for augmented reality applications}. In \bibinfo{booktitle}{\emph{Proceedings of the 23rd ACM international conference on Multimedia}}. \bibinfo{pages}{743--744}.
\newblock


\bibitem[Lindgren and Johnson-Glenberg(2013)]%
        {lindgren2013emboldened}
\bibfield{author}{\bibinfo{person}{Robb Lindgren} {and} \bibinfo{person}{Mina Johnson-Glenberg}.} \bibinfo{year}{2013}\natexlab{}.
\newblock \showarticletitle{Emboldened by embodiment: Six precepts for research on embodied learning and mixed reality}.
\newblock \bibinfo{journal}{\emph{Educational researcher}} \bibinfo{volume}{42}, \bibinfo{number}{8} (\bibinfo{year}{2013}), \bibinfo{pages}{445--452}.
\newblock


\bibitem[Lindgren et~al\mbox{.}(2022)]%
        {lindgren2022learning}
\bibfield{author}{\bibinfo{person}{Robb Lindgren}, \bibinfo{person}{Jason~W Morphew}, \bibinfo{person}{Jina Kang}, \bibinfo{person}{James Planey}, {and} \bibinfo{person}{Jos{\'e}~P Mestre}.} \bibinfo{year}{2022}\natexlab{}.
\newblock \showarticletitle{Learning and transfer effects of embodied simulations targeting crosscutting concepts in science.}
\newblock \bibinfo{journal}{\emph{Journal of Educational Psychology}} \bibinfo{volume}{114}, \bibinfo{number}{3} (\bibinfo{year}{2022}), \bibinfo{pages}{462}.
\newblock


\bibitem[L{\"o}ffler(2014)]%
        {loffler2014population}
\bibfield{author}{\bibinfo{person}{Diana L{\"o}ffler}.} \bibinfo{year}{2014}\natexlab{}.
\newblock \showarticletitle{Population stereotypes of color attributes for tangible interaction design}. In \bibinfo{booktitle}{\emph{Proceedings of the 8th International Conference on Tangible, Embedded and Embodied Interaction}}. \bibinfo{pages}{285--288}.
\newblock


\bibitem[Long and Magerko(2020)]%
        {long2020ai}
\bibfield{author}{\bibinfo{person}{Duri Long} {and} \bibinfo{person}{Brian Magerko}.} \bibinfo{year}{2020}\natexlab{}.
\newblock \showarticletitle{What is AI literacy? Competencies and design considerations}. In \bibinfo{booktitle}{\emph{Proceedings of the 2020 CHI conference on human factors in computing systems}}. \bibinfo{pages}{1--16}.
\newblock


\bibitem[Long et~al\mbox{.}(2023)]%
        {long2023fostering}
\bibfield{author}{\bibinfo{person}{Duri Long}, \bibinfo{person}{Sophie Rollins}, \bibinfo{person}{Jasmin Ali-Diaz}, \bibinfo{person}{Katherine Hancock}, \bibinfo{person}{Samnang Nuonsinoeun}, \bibinfo{person}{Jessica Roberts}, {and} \bibinfo{person}{Brian Magerko}.} \bibinfo{year}{2023}\natexlab{}.
\newblock \showarticletitle{Fostering AI Literacy with Embodiment \& Creativity: From Activity Boxes to Museum Exhibits}. In \bibinfo{booktitle}{\emph{Proceedings of the 22nd Annual ACM Interaction Design and Children Conference}}. \bibinfo{pages}{727--731}.
\newblock


\bibitem[Lops et~al\mbox{.}(2011)]%
        {lops2011content}
\bibfield{author}{\bibinfo{person}{Pasquale Lops}, \bibinfo{person}{Marco De~Gemmis}, {and} \bibinfo{person}{Giovanni Semeraro}.} \bibinfo{year}{2011}\natexlab{}.
\newblock \showarticletitle{Content-based recommender systems: State of the art and trends}.
\newblock \bibinfo{journal}{\emph{Recommender systems handbook}} (\bibinfo{year}{2011}), \bibinfo{pages}{73--105}.
\newblock


\bibitem[Lops et~al\mbox{.}(2019)]%
        {lops2019trends}
\bibfield{author}{\bibinfo{person}{Pasquale Lops}, \bibinfo{person}{Dietmar Jannach}, \bibinfo{person}{Cataldo Musto}, \bibinfo{person}{Toine Bogers}, {and} \bibinfo{person}{Marijn Koolen}.} \bibinfo{year}{2019}\natexlab{}.
\newblock \showarticletitle{Trends in content-based recommendation: Preface to the special issue on Recommender systems based on rich item descriptions}.
\newblock \bibinfo{journal}{\emph{User Modeling and User-Adapted Interaction}}  \bibinfo{volume}{29} (\bibinfo{year}{2019}), \bibinfo{pages}{239--249}.
\newblock


\bibitem[Lortie-Forgues et~al\mbox{.}(2015)]%
        {lortie2015learning}
\bibfield{author}{\bibinfo{person}{Hugues Lortie-Forgues}, \bibinfo{person}{Jing Tian}, {and} \bibinfo{person}{Robert~S Siegler}.} \bibinfo{year}{2015}\natexlab{}.
\newblock \showarticletitle{Why is learning fraction and decimal arithmetic so difficult?}
\newblock \bibinfo{journal}{\emph{Developmental Review}}  \bibinfo{volume}{38} (\bibinfo{year}{2015}), \bibinfo{pages}{201--221}.
\newblock


\bibitem[Lu et~al\mbox{.}(2015)]%
        {lu2015recommender}
\bibfield{author}{\bibinfo{person}{Jie Lu}, \bibinfo{person}{Dianshuang Wu}, \bibinfo{person}{Mingsong Mao}, \bibinfo{person}{Wei Wang}, {and} \bibinfo{person}{Guangquan Zhang}.} \bibinfo{year}{2015}\natexlab{}.
\newblock \showarticletitle{Recommender system application developments: a survey}.
\newblock \bibinfo{journal}{\emph{Decision support systems}}  \bibinfo{volume}{74} (\bibinfo{year}{2015}), \bibinfo{pages}{12--32}.
\newblock


\bibitem[Lundberg and Lee(2017)]%
        {lundberg2017unified}
\bibfield{author}{\bibinfo{person}{Scott~M Lundberg} {and} \bibinfo{person}{Su-In Lee}.} \bibinfo{year}{2017}\natexlab{}.
\newblock \showarticletitle{A unified approach to interpreting model predictions}.
\newblock \bibinfo{journal}{\emph{Advances in neural information processing systems}}  \bibinfo{volume}{30} (\bibinfo{year}{2017}).
\newblock


\bibitem[Ma(2016)]%
        {ma2016designing}
\bibfield{author}{\bibinfo{person}{Jasmine~Y Ma}.} \bibinfo{year}{2016}\natexlab{}.
\newblock \showarticletitle{Designing disruptions for productive hybridity: The case of walking scale geometry}.
\newblock \bibinfo{journal}{\emph{Journal of the Learning Sciences}} \bibinfo{volume}{25}, \bibinfo{number}{3} (\bibinfo{year}{2016}), \bibinfo{pages}{335--371}.
\newblock


\bibitem[Manches and O’malley(2012)]%
        {manches2012tangibles}
\bibfield{author}{\bibinfo{person}{Andrew Manches} {and} \bibinfo{person}{Claire O’malley}.} \bibinfo{year}{2012}\natexlab{}.
\newblock \showarticletitle{Tangibles for learning: a representational analysis of physical manipulation}.
\newblock \bibinfo{journal}{\emph{Personal and ubiquitous computing}}  \bibinfo{volume}{16} (\bibinfo{year}{2012}), \bibinfo{pages}{405--419}.
\newblock


\bibitem[Markova et~al\mbox{.}(2012)]%
        {markova2012tangible}
\bibfield{author}{\bibinfo{person}{Milena~S Markova}, \bibinfo{person}{Stephanie Wilson}, {and} \bibinfo{person}{Simone Stumpf}.} \bibinfo{year}{2012}\natexlab{}.
\newblock \showarticletitle{Tangible user interfaces for learning}.
\newblock \bibinfo{journal}{\emph{International Journal of Technology Enhanced Learning}} \bibinfo{volume}{4}, \bibinfo{number}{3-4} (\bibinfo{year}{2012}), \bibinfo{pages}{139--155}.
\newblock


\bibitem[Morais et~al\mbox{.}(2024)]%
        {morais2024exploration}
\bibfield{author}{\bibinfo{person}{Luiz Morais}, \bibinfo{person}{Georgia Panagiotidou}, \bibinfo{person}{Sarah Hayes}, \bibinfo{person}{Tatiana Losev}, \bibinfo{person}{Rebecca Noonan}, {and} \bibinfo{person}{Uta Hinrichs}.} \bibinfo{year}{2024}\natexlab{}.
\newblock \showarticletitle{From exploration to end of life: Unpacking sustainability in physicalization practices}. In \bibinfo{booktitle}{\emph{Proceedings of the CHI Conference on Human Factors in Computing Systems}}. \bibinfo{pages}{1--17}.
\newblock


\bibitem[Murray-Rust et~al\mbox{.}(2024)]%
        {murray2024metaphor}
\bibfield{author}{\bibinfo{person}{Dave Murray-Rust}, \bibinfo{person}{Maria Luce~Lupetti}, {and} \bibinfo{person}{Iohanna Nicenboim}.} \bibinfo{year}{2024}\natexlab{}.
\newblock \showarticletitle{Metaphor Gardening: Experiential engagements for designing AI interactions}.
\newblock  (\bibinfo{year}{2024}).
\newblock


\bibitem[Nathan(2012)]%
        {nathan2012rethinking}
\bibfield{author}{\bibinfo{person}{Mitchell~J Nathan}.} \bibinfo{year}{2012}\natexlab{}.
\newblock \showarticletitle{Rethinking formalisms in formal education}.
\newblock \bibinfo{journal}{\emph{Educational Psychologist}} \bibinfo{volume}{47}, \bibinfo{number}{2} (\bibinfo{year}{2012}), \bibinfo{pages}{125--148}.
\newblock


\bibitem[Nicholson et~al\mbox{.}(2021)]%
        {nicholson2021tangible}
\bibfield{author}{\bibinfo{person}{Rebecca Nicholson}, \bibinfo{person}{David Kirk}, {and} \bibinfo{person}{Tom Bartindale}.} \bibinfo{year}{2021}\natexlab{}.
\newblock \showarticletitle{Tangible Lighting Proxies: Brokering the Transition from Classroom to Stage}. In \bibinfo{booktitle}{\emph{Proceedings of the Fifteenth International Conference on Tangible, Embedded, and Embodied Interaction}}. \bibinfo{pages}{1--13}.
\newblock


\bibitem[Olympiou and Zacharia(2012)]%
        {olympiou2012blending}
\bibfield{author}{\bibinfo{person}{Georgios Olympiou} {and} \bibinfo{person}{Zacharias~C Zacharia}.} \bibinfo{year}{2012}\natexlab{}.
\newblock \showarticletitle{Blending physical and virtual manipulatives: An effort to improve students' conceptual understanding through science laboratory experimentation}.
\newblock \bibinfo{journal}{\emph{Science Education}} \bibinfo{volume}{96}, \bibinfo{number}{1} (\bibinfo{year}{2012}), \bibinfo{pages}{21--47}.
\newblock


\bibitem[Papert and Harel(1991)]%
        {papert1991situating}
\bibfield{author}{\bibinfo{person}{Seymour Papert} {and} \bibinfo{person}{Idit Harel}.} \bibinfo{year}{1991}\natexlab{}.
\newblock \showarticletitle{Situating constructionism}.
\newblock \bibinfo{journal}{\emph{constructionism}} \bibinfo{volume}{36}, \bibinfo{number}{2} (\bibinfo{year}{1991}), \bibinfo{pages}{1--11}.
\newblock


\bibitem[Pazzani and Billsus(2007)]%
        {pazzani2007content}
\bibfield{author}{\bibinfo{person}{Michael~J Pazzani} {and} \bibinfo{person}{Daniel Billsus}.} \bibinfo{year}{2007}\natexlab{}.
\newblock \showarticletitle{Content-based recommendation systems}.
\newblock In \bibinfo{booktitle}{\emph{The adaptive web}}. \bibinfo{publisher}{Springer}, \bibinfo{pages}{325--341}.
\newblock


\bibitem[Pedaste et~al\mbox{.}(2015)]%
        {pedaste2015phases}
\bibfield{author}{\bibinfo{person}{Margus Pedaste}, \bibinfo{person}{Mario M{\"a}eots}, \bibinfo{person}{Leo~A Siiman}, \bibinfo{person}{Ton De~Jong}, \bibinfo{person}{Siswa~AN Van~Riesen}, \bibinfo{person}{Ellen~T Kamp}, \bibinfo{person}{Constantinos~C Manoli}, \bibinfo{person}{Zacharias~C Zacharia}, {and} \bibinfo{person}{Eleftheria Tsourlidaki}.} \bibinfo{year}{2015}\natexlab{}.
\newblock \showarticletitle{Phases of inquiry-based learning: Definitions and the inquiry cycle}.
\newblock \bibinfo{journal}{\emph{Educational research review}}  \bibinfo{volume}{14} (\bibinfo{year}{2015}), \bibinfo{pages}{47--61}.
\newblock


\bibitem[Piaget(1976)]%
        {piaget1976piaget}
\bibfield{author}{\bibinfo{person}{Jean Piaget}.} \bibinfo{year}{1976}\natexlab{}.
\newblock \bibinfo{title}{Piaget’s theory}.
\newblock
\newblock


\bibitem[Presmeg(1998)]%
        {PRESMEG199825}
\bibfield{author}{\bibinfo{person}{Norma~C. Presmeg}.} \bibinfo{year}{1998}\natexlab{}.
\newblock \showarticletitle{Metaphoric and metonymic signification in mathematics}.
\newblock \bibinfo{journal}{\emph{The Journal of Mathematical Behavior}} \bibinfo{volume}{17}, \bibinfo{number}{1} (\bibinfo{year}{1998}), \bibinfo{pages}{25--32}.
\newblock
\showISSN{0732-3123}
\urldef\tempurl%
\url{https://doi.org/10.1016/S0732-3123(99)80059-5}
\showDOI{\tempurl}
\newblock
\shownote{Representations and the Psychology of Mathematics Education: Part I}.


\bibitem[Price et~al\mbox{.}(2009)]%
        {price2009effect}
\bibfield{author}{\bibinfo{person}{Sara Price}, \bibinfo{person}{Taciana~Pontual Falc{\~a}o}, \bibinfo{person}{Jennifer~G Sheridan}, {and} \bibinfo{person}{George Roussos}.} \bibinfo{year}{2009}\natexlab{}.
\newblock \showarticletitle{The effect of representation location on interaction in a tangible learning environment}. In \bibinfo{booktitle}{\emph{Proceedings of the 3rd International Conference on Tangible and Embedded Interaction}}. \bibinfo{pages}{85--92}.
\newblock


\bibitem[Qin(2024)]%
        {qin2024kaleidolight}
\bibfield{author}{\bibinfo{person}{Xiaoyan Qin}.} \bibinfo{year}{2024}\natexlab{}.
\newblock \showarticletitle{Kaleidolight: An Interactive Educational Device for Children to Explore Additive Color Theory and Create Visual Art with Light, Color, and Shapes}. In \bibinfo{booktitle}{\emph{Proceedings of the 23rd Annual ACM Interaction Design and Children Conference}}. \bibinfo{pages}{955--959}.
\newblock


\bibitem[Reid~Chassiakos et~al\mbox{.}(2016)]%
        {reid2016children}
\bibfield{author}{\bibinfo{person}{Yolanda~Linda Reid~Chassiakos}, \bibinfo{person}{Jenny Radesky}, \bibinfo{person}{Dimitri Christakis}, \bibinfo{person}{Megan~A Moreno}, \bibinfo{person}{Corinn Cross}, \bibinfo{person}{David Hill}, \bibinfo{person}{Nusheen Ameenuddin}, \bibinfo{person}{Jeffrey Hutchinson}, \bibinfo{person}{Alanna Levine}, \bibinfo{person}{Rhea Boyd}, {et~al\mbox{.}}} \bibinfo{year}{2016}\natexlab{}.
\newblock \showarticletitle{Children and adolescents and digital media}.
\newblock \bibinfo{journal}{\emph{Pediatrics}} \bibinfo{volume}{138}, \bibinfo{number}{5} (\bibinfo{year}{2016}).
\newblock


\bibitem[Rendle et~al\mbox{.}(2020)]%
        {rendle2020neural}
\bibfield{author}{\bibinfo{person}{Steffen Rendle}, \bibinfo{person}{Walid Krichene}, \bibinfo{person}{Li Zhang}, {and} \bibinfo{person}{John Anderson}.} \bibinfo{year}{2020}\natexlab{}.
\newblock \showarticletitle{Neural collaborative filtering vs. matrix factorization revisited}. In \bibinfo{booktitle}{\emph{Proceedings of the 14th ACM Conference on Recommender Systems}}. \bibinfo{pages}{240--248}.
\newblock


\bibitem[Riehmann et~al\mbox{.}(2014)]%
        {riehmann2014visualizing}
\bibfield{author}{\bibinfo{person}{Patrick Riehmann}, \bibinfo{person}{Wieland M{\"o}bus}, {and} \bibinfo{person}{Bernd Froehlich}.} \bibinfo{year}{2014}\natexlab{}.
\newblock \showarticletitle{Visualizing food ingredients for children by utilizing glyph-based characters}. In \bibinfo{booktitle}{\emph{Proceedings of the 2014 International Working Conference on Advanced Visual Interfaces}}. \bibinfo{pages}{133--136}.
\newblock


\bibitem[Ryokai et~al\mbox{.}(2014)]%
        {ryokai2014energybugs}
\bibfield{author}{\bibinfo{person}{Kimiko Ryokai}, \bibinfo{person}{Peiqi Su}, \bibinfo{person}{Eungchan Kim}, {and} \bibinfo{person}{Bob Rollins}.} \bibinfo{year}{2014}\natexlab{}.
\newblock \showarticletitle{Energybugs: Energy harvesting wearables for children}. In \bibinfo{booktitle}{\emph{Proceedings of the SIGCHI conference on human factors in computing systems}}. \bibinfo{pages}{1039--1048}.
\newblock


\bibitem[Schaper et~al\mbox{.}(2023)]%
        {schaper2023five}
\bibfield{author}{\bibinfo{person}{Marie-Monique Schaper}, \bibinfo{person}{Mariana~Aki Tamashiro}, \bibinfo{person}{Rachel~Charlotte Smith}, \bibinfo{person}{Maarten Van~Mechelen}, {and} \bibinfo{person}{Ole~Sejer Iversen}.} \bibinfo{year}{2023}\natexlab{}.
\newblock \showarticletitle{Five design recommendations for teaching teenagers’ about artificial intelligence and machine learning}. In \bibinfo{booktitle}{\emph{Proceedings of the 22nd Annual ACM Interaction Design and Children Conference}}. \bibinfo{pages}{298--309}.
\newblock


\bibitem[Scheidt and Pulver(2019)]%
        {scheidt2019any}
\bibfield{author}{\bibinfo{person}{Alexander Scheidt} {and} \bibinfo{person}{Tim Pulver}.} \bibinfo{year}{2019}\natexlab{}.
\newblock \showarticletitle{Any-cubes: A children's toy for learning ai: Enhanced play with deep learning and mqtt}.
\newblock In \bibinfo{booktitle}{\emph{Proceedings of mensch und computer 2019}}. \bibinfo{pages}{893--895}.
\newblock


\bibitem[Shu et~al\mbox{.}(2018)]%
        {shu2018content}
\bibfield{author}{\bibinfo{person}{Jiangbo Shu}, \bibinfo{person}{Xiaoxuan Shen}, \bibinfo{person}{Hai Liu}, \bibinfo{person}{Baolin Yi}, {and} \bibinfo{person}{Zhaoli Zhang}.} \bibinfo{year}{2018}\natexlab{}.
\newblock \showarticletitle{A content-based recommendation algorithm for learning resources}.
\newblock \bibinfo{journal}{\emph{Multimedia Systems}} \bibinfo{volume}{24}, \bibinfo{number}{2} (\bibinfo{year}{2018}), \bibinfo{pages}{163--173}.
\newblock


\bibitem[Solyst et~al\mbox{.}(2023a)]%
        {solyst2023investigating}
\bibfield{author}{\bibinfo{person}{Jaemarie Solyst}, \bibinfo{person}{Alexis Axon}, \bibinfo{person}{Angela~EB Stewart}, \bibinfo{person}{Motahhare Eslami}, {and} \bibinfo{person}{Amy Ogan}.} \bibinfo{year}{2023}\natexlab{a}.
\newblock \showarticletitle{Investigating girls' perspectives and knowledge gaps on ethics and fairness in Artificial Intelligence in a Lightweight workshop}.
\newblock \bibinfo{journal}{\emph{arXiv preprint arXiv:2302.13947}} (\bibinfo{year}{2023}).
\newblock


\bibitem[Solyst et~al\mbox{.}(2023b)]%
        {solyst2023would}
\bibfield{author}{\bibinfo{person}{Jaemarie Solyst}, \bibinfo{person}{Shixian Xie}, \bibinfo{person}{Ellia Yang}, \bibinfo{person}{Angela~EB Stewart}, \bibinfo{person}{Motahhare Eslami}, \bibinfo{person}{Jessica Hammer}, {and} \bibinfo{person}{Amy Ogan}.} \bibinfo{year}{2023}\natexlab{b}.
\newblock \showarticletitle{“I Would Like to Design”: Black Girls Analyzing and Ideating Fair and Accountable AI}. In \bibinfo{booktitle}{\emph{Proceedings of the 2023 CHI Conference on Human Factors in Computing Systems}}. \bibinfo{pages}{1--14}.
\newblock


\bibitem[Speer et~al\mbox{.}(2023)]%
        {speer2023speerloom}
\bibfield{author}{\bibinfo{person}{Samantha Speer}, \bibinfo{person}{Ana~P Garcia-Alonzo}, \bibinfo{person}{Joey Huang}, \bibinfo{person}{Nickolina Yankova}, \bibinfo{person}{Carolyn Rose}, \bibinfo{person}{Kylie~A Peppler}, \bibinfo{person}{James McCann}, {and} \bibinfo{person}{Melisa Orta~Martinez}.} \bibinfo{year}{2023}\natexlab{}.
\newblock \showarticletitle{SPEERLoom: An Open-Source Loom Kit for Interdisciplinary Engagement in Math, Engineering, and Textiles}. In \bibinfo{booktitle}{\emph{Proceedings of the 36th Annual ACM Symposium on User Interface Software and Technology}}. \bibinfo{pages}{1--15}.
\newblock


\bibitem[Sui et~al\mbox{.}(2020)]%
        {sui2020review}
\bibfield{author}{\bibinfo{person}{Xiubao Sui}, \bibinfo{person}{Qiuhao Wu}, \bibinfo{person}{Jia Liu}, \bibinfo{person}{Qian Chen}, {and} \bibinfo{person}{Guohua Gu}.} \bibinfo{year}{2020}\natexlab{}.
\newblock \showarticletitle{A review of optical neural networks}.
\newblock \bibinfo{journal}{\emph{IEEE Access}}  \bibinfo{volume}{8} (\bibinfo{year}{2020}), \bibinfo{pages}{70773--70783}.
\newblock


\bibitem[Sun and Peng(2022)]%
        {sun2022survey}
\bibfield{author}{\bibinfo{person}{Anchen Sun} {and} \bibinfo{person}{Yuanzhe Peng}.} \bibinfo{year}{2022}\natexlab{}.
\newblock \showarticletitle{A survey on modern recommendation system based on big data}.
\newblock \bibinfo{journal}{\emph{arXiv e-prints}} (\bibinfo{year}{2022}), \bibinfo{pages}{arXiv--2206}.
\newblock


\bibitem[Taher et~al\mbox{.}(2015)]%
        {taher2015exploring}
\bibfield{author}{\bibinfo{person}{Faisal Taher}, \bibinfo{person}{John Hardy}, \bibinfo{person}{Abhijit Karnik}, \bibinfo{person}{Christian Weichel}, \bibinfo{person}{Yvonne Jansen}, \bibinfo{person}{Kasper Hornb{\ae}k}, {and} \bibinfo{person}{Jason Alexander}.} \bibinfo{year}{2015}\natexlab{}.
\newblock \showarticletitle{Exploring interactions with physically dynamic bar charts}. In \bibinfo{booktitle}{\emph{Proceedings of the 33rd annual acm conference on human factors in computing systems}}. \bibinfo{pages}{3237--3246}.
\newblock


\bibitem[Tancredi et~al\mbox{.}(2022)]%
        {tancredi2022balance}
\bibfield{author}{\bibinfo{person}{Sofia Tancredi}, \bibinfo{person}{Julia Wang}, \bibinfo{person}{Helen~Tong Li}, \bibinfo{person}{Carissa~Jiayuan Yao}, \bibinfo{person}{Genna Macfarlan}, {and} \bibinfo{person}{Kimiko Ryokai}.} \bibinfo{year}{2022}\natexlab{}.
\newblock \showarticletitle{Balance Board Math:“Being the graph” through the sense of balance for embodied self-regulation and learning}. In \bibinfo{booktitle}{\emph{Proceedings of the 21st Annual ACM Interaction Design and Children Conference}}. \bibinfo{pages}{137--149}.
\newblock


\bibitem[Touretzky and Gardner-McCune(2021)]%
        {touretzky2021artificial}
\bibfield{author}{\bibinfo{person}{David~S Touretzky} {and} \bibinfo{person}{Christina Gardner-McCune}.} \bibinfo{year}{2021}\natexlab{}.
\newblock \showarticletitle{Artificial Intelligence Thinking in K-12}.
\newblock \bibinfo{journal}{\emph{Retrieved February}}  \bibinfo{volume}{19} (\bibinfo{year}{2021}).
\newblock


\bibitem[Trichopoulos et~al\mbox{.}(2023)]%
        {trichopoulos2023large}
\bibfield{author}{\bibinfo{person}{Georgios Trichopoulos}, \bibinfo{person}{Markos Konstantakis}, \bibinfo{person}{Georgios Alexandridis}, {and} \bibinfo{person}{George Caridakis}.} \bibinfo{year}{2023}\natexlab{}.
\newblock \showarticletitle{Large language models as recommendation Systems in Museums}.
\newblock \bibinfo{journal}{\emph{Electronics}} \bibinfo{volume}{12}, \bibinfo{number}{18} (\bibinfo{year}{2023}), \bibinfo{pages}{3829}.
\newblock


\bibitem[Wang et~al\mbox{.}(2024)]%
        {wang2024chaitok}
\bibfield{author}{\bibinfo{person}{Ge Wang}, \bibinfo{person}{Jun Zhao}, \bibinfo{person}{Samantha-Kaye Johnston}, \bibinfo{person}{Zhilin Zhang}, \bibinfo{person}{Max Van~Kleek}, {and} \bibinfo{person}{Nigel Shadbolt}.} \bibinfo{year}{2024}\natexlab{}.
\newblock \showarticletitle{CHAITok: A Proof-of-Concept System Supporting Children's Sense of Data Autonomy on Social Media}. In \bibinfo{booktitle}{\emph{Proceedings of the CHI Conference on Human Factors in Computing Systems}}. \bibinfo{pages}{1--19}.
\newblock


\bibitem[Wang et~al\mbox{.}(2022)]%
        {wang2022don}
\bibfield{author}{\bibinfo{person}{Ge Wang}, \bibinfo{person}{Jun Zhao}, \bibinfo{person}{Max Van~Kleek}, {and} \bibinfo{person}{Nigel Shadbolt}.} \bibinfo{year}{2022}\natexlab{}.
\newblock \showarticletitle{'Don't make assumptions about me!': Understanding Children's Perception of Datafication Online}.
\newblock \bibinfo{journal}{\emph{Proceedings of the ACM on Human-Computer Interaction}} \bibinfo{volume}{6}, \bibinfo{number}{CSCW2} (\bibinfo{year}{2022}), \bibinfo{pages}{1--24}.
\newblock


\bibitem[Wang et~al\mbox{.}(2023)]%
        {wang2023treat}
\bibfield{author}{\bibinfo{person}{Ge Wang}, \bibinfo{person}{Jun Zhao}, \bibinfo{person}{Max Van~Kleek}, {and} \bibinfo{person}{Nigel Shadbolt}.} \bibinfo{year}{2023}\natexlab{}.
\newblock \showarticletitle{‘Treat me as your friend, not a number in your database’: Co-designing with Children to Cope with Datafication Online}. In \bibinfo{booktitle}{\emph{Proceedings of the 2023 CHI Conference on Human Factors in Computing Systems}}. \bibinfo{pages}{1--21}.
\newblock


\bibitem[Wimer et~al\mbox{.}(2024)]%
        {wimer2024beyond}
\bibfield{author}{\bibinfo{person}{Brianna~L Wimer}, \bibinfo{person}{Annalisa Szymanski}, {and} \bibinfo{person}{Ronald~A Metoyer}.} \bibinfo{year}{2024}\natexlab{}.
\newblock \showarticletitle{Beyond Static Labels: Unpacking Nutrition Comprehension in the Digital Age}. In \bibinfo{booktitle}{\emph{Proceedings of the CHI Conference on Human Factors in Computing Systems}}. \bibinfo{pages}{1--15}.
\newblock


\bibitem[Yannier et~al\mbox{.}(2016)]%
        {yannier2016adding}
\bibfield{author}{\bibinfo{person}{Nesra Yannier}, \bibinfo{person}{Scott~E Hudson}, \bibinfo{person}{Eliane~Stampfer Wiese}, {and} \bibinfo{person}{Kenneth~R Koedinger}.} \bibinfo{year}{2016}\natexlab{}.
\newblock \showarticletitle{Adding physical objects to an interactive game improves learning and enjoyment: Evidence from EarthShake}.
\newblock \bibinfo{journal}{\emph{ACM Transactions on Computer-Human Interaction (TOCHI)}} \bibinfo{volume}{23}, \bibinfo{number}{4} (\bibinfo{year}{2016}), \bibinfo{pages}{1--31}.
\newblock


\bibitem[Yuen et~al\mbox{.}(2011)]%
        {yuen2011augmented}
\bibfield{author}{\bibinfo{person}{Steve Chi-Yin Yuen}, \bibinfo{person}{Gallayanee Yaoyuneyong}, {and} \bibinfo{person}{Erik Johnson}.} \bibinfo{year}{2011}\natexlab{}.
\newblock \showarticletitle{Augmented reality: An overview and five directions for AR in education}.
\newblock \bibinfo{journal}{\emph{Journal of Educational Technology Development and Exchange (JETDE)}} \bibinfo{volume}{4}, \bibinfo{number}{1} (\bibinfo{year}{2011}), \bibinfo{pages}{11}.
\newblock


\bibitem[Zhou et~al\mbox{.}(2024)]%
        {zhou2024bee}
\bibfield{author}{\bibinfo{person}{Xiaofei Zhou}, \bibinfo{person}{Yushan Zhou}, \bibinfo{person}{Yunfan Gong}, \bibinfo{person}{Zhenyao Cai}, \bibinfo{person}{Annie Qiu}, \bibinfo{person}{Qinqin Xiao}, \bibinfo{person}{Alissa~N Antle}, {and} \bibinfo{person}{Zhen Bai}.} \bibinfo{year}{2024}\natexlab{}.
\newblock \showarticletitle{" Bee and I need diversity!" Break Filter Bubbles in Recommendation Systems through Embodied AI Learning}. In \bibinfo{booktitle}{\emph{Proceedings of the 23rd Annual ACM Interaction Design and Children Conference}}. \bibinfo{pages}{44--61}.
\newblock


\end{thebibliography}

\appendix
\section{Appendix}
\begin{table*}[]
    \centering
    \caption{Study 1: Evidence that students' embodied interaction with DARK-BRIGHT is effective for learning recommendation output.}
    \label{tab:embodied-data1-1}
    \begin{tabular}{p{0.91\textwidth}}
    \toprule
       Explore the Recommendation System: Recommendation output \& DARK-BRIGHT \\
    \midrule
        \begin{description}
            \item[A1 \& A2] read a light dot with a medium level of intensity: ``I think it's over the middle.''
            \item[A3 \& A4] applied trial and error by rotating the knob from left to right and observing the changes in the final light dot.
            \item[A5 \& A6] shared the thought during the interaction with DARK-BRIGHT: ``It's interesting that you are actually using the shadow of the object as the pointer to the preference prediction.'' Then A5 verbalized the low value in the recommendation output: ``I barely see the light at all, so he probably dislikes that.'' and the medium value: ``It's kind of in the middle - not too bright and not too dark.'' A5 developed further sense-making of how DARK-BRIGHT of the final light dot could be influenced by the two data bars: ``If these two (corresponding) attributes agree with each other, both having ten and ten, they will have very bright light; if they disagree with each other, they will have very dark light in the end. I guess you can call it lights' happiness level.''
            \item[A7 \& A8]: ``The darker shade means zero. It won't make any changes to the end.''
            \item[A9 \& A10] read each prediction result into high, medium, low correctly.
        \end{description} \\
    \bottomrule
    \end{tabular}
\end{table*}

\begin{table*}[]
    \centering
    \caption{Study 1: Evidence that students' embodied interaction with BLOCKAGE is effective for learning data vector values.}
    \label{tab:embodied-data1-2}
    \begin{tabular}{p{0.91\textwidth}}
    \toprule
       Explore the Recommendation System: Modifying values in data vectors \& BLOCKAGE \\
    \midrule
       \begin{description}
           \item[A1 \& A2] A1: ``The light goes through this (the knob). Each time it blocks part of the light.'' A2: ``It (the knob) cuts out the light.'' A1 and A2 further demonstrated their understanding by unmounting the convex lens to turn the one focused light dot into three larger light dots, rotating the knobs from the left side to the right, and observing the changes in the intensity of the specific light dot on the final projection screen. 
           \item[A3 \& A4] rotated the knob to no light: ``it's zero here.''
           \item[A5 \& A6] described the function of the knobs: ``If you rotate it, one way makes it brighter and the other way makes it darker.''
           \item[A7 \& A8] described the function of the knobs: ``You use them to make the light circles brighter or darker.''
           \item[A9 \& A10] explained how the BLOCKAGE of knobs influences the value of the output: ``The smallest number gets ruled out.''
       \end{description} \\
    \bottomrule
    \end{tabular}
\end{table*}

\begin{table*}[]
    \centering
    \caption{Study 1: Evidence that students' embodied interaction with MERGING is effective for learning the addition operation.}
    \label{tab:embodied-data1-3}
    \begin{tabular}{p{0.91\textwidth}}
    \toprule
       Explore the Recommendation System: Addition operation in the dot product \& MERGING \\
    \midrule
       \begin{description}
           \item[A1 \& A2] unmounted and re-mounted the lens to switch between one focused light dot and three larger light dots: ``This (the convex lens) adds the light together.''
           \item[A3 \& A4] mounted the convex lens: ``It put the color together.''
           \item[A5 \& A6] reported: ``I think what this (convex lens) does is to focus all the lights in the end.''
           \item[A7 \& A8] explicitly described the change in the light intensity by merging three light beams: ``It makes the circles all go to the center; it makes the circle brighter'' and ``It piles all the light together''. 
           \item[A9 \& A10] directly elaborated in the context of recommendation systems: ``[Taking out the lens] Oh it (the individual three separated light dots) represents how much the person likes each attribute. [Placing the lens back] Now it combines them and see how much you like it overall.'' 
       \end{description} \\
    \bottomrule
    \end{tabular}
\end{table*}

\begin{table*}[]
    \centering
    \caption{Study 1: Evidence that students' embodied interaction with LINKAGE is effective for learning the corresponding relationships between attributes.}
    \label{tab:embodied-data1-4}
    \begin{tabular}{p{0.91\textwidth}}
    \toprule
       Explore the Recommendation System: Corresponding relationships between values in the algorithm \& LINKAGE \\
    \midrule
       \begin{description}
           \item[A1 \& A2] unmounted the convex lens and traced how all three light beams passed through two data bars and ended up on the final projection screen.
           \item[A3 \& A4] verbalized the AI mechanism: ``As long as the attributes match, changing the order won't change the result.''
           \item[A5 \& A6] explained the change in the recommendation output: ``This person likes salty food and this snack has salt so this person will like this snack... [changing the user vector to Jessy and observing the change in the output] The snack stays the same, so it must be something related to the user; Jessy likes salty food a lot but this snack has high sugar which is totally not the attribute.''
           \item[A7 \& A8]: ``(The light beams) make the connection between user preference and the item attribute really clear.''
           \item[A9 \& A10] illustrated the connections between the corresponding attributes by using their fingers to trace the (invisible) light beams.
       \end{description} \\   
    \bottomrule
    \end{tabular}
\end{table*}

\begin{table*}[]
    \centering
    \caption{Study 1: Evidence that students' embodied interaction with linking metaphors is effective for learning the multiplication operation.}
    \label{tab:embodied-data2-1}
    \begin{tabular}{p{0.91\textwidth}}
    \toprule
       Control the Recommendation System: The multiplication operation in the dot product \& Linking Metaphors\\
    \midrule
       \begin{description}
           \item[A1 \& A2] experimented to unmount the convex lens, rotate individual knobs (PART-WHOLE), and observe the corresponding light dots (LINKAGE and DARK-BRIGHT): ``[Taking out the convex lens; observing that one focused light dot turning into three separated light dots with two of them being dimmer; rotating the corresponding knobs to increase the intensity for the two dimmer light dots.] The scales (of the knobs) are multiplied together---the light goes through this (set of knobs) and then this (set of knobs). Each time it blocks part of the light. This is also like a division---it cuts out part of the light.'' Then they placed the color filter (HUE) in front of each knob to further prove the multiplication operation.
           \item[A3 \& A4] set the corresponding attributes (LINKAGE) to 1 and 0 (PART-WHOLE \& DARK-BRIGHT) iteratively: ``[Setting one knob in the second data vector into 0 and pointing to this knob,] it's zero here. [Rotating the corresponding knob in the first data vector; observing the changes in the focused light dot,] there is no light change in the end because this [pointing to the second set of knobs] is blocking the light.'' Through contrastive learning, they switched the values of 1 and 0 between the two sets of knobs, observed the changes, and realized: ``one of it (a knob set) is depending on the other (knob set). If one of them is high, the other one can make a big difference to the final output.''
           \item[A5 \& A6] experimented with PART-WHOLE, LINKAGE, and DARK-BRIGHT to figure out the multiplication and correct their own misunderstanding. First, they applied a similar strategy as A3 and A4: ``If you turn this one (the first set of knobs) to zero, then this one (the second set) doesn't get any light at all.'' Then A5 generated a hypothesis of the dot product: ``The first set of knobs has more power than the second set because the light hits the first set first.'' Then A5 turned two flashlights off, observed how the only flashlight left contributed to the output, and iterated the previous hypothesis: ``If the second one is zero, the first is not zero, still there will not make any light.
Now I have changed my hypothesis. I think both values need to be larger. Both are the highest value and will be the most preferable.
I guess they work the same. They work together.'' 
With the support of PART-WHOLE and DARK-BRIGHT, A5 understood the direction of the multiplication is decreasing: ``The maximum value is one, so multiplying a value less than one leads to decreasing.''
           \item[A7 \& A8] confirmed their hypothesis about multiplication by ruling out the potential of addition: ``[Setting one knob as one and the corresponding knob as zero] the result is zero. But one plus zero is one. So it's not an addition. It's multiplication.''
           \item[A9 \& A10] Instead of only experimenting with the extreme values such as one and zero, they also experimented with medium values (DARK-BRIGHT) to understand the multiplication operation (PART-WHOLE and LINKAGE) along with color filters (HUE): ``[Placing color filters in front of a pair of knobs with low values], the color doesn't work. The small numbers, after you multiply them together, the less vibrant. The knobs ruled it out.'' 
       \end{description} \\   
    \bottomrule
    \end{tabular}
\end{table*}

\begin{table*}[]
    \centering
    \caption{Study 1: Evidence that students' embodied interaction with linking metaphors is effective for debugging.}
    \label{tab:embodied-data2-2}
    \begin{tabular}{p{0.91\textwidth}}
    \toprule
       Control the Recommendation System: The efficient debugging \& Linking Metaphors\\
    \midrule
       \begin{description}
           \item[A1 \& A2] at the very beginning, A1 set the user vector data bar to represent his own preference for ``Sugar'', ``Salt'', and ``Fat'' and expected a high prediction value of his preference for the chocolate bar; but the focused light dot turned out to be very dim; without any hesitation, A1 quickly identified that he set one of his preference to the wrong direction: ``instead of a teacher telling you, I can figure it out myself.''
           \item[A3 \& A4] A4 rotated the knob representing the preference for ``Fat'' in the user vector data bar to one and placed the yellow filter in front of it; A4 thought the focused light dot would turn into yellow, but it didn't; then A3 quickly rotated the ``Fat'' attribute knob in the item vector data bar to a higher value which resulted in the desired outcome.
           \item[A5 \& A6] rotated the knobs and read the final light dot: ``The snack stays the same, so it must be something related to the user; Jessy likes salty food a lot but this snack has high sugar which is totally not the attribute;'' 
           \item[A7 \& A8] made one change and observed no change in the final output, and then they quickly realized that it's the last attribute that should be changed to make the final result change.
       \end{description} \\   
    \bottomrule
    \end{tabular}
\end{table*}

\begin{table*}[]
    \centering
    \caption{Study 1: Evidence that students' embodied interaction with linking metaphors supports learning the dot product.}
    \label{tab:embodied-data2-3}
    \begin{tabular}{p{0.91\textwidth}}
    \toprule
       Control the Recommendation System: Exploration, experimentation, and explanation \& Linking Metaphors\\
    \midrule
       \begin{description}
           \item[A1 \& A2] unmounted the convex lens, placed a color filter, and read the color of the three unfocused final light dots to prove that the light passing through two sets of knobs is multiplication.
           \item[A3 \& A4] first placed one color filter to identify the attribute that has the highest impact on the final result; then they rotated the knob behind the color filter to turn the light brighter and dimmer to show how this light beam is contributing to the final light dot: ``If you turn it higher, it will work better.'' They further tried two hues for the explanation by placing the red color filter in front of a high-impact attribute and the blue color filter in front of a low-impact attribute and observed the color of the final light dot; then they placed three different color filters in front of three flashlights and observed the final light dot to prove that all the light beams were put together.
           \item[A5 \& A6] stacked and placed all filters together in front of a light to test if it made the final light black; rotated the knots to make the lights dimmer, middle, and brighter one by one: ``if the knob was 0 then it made no light''
           \item[A7 \& A8]: ``The darker shades mean zero, so the color won't make any changes to the end.'' 
           \item[A9 \& A10]: ``[the left and the right knobs being set to zero,] now the color can be only added to the middle one.'' 
       \end{description} \\   
    \bottomrule
    \end{tabular}
\end{table*}

\begin{table*}[]
    \centering
    \caption{Study 2: Students’ exploration with Briteller (B1-B7).}
    \label{tab:embodied2-data1}
    \begin{tabular}{p{0.91\textwidth}}
    \toprule
       Evidence for exploration, experimentation, and explanation with Linking Metaphors\\
    \midrule
       \begin{description}
           \item[B1 \& B2] tried different values to observe the final light dot. B1 and B2 set the first vector as (1, 0.3, 0) and the vector as (1, 1, 1). B2 placed a pink color filter in front of the first data bar valued at 1, and B1 and B2 noticed that the final light dot became pink. But when they placed the color filter in front of the 0.3 and 0, they noticed the final light dot didn't change. When the final light dot became pink, B2 said, ``Because this one is set to 1.'' B2 concluded that the knob value was ``0'', so that's why the color filter didn't work. But they didn't realize it was multiplication.
           \item[B3 \& B4] tried to rotate the knob values randomly and said, ``It's lighter and darker.''B4 mounted and unmounted the convex lens and then said, ``These two going through them. Combine together.'' B4 used ``add together'' to describe the math operation involving the convex lens. B3 and B4 tried to adjust the three light dots to light, medium, and bright levels. B3 and B4 put the color filter in front of the 0.3 value knob and said, ``It's lighter'', noticing the color difference. Then B3 and B4 placed a color filter in front of the 0 value knob, and the final light dot color didn't change. B3 explained, ``Because the light isn't going through it, the value is 0.'' B3 said it was a subtraction. Then B3 rotated one knob to 0 and another to 0.5, then he realized it was multiplication. B4 was silent.
           \item[B5 \& B6] rotated the knobs to the brightest and observed the change of the light dot. B6 pointed out the link between the brightness and the knob values. B6 mounted and unmounted the convex lens to check the effect of the lens, he validated the effect of light dots from 3 to 1. Beyond noticing that the convex lens combined all the light dots, B6 identified a ``crossover'' relationship. They tried to rotate one knob value and found that the middle light beam remained straight while the left and right beams were swapped by the convex lens on the final screen. B5 and B6 observed the result of placing a color filter at the back of a knob of 0. When trying color filters, B6 realized the color behind of knobs didn't show up because the value of the knob was 0. B6 identified the operation as multiplication, and he said, ``The intensity would be multiplied.'' B5 was silent.
           \item[B7] tried to rotate the knob and concluded, ``I think 1 is the brightest, and 0 is the darkest.''. B7 correctly set the given user's attributes and interpreted the brightness as the prediction result correctly. B7 observed the convex lens and said, ``It brings the dark light to the paper and makes it brighter. Makes it center. Add them all.'' B7 used the color filter to explore, then he pointed to one knob and said, ``This one is dimmer, less light to the white paper'', and he pointed at the final screen. He found the brighter one would contribute more to the final screen. B7 initially thought the operation between two knobs was subtraction. After experimentation by rotating the knobs, B7 pointed to the two knobs and said, ``it is multiplication. Because this (knob) is 0.5, so less light.'' He explained the first knob was 1, and the second was 0.5, so the final light dot was a little bit dimmer. 
       \end{description} \\   
    \bottomrule
    \end{tabular}
\end{table*}

\begin{table*}[]
    \centering
    \caption{Study 2: Students’ exploration with Briteller (B8-B9).}
    \label{tab:embodied2-data2}
    \begin{tabular}{p{0.91\textwidth}}
    \toprule
       Evidence for exploration, experimentation, and explanation with Linking Metaphors\\
    \midrule
       \begin{description}
           \item[B8] rotated all the knob values to 0 and said, ``If we put this one to 0, you can’t see it''. B8 pointed to the first knob data bar, then the second data bar, describing the process of light passing through. He elaborated by pointing at the knobs, stating, ``It is rejecting the light.''. B8 used ``open it up for a little bit'' to describe the behavior of raising the second knob's value. He then rotated all the values to 1, demonstrating how he could maximize the brightness of the final light dot. He pointed at the flashlight, showing this was the light source, and he pointed at the first data bar and the second data bar, ``This is the measurement of light.'' B8 correctly set the attributes and interpreted the recommendation results, and he found the light dot was quite bright. He explained, ``How I did this was I used different brightness for each one. Everything was different on their own.'' And he finished the task by rotating the sugar's knob value to 0, and he pointed at the light dot, and explained, ``It means she likes sugar because she put sugar as 1. If I lower the sugar to 0, she probably doesn't like sugar anymore. It's getting darker.'' After placing the color filter in front of different knobs, B8 quickly discerned the system's inner mechanics. He pointed to the knob with a higher value and remarked, ``Because this one added more reflection from the blue''. B8 placed the blue color filter in front of the knob with one hand while rotating the knob value with the other. He explained, ``If you can see it [blue light dot] at 1, if 0.5, you can see less, and 0, we can't see it at all. Because it has less light going to reflect on this [color filter]''. B8 then realized that although the left-side knob value was rotated to 1 when he placed the color filter in front of it, the final light dot did not change. After further exploration, he understood this was because the second knob value was rotated to 0: ``Because this one is 0 [pointed to the knob] even if this one is 1, the light won't reflect to hit it''. He then demonstrated this concept using another pair of knob data bars, rotating the first knob value to 1 and the second to 0. He explained to the researcher, ``We can't see the blue. If we increase this to 1 [adjusted the second knob to 1], we can see the blue''. B8 said the operation between two knob values was the addition. B8 said, ``These two had added on each and made a brighter. Added on track.'' B8 was confused here. But then he rotated one knob value to 0.5, and another value to 0, and he realized the final light dot was gone so the operation was not an addition. B8 finally concluded that rotating one of the knob values to 0 would result in a final value of 0, realizing that this was due to multiplication: ``Because 1 * 0, 0 times anything is 0, so 0 times 0.5 is going to be 0.'' To confirm his understanding, he turned on another flashlight and rotated some values to 0, further proving his idea.
           \item[B9] rotated all the values to 1 to maximize the brightness and said, ``I mean, one is the highest, so I thought I might be able to see the most amount of light from each one.'' B9 pointed at the knobs, then the convex lens and white paper, saying, ``It converts the brightness from here all the way down to there.'' B9 interpreted the result as ``Nia likes cheesecake'' and explained, ``Because if she didn't, the light would've been dimmer. And since it's obviously visible and pretty light, you could tell that she likes cheesecake.'' B9 changed the user's sugar value to 0 to change the prediction to ``dislike''. He explained, ``Changing her sugar likeness down to here, which made a very big change. Well, not that much of a change, but you can see the light is smaller and it's way less; I mean, not way less, but it's dimmer. That means she doesn't really like cheesecake that much.'' B9 stood up and pointed at the final screen and said, ``I know because sugar is the most out of salt and fat. So that takes them more. So this has the highest number; it must have had the most help in making it the lightest since it had the most.'' B9 thought the operation between two knobs was subtraction, and he rotated the knobs and concluded, ``When I do this, I'm taking away from my light. When I do this, I'm adding to.''After experimentation, B9 said, ``Oh wait, is it like multiplication? It's multiplication because you're doubling it between these two.'' Then he changed the corresponding values to 0 and 0.5, ``One times 0.5, point five''. Then he adjusted one number as 1 and found it was brighter. But he made the mistake, of saying, ``One times one will equal to 2''.
       \end{description} \\   
    \bottomrule
    \end{tabular}
\end{table*}

\begin{table*}[]
    \centering
    \caption{Study 2: Students’ exploration with Briteller (B10).}
    \label{tab:embodied2-data3}
    \begin{tabular}{p{0.91\textwidth}}
    \toprule
       Evidence for exploration, experimentation, and explanation with Linking Metaphors\\
    \midrule
       \begin{description}
           \item[B10] rotated all the knobs to 1 to maximize the brightness. He used his hand to point at the different parts, including the flashlight, knobs, the convex lens, and white paper, ``I turned on the flashlight, I adjusted this to one, and then went through here and then it came here and here.'' He correctly interpreted the brightness as prediction results and finished the task by rotating the sugar's value to 0. B10 mentioned he did that before because he liked to use flashlights to do different things at home. B10 unmounted and held the convex lens, and moved it to show that the light dot would also move with it. Then he mentioned the 3 knobs gave different light beams but the lens made only one dot finally. ``So you see because three flashlights and there's three different types of lights that put over here. You can see it all either, but if you just have it (convex lens) sitting down so I can get it. Yeah, it's just one light. All those lights are just coming into one''. B10 rotated the first data bar of knobs as (0.5, 1, 0) second data bar of knobs as (1, 1, 1), and he used the color filter and observed the final light dot. and he found when he placed the color filter in front of the 0, nothing changed, he explained, ``Because if I put this as 0, nothing there (pointed at the final screen).'' And he said the operation between two knobs was subtraction. He rotated different values and insisted it was subtraction, ``It's still subtraction. It's taking it away.'' He then tried another pair of values (0 and 1), and said, ``0 and 1 will get 0 by subtraction''. Then he guessed it was an addition. B10 didn't understand it was multiplication.
       \end{description} \\   
    \bottomrule
    \end{tabular}
\end{table*}

\begin{table*}[]
    \centering
    \caption{Study 2: Students' explorations with AR-enhanced Briteller for understanding dot product.}
    \label{tab:ar-data1-1}
    \begin{tabular}{p{0.91\textwidth}}
    \toprule
       Evidence for that visible numerical values in AR support students' understanding of the dot product\\
    \midrule
       \begin{description}
           \item[B1 \& B2] randomly slid the values. B1 and B2 set the first data bar of knobs to (1, 1, 1) and the second to (0, 1, 1), with B1 explaining the final numerical result: ``0, 1, and 0, added together equals 1.''
           \item[B3] dragged the slider to change the value for individual knobs. He explained why the final result value was 2. 
           \item[B4] pressed two flashlights to turn off the light, keeping only one. B4 understood it was multiplication and did the calculation, ``1 times 1 is one.'' He tried different values, saying, ``Because 1 times 0.5 is 0.5.''
           \item[B5] tried different values and slid the value slider of the first data bar of knobs to adjust values to (1, 0.5, 1) and the second to (1, 0, 1). She found the final result was 2 and explained step by step: ``One multiply one is 1. And the second row of light is anything times 0 is 0. And the third row, then one. So then it's multiplication, then addition. So you add those two, which makes it two.''
           \item[B6] slid value sliders to adjust all the values to 1 and he found the final numerical result was 3. Then he slid the value slider to adjust one knob value to 0 and noticed the final numerical result was changed to 2, and he explained, ``Zero times one will get zero. Okay, so multiply first and then the adding.''
           \item[B7] tried to slide value sliders to  change the values and found the final result was 1.5 and he explained, ``because 1 and 1 equals 1, 1 times 0.5 equals 0.5, so it is 1.5." B7 pointed to one knob and connected it with light, "Now it's 0, it's only dark involved here.'' 
           \item[B8] viewed the light beam from different angles, and noticed the light change. B8 only pressed one flashlight to turn on the light and set one knob value as 1 and another one as 0.2, and explained, ``1 timed 0.2 is 0.2'' B8 just tried to set values randomly and got 2. Then he explained, ``Because 1 times 1 all equals to 1, and then plus the addition, then you adding, through this (pointed at convex len) this is going to be 2''. B8 slid all value sliders to change all the values to one, ``To get three, I just want to put all them to 1''. 
           \item[B9] viewed the light beam from different angles and noticed the light change. B9 tried different knob values, and he set one as 0.5 and another as 0.5, then he observed the final screen number, and he realized, ``Oh, because 0.5 times 0.5 is 0.25''.
           \item[B10] was confused at first and just clicked on the screen randomly to get familiar with it. Then he tried to slide all value sliders to adjust all the values to 1.
       \end{description} \\   
    \bottomrule
    \end{tabular}
\end{table*}

\begin{table*}[]
    \centering
    \caption{Study 2: Students read and control food recommendation in AR-enhanced Briteller (B1-B5).}
    \label{tab:ar-data2-1}
    \begin{tabular}{p{0.91\textwidth}}
    \toprule
       Evidence for that item visibility contextualized the light-based tangible system\\
    \midrule
       \begin{description}
           \item[B1 \& B2] noticed the button's differences and said, ``Sugar, salt, fat.'' B1 slid the value sliders to input her flavor preferences and observed the final result of the screen, saying ``Lollipop.'' B1 said she likes lollipop. After sliding value sliders to input the given user's user vector data bar, B1 observed the item image firstly and immediately said ``Cake.'' B2 noticed there were some invisible items, and she explained, ``Because she eats those the least.'' After understanding the task, B1 immediately slid value sliders to lower Nia's sugar preference knob value (0.9 -> 0.2), increased the salt knob value (0.5 -> 0.6), and increased the fat knob value (0.5 -> 0.7). After observing the changed result, with French fries as the top recommendation, they laughed and said, ``It's not sweet anymore.''
           \item[B3] slid the value sliders to adjust values to input his own preferences, and observed a lollipop showing on the screen, B3 showed his interpretation that it was based on the feature of the food, i.e., the sugar, and he pointed that the system knew he liked the food. After clicking ``next'', B3 said he noticed the light was changing, ``It got darker.'' B3 noticed the lettuce was invisible, and interpreted, ``Because it has it has zero fat and low sugar and salt. Because I didn't put that in the thing, so it knows I don't like it.'' B3 quickly finished the task by lowering the user's data bar sugar value. He explained, ``It takes information from the salt, fat, and sugar. And recommends you stuff.''
           \item[B4] observed the buttons changed and named them ``sugar, salt, fat.'' B4 slid the sugar value sliders to adjust the value to 0.9 and said he liked sweet food. After sliding the value sliders to adjust values for his user vector data bar for the first time, he clicked ``recommend'' to generate recommendations and saw the fries. He said he wanted to switch a little bit. When reviewing the recommended food list, B4 commented on his real preferences for the foods. For example, he said he liked steak and pizza, using yes or no to give his feedback. B4 interpreted the steak's item vector (0.2, 0.6, 0.8) and said, ``There's less sugar. There's some fat and some salt and a lot of fat.'' He added, ``And that doesn't really add up because I set some fat, a lot of salt.'' When B4 saw the lettuce disappear in the final projection, he first asked why and then realized it was because the value was so low. He understood the result, saying, ``Keep decreasing.'' B4 asked the researcher, ``I mean this is the ones I don't like, right? When they keep going low.'' B4 found he couldn't see the broccoli, and he said, ``It's low. There's a small amount of sugar in it and a small amount of salt. It has no fat. Yeah. And my recommendation was a little bit of sugar, a lot of salt, and some fat. It didn't add up.'' He added, ``That's right, I don't like that.'' B4 said the machine could kind of read his mind. And B4 also quickly finished the task. Then he noticed the recommendation list changed and said it was because the sugar value is changed.
           \item[B5] slid the value sliders to input her own preference and clicked to navigate between different items using the AR to see the food recommended to her. She noticed that the visibility of the items was changing. When she observed some dimmer food, she immediately said, ``Those were the ones that aren't recommended.'' Then, she looked at her user vector data bar and the pizza's item vector data bar and observed, ``So that side doesn't contain much sugar.'' Curious, she slid the value sliders to adjust her user vector data bar, increasing her sugar value to see what would change. She noticed the food order shifted and explained, ``You change your flavor. You're more craving for sugary food right now.'' B5 understood what the values of the knobs meant for Nia and her food preferences, as well as the prediction results that recommended certain foods. When doing the task, B5 said she could change her preference toward sugar. B5 used the dot product to explain, ``If we switched up the, I'm sorry, the decimals. So if it was 1.5 and then 2.9, that would be the most recommended.'' And she pointed at the AR’s item picture, and she added, ``The recommendation of the person's preference.''
       \end{description} \\   
    \bottomrule
    \end{tabular}
\end{table*}

\begin{table*}[]
    \centering
    \caption{Study 2: Students read and control food recommendation in AR-enhanced Briteller (B6-B10).}
    \label{tab:ar-data2-2}
    \begin{tabular}{p{0.91\textwidth}}
    \toprule
       Evidence for that item visibility contextualized the light-based tangible system\\
    \midrule
       \begin{description}
           \item[B6] slid the value sliders to input his own preferences and said, ``So it's just based on my preference.'' After clicking ``recommendation''  and observing the result, he noticed the item vector data bar values changed and remarked, ``Oh, I just noticed they're changing based on the item. Better let they recommend.'' He also noticed the lights, saying, ``Some are dimmer. Some are brighter.'' He quickly clicked the button to navigate between different items, observed the numerical values, and said, ``Percentages based on. Less recommended, higher recommended. I'm just thinking this is a higher percentage of what this will get recommended to me based on the values I've put in.'' B6 continued using the AR to navigate between items and noted, ``Yeah, because as the numbers start to get low and it starts disappearing, that means it's basically not recommended based on what I've put in.'' B6 noticed the broccoli was invisible, and he explained using the dot product, ``Because it's low value it's multiply by. Multiplied by the item. Low preference.'' He continued to explain, ``If you can see the item, that basically means it's a high recommendation based on the, I know it but I can't really put it to words. It tried to match it.'' When asked to do the task, he said, ``So we just changed it to less sugar.'' After clicking the recommendation button, he said, ``I just changed the recommendation. I just only changed it down from the sugar, that's why.'' B6 compared the original recommendation order to the current order and noticed the first food was no longer cake. He commented, ``So it moved it down based on the value. So less sugar, it moved down and changed the order of recommendations.'' He interpreted that the reason the order of food changed was due to adjusting Nia's preference to less sugar. Finally, B6 explained how the prediction was generated: ``User vector means preference. Values. And then the AI recommends it similar to the value I've put in.''
           \item[B7] slid the value sliders to input his own preference values and commented on whether he liked the items recommended to him. When asked why those foods were recommended to him first, he pointed to his user vector values and explained, ``Because I had put a lot of sugar in it, and I put not too much salt, and I put a lot of fat.'' He observed the final screen changing and said, ``It's becoming lower.'' B7 noticed the broccoli was invisible and explained, ``Because it's a vegetable and it has less sugar, and mine has a lot of sugar. So this one like multiplies and makes it lower. And the sugar, you don't have a lot of sugar in broccoli. That's why the one is the lowest.'' B7 quickly finished the task and explained the reason using attribute values.
           \item[B8] clicked ``recommend'' and saw French fries, saying, ``I love fries.'' As he went through the recommendations, he commented on whether he liked each item. When he noticed he couldn't see the broccoli, B8 said, ``Probably it is not recommended. This is like the last thing recommended.'' He explained, ``Because there is no sugar or salt, and fat,'' which is why the broccoli was not recommended to him. B8 mentioned that he likes Coke but noticed it was a little bit dimmer. He explained, ``Because my lightness is dimmer. If I turn this light... (then he adjusted the sugar and salt to 1) This is how the Coke will be recommended.'' When he noticed that the results were different from the given user's, B8 said, ``Because she likes non-fat food, I like a little bit more fatty food.'' B8 noticed the broccoli was invisible, then picked up the broccoli card and said, ``It's 0.5 and 0.5, and my values are 1 and 1. I have 0 fat, so maybe that may be why. But those two are 0.5, 0.5. And 1, 1, so they probably have greater value.''
           \item[B9] went through the recommendation list, commenting on whether he liked each item. When he saw the egg was invisible on the screen, he said, ``Oh, because they wouldn't recommend that for me.'' Then, when he saw the lettuce and broccoli were also invisible, he noted they wouldn't be recommended either. B9 explained why the broccoli was not recommended to him: ``Because it really doesn't match my levels. I got 0.7 sugar, 0.5 salt, and 0.3 fat. Broccoli has 0.1 sugar, 0.1 salt, and 0 fat. They are low values while mine are high.'' He navigated between different items and observed the light was changing. He realized that the more visible items meant those things would be recommended to him, and he'd like them more. ``For the other stuff, they wouldn't recommend to me because they think I don't like it. The brighter, the more I like it.'' ``It's getting darker. Because you may not like it,'' B9 observed. When comparing the results, B9 explained, ``We both like different things. The measurement we put is different. The recommendation system looked at what we like.'' He explained how the light-based recommendation system worked, turning on the flashlight and pointing at the knobs: ``The more you like, the higher the number goes. If I like sugar, I put sugar as 1. If I hate fat, put that as 0.'' Then, he slid the value sliders to adjust the knob values accordingly.
           \item[B10] \  ~ \ ~ \
       \end{description} \\   
    \bottomrule
    \end{tabular}
\end{table*}

\begin{table*}[]
    \centering
    \caption{Study 2: Students expand attributes for better recommendation (B1-B5).}
    \label{tab:ar-data3-1}
    \begin{tabular}{p{0.91\textwidth}}
    \toprule
       Evidence for that allowing students to expand attributes in AR-enhanced Briteller increased their understanding of data representation\\
    \midrule
       \begin{description}
           \item[B1 \& B2] After reading the David task card, B2 said he probably likes ``like fruit.'' B2 tried to slide the value sliders to set David's user vector data bar values to (0.4, 0.3, 0.4), which was a lower value. When thinking about Apple's other attributes, B2 said ``Sour.'' B2 added a ``sour'' attribute. B2 said, ``Adding more interests can make the recommendation more accurate.''
           \item[B3] explained the incomplete item and said, ``Because that's not right.'' He designed one new attribute, ``Nutrition.'' After adding one more flashlight to expand data bars, pressing data bar labels to change the name of the attribute to ``Nutrition'', and sliding the value sliders to adjust the values, he realized, ``The reason to add an attribute was that the system knows more now.'' He mentioned that the information the recommendation system wanted to know was about ``what I eat. What I do.'' He explained that the user's preferences would be sent to the machine so it knows what the user likes and recommends accordingly. B3 said, ``So to know what you like and recommend it to you.''
           \item[B4] After reading David's description, B4 said, ``He wants to stay healthy, so,'' and then he designed David's user vector. B4 listed a few new attributes, ``Taste, nutrition.'' B4 added one more flashlight to expand data bars and pressed data bar labels to change the name of the attribute to ``Protein''. Then B4 realized that apples contained a lot of vitamins and said he wanted to switch the attribute name to ``Vitamins''. 
           \item[B5] read David's description and she said, ``Ok, first, let's first. So it's less sugar, more fat.'' And she slid the value sliders to adjust David's user vector data bar values as (0, 0.3, 0.7), and she explained, ``Maybe list probably up 0.3 in salts and fat leading to protein.'' B5 was confused here and it was difficult for her to come up with new attributes. Then she said, ``Taste, vitamin''. B5 added one more flashlight to expand data bars and pressed data bar labels to change the name of the attribute to ``Vitamin'' to complete the task. 
      \end{description} \\
    \bottomrule
    \end{tabular}
\end{table*}

\begin{table*}[]
    \centering
    \caption{Study 2: Students expand attributes for better recommendation (B6-B10).}
    \label{tab:ar-data3-2}
    \begin{tabular}{p{0.91\textwidth}}
   \toprule
       Evidence for that allowing students to expand attributes in AR-enhanced Briteller supported their understanding of data representation\\
    \midrule
       \begin{description}
           \item[B6] added one flashlight to expand data bars, and B6 listed several things people cared about for apples ``calories could be one. Ingredients. It could be. Vitamins.'' and he said, ``Could just put in vitamin. Could that work?'' And he pressed data bar labels to change the name of the attribute to ``Vitamin'' and he slid the value slider to set David's user vector data bar value as 0.7 and the Apple item vector data bar value as 1 to finish the task. B6 said, ``You will consider other things other than salt, sugar, fat. By adding other attributes or values can make the recommendation more accurate.'' And B6 drew a parallel with TikTok’s data filtering, explaining that TikTok asks for a user’s age to filter content: ``It filters out if users are underage or show similar content by clicking `like'.'' He explained, ``It basically shows that you like the content or similar to it. So it would try to show something similar more often than the other. So they actually are getting your attribute and build the vector.'' He explained, ``The vector can describe yourself, your preference, your likes, and what you're into.''
           \item[B7] said he thought David would like broccoli, but it was not fully complete because there were no other parts (attributes) here. B7 said, ``Because more parts in broccoli than sugar, salt, fat. It means it didn't recommend enough stuff to make the broccoli full. It's like you didn't give low fat, or fiber.'' B7 added that it was inaccurate without adding more flashlights, explaining, ``It's inaccurate because there are more than 3 parts for the thing.'' He explained, ``For example, maybe 10 parts, but there are only 3 here, so it is not fully complete to represent the broccoli.'' B7 added some new flashlights to expand data bars and pressed data bar labels to change the names of the attributes to ``low salt'', ``low fat'', and ``fiber''.
           \item[B8] designed good attribute values for broccoli and explained, ``Broccoli is a healthy vegetable. So it probably has low sugar, could be good for you, it probably has very low salt, and low fat. And it has a lot of vitamins and is healthy and fresh. And it's low cost.'' He also added a new flashlight to expand data bars and pressed data bar labels to change the name of the attribute to ``color green'' . B8 then said the recommendation was right since he had added those new attributes. But after a moment, he corrected himself, saying it wasn't complete yet since he hadn't added all the features, and asked if he could add more flashlights.
           \item[B9] added some new flashlights to expand data bars and pressed data bar labels to change the name of the attribute to ``caffeine'', ``sugar'', and ``water'', because he wanted the system to recommend caffeine to him. B9 thought David was a healthy guy and wanted to eat healthy food. He designed the attribute values accordingly and added a new attribute—caffeine—saying, ``Something to boost him up.'' B9 explained why he designed those attributes: ``Low sugar, he doesn't like sugar, so it should be more.'' When he noticed the left side and right side were not matching, B9 said, ``It means David likes broccoli. The recommendation system is wrong.'' B9 then added that if more attributes were included, the light for broccoli would expand, while the recommendation system added more things, which turned out the user would like the stuff. 
           \item[B10] tried to slide the value sliders to adjust the current values but found that didn't work. Then B10 added a new flashlight to expand data bars and pressed data bar labels to change the name of the attribute to ``Vitamin''.
       \end{description} \\
    \bottomrule
    \end{tabular}
\end{table*}

\end{document}